\documentclass[12pt,a4paper]{article}
\pdfoutput=1
\usepackage{dsfont}
\usepackage{slashed}
\usepackage{amsmath,amssymb,amscd,amsfonts,mathtools}

\DeclareMathOperator{\Tr}{Tr}

\usepackage[toc,page]{appendix}
\usepackage{epsfig}
\usepackage{epstopdf}
\usepackage{latexsym}
\usepackage{graphicx}
\usepackage{placeins}
\usepackage{floatrow}
\usepackage{subfig}
\usepackage{caption}
\usepackage{booktabs}
\usepackage{bbm}
\usepackage{color}
\usepackage{physics}
\usepackage{tensor}
\usepackage{tikz}
\usetikzlibrary{matrix}
\usetikzlibrary{decorations.markings,calc,shapes,decorations.pathmorphing,patterns,decorations.pathreplacing}
\usetikzlibrary{positioning}

\makeatletter
\def\@fpheader{\relax}
\makeatother

\def\E{{\cal E}}
\def\F{{\cal F}}

\def\N{{\cal N}}

\def\S{{\cal S}}
\def\A{{\cal A}}

\def\J{{\cal J}}
\def\F{{\cal F}}

\def\D{{\cal D}}
\def\H{{\cal H}}
\def\I{{\cal I}}
\def\S{{\cal S}}

\def\bv{\bar v}
\def\bu{\bar u}
\def\bx{\bar x}
\def\by{\bar y}

\def\be{\begin{equation}}
\def\ee{\end{equation}}
\def\ba{\begin{split}}
\def\ea{\end{split}}
\def\beq{\begin{eqnarray}}
\def\eeq{\end{eqnarray}}

\def\D{\mathcal D}
\def\E{\mathcal E}
\def\M{\mathcal M}
\def\O{\mathcal O}

\begin{document}

\begin{titlepage}

\bigskip
\bigskip
\bigskip
\centerline{\Large \bf Wormholes \& Holography: An Introduction}
\bigskip
\bigskip
\centerline{\bf Arnab Kundu}
\bigskip
\bigskip
\centerline{Theory Division, Saha Institute of Nuclear Physics,}
\centerline{1/AF, Bidhannagar, Kolkata 700064, India.}
\bigskip
\centerline{Homi Bhaba National Institute,}
\centerline{Training School Complex, Anushaktinagar, Mumbai 400094, India.}
\bigskip
\bigskip
\bigskip
\centerline{arnab.kundu@saha.ac.in}
\bigskip
\bigskip
\bigskip

\begin{abstract}

\noindent Wormholes are intriguing classical solutions in General Relativity, that have fascinated theoretical physicists for decades. In recent years, especially in Holography, gravitational Wormhole geometries have found a
new life in many theoretical ideas related to quantum aspects of gravity. These ideas primarily revolve around aspects of quantum entanglement and quantum information in (semi-classical) gravity. This is an introductory and pedagogical review of Wormholes and their recent applications in Gauge-Gravity duality and related ideas.

\end{abstract}

\newpage


\end{titlepage}
\tableofcontents

\section{Introduction}

General Relativity allows for warping of the spacetime. This key feature widely opens up a plethora of rather interesting geometries with curious properties. Of these, Black Holes are an extremely interesting and ubiquitous class of geometries, that has recently been directly detected by the Event-Horizon Telescope experiments\cite{EventHorizonTelescope:2021bee, EventHorizonTelescope:2021srq}, as well as by the gravitational wave based experiments\cite{LIGOScientific:2016aoc}. From early theoretical studies of Black Holes, specially by Einstein and Rosen in \cite{Einstein:1935tc}, it was suggestive that a special geometric structure that connects to asymptotic regions can exist for Black Holes, and beyond.

In \cite{Misner:1957mt}, such geometries structures were termed as ``Wormhole". Since then, such geometries have been a constant source of inspiration and imagination, both in science and science-fiction. In particular, since Wormholes connect to two (or more) asymptotic geometries by a ``throat region", it has long inspired extremely fast travel across remarkably large distances of the Universe. However, upon further scrutiny, distinctions can be drawn between Wormholes that generally tend to be either unstable for such travels or need to be supported by some exotic matter field for them to be humanly traversable or the traversable ones that can supported by standard matter field but do not provide the shortest path between two points. Nonetheless, these geometries bring together foundational concepts in theoretical physics {\it e.g.}~causality, locality, chronology-protection and so on, and helps us sharpen them further. This is a good point to refer the Reader to other reviews on Wormholes from complementary and different perspectives in {\it e.g.}~\cite{Hebecker:2018ofv, VanRiet:2020pcn}.

These ideas and the corresponding technical history of the subject is rather long, which we do not intend to visit here. Instead, in this review, we will briefly touch upon a range of recent ideas, along with basic technical discussions, where Wormholes play a crucial role. All of these recent advancements are based on the framework of Holography\footnote{Early ideas of Holography were proposed in \cite{Stephens:1993an, Susskind:1994vu}. This duality takes a particularly sharp and precise form in \cite{Maldacena:1997re}, which is known as the AdS/CFT correspondence.} which posits an equivalence between a quantum-gravitational system in $(d+1)$-dimensions with a Quantum Field Theory (QFT) in $d$-dimensions. Typically, at least in the well-understood examples, the $d$-dimensional QFT is defined at the asymptotic boundary of the $(d+1)$-dimensional quantum-gravitational description. In a semi-classical limit, in which the quantum-gravitational system can be approximated by a classical geometry with quantum fields propagating inside it, a putative Wormhole can connect otherwise disjoint asymptotic regions of the spacetime. In the Holographic dual description, this implies a highly non-local interaction between two identical QFTs, which are {\it a priori} defined at two disjoint asymptotia.

A crucial aspect of the modern perspective on quantum gravitational dynamics comes from a (quantum) information theoretic framework. Within AdS/CFT, this idea essentially stems from the so-called Ryu-Takayanagi proposal\cite{Ryu:2006bv, Ryu:2006ef}, in which a sharp statement was made connecting a geometric object in the bulk gravitational description to an inherently quantum mechanical concept of entanglement in the boundary gauge theory. Not only this idea connects quantum entanglement with the structure (emergence) of spacetime\footnote{For a recent set of such ideas, referred to as the ER=EPR conjecture, see \cite{Maldacena:2013xja}. For more general aspect of this connection and recent progress, see {\it e.g.}~\cite{Harlow:2016vwg, Harlow:2018fse}.}, it allows us to extract fine-grained physical observables of the strongly coupled gauge-dynamics at the boundary, by performing entirely geometric computations. Wormholes play remarkably crucial roles in Holography, specially from the quantum information theoretic perspective.

Particularly, in the context of quantum dynamics of Black Holes in AdS, extracting fine-grained physics is an extremely important problem. From an entropic perspective, a radiating Black Hole keeps emitting Hawking quanta which keeps unboundedly increasing the corresponding entanglement entropy of the radiation. This is in stark contrast with an upper bound of the total entropy of the Black Hole, which is given by the famous Bekenstein-Hawking formula\cite{Bekenstein:1973ur, Hawking:1975vcx}. In a nutshell, this is the essence of the (in)-famous information paradox\footnote{For more detailed account of this, see {\it e.g.}~\cite{Mathur:2009hf}.} that a consistent theory of quantum gravity is expected to resolve.

It is expected that a fine-grained notion of entropy is able to probe deeper into this dynamics. In fact, as argued by Page in \cite{Page:1993wv}, entanglement entropy is capable of capturing such fine-grained physics: For any unitary dynamics entanglement entropy can increase with time only up to a point --- known as the Page time --- before it begins decreasing again. A similar physics has long been desired for the quantum dynamics of Black Holes. In a Holographic context, on one hand, this would clarify how quantum information is encoded in the quantum regime of gravity and equivalently in the quantum dynamics of the dual strong gauge-dynamics; on the other, it will shed light on the infamous Black Hole information paradox. Recent progress in \cite{Almheiri:2018xdw, Hayden:2018khn, Penington:2019npb, Almheiri:2019psf, Almheiri:2019hni} has precisely extracted the desired Page-curve time-dependence of entanglement entropy, as a fine-grained entropy, of the Black Hole radiation degrees of freedom.

The above-mentioned models involve $2$-dimensional (quantum) gravity, in which much of the explicit calculations are under analytic control. Wormholes play an explicit and important role in producing the Page-curve dynamics for such models. Perhaps more broadly, these Wormholes {\it generalize} the Ryu-Takayanagi prescription to include bulk quantum corrections to a generic class of fine-grained entropies, known as the Renyi-entropy. To construct a specific state ({\it i.e.}~the corresponding density matrix) in any quantum system, one requires the knowledge of all Renyi-entropies. Thus, Wormholes emerge as an integral aspect in such state-construction.

In keeping with the theme, generic (Euclidean) Wormhole geometries in AdS encode multi-partite entanglement properties in quantum gravitational states. Geometrically, such Wormholes asymptote to multiple conformal boundaries on which the dual CFTs are defined. Furthermore, upon carefully introducing a direct coupling between these copies of CFTs, geometrically, renders the Wormholes traversable. In turn, this traversability can be viewed as quantum teleportation protocols. For recent progress in connections between Holography and quantum simulators, we refer the Reader to \cite{Lata}.

We simply want to impress upon the fact that Wormholes have found a wide range of interesting applications, in the context of Holography, and deserve further attention. In this article, we do not intend to review each such application in detail. This is largely because it is a highly evolving field of active research and each such topic deserves a review of its own. Rather, we will focus on the basic features of Wormhole geometries as saddles of Einstein-gravity and construction of such Wormholes in both Euclidean and Lorentzian frameworks. In parallel, we will review key aspects of various applications of Wormhole geometries in the contexts mentioned above. We hope that this review will serve as a bridge between earlier literature on Wormhole geometries in gravity and future work in quantum gravity from a quantum information theoretic perspective. We will review technical aspects of Wormholes in detail, as well as quantitatively motivate new ideas in quantum gravity that makes explicit use of them.

Before concluding this section let us note that there are three broad categories of Wormholes, which we will review in this article. First, spacetime Wormholes: these Wormholes lead to effective non-local interaction between two asymptotic regions of the spacetime and affect a naive factorization property of a quantum field theory which is dual to the geometric structure. Secondly, Einstein-Rosen bridges and their generalizations: these Wormholes appear whenever there is a black hole in the geometry and, typically, these are associated to the entanglement structure of the state in the dual QFT. For example, the Thermofield Double (TFD) state is a maximally entangled state whose entanglement is encoded in the Einstein-Rosen bridge connecting the left and the right boundaries of the corresponding Penrose diagram of an AdS-Schwarzschild geometry. Such Wormholes can be made traversable by adding a suitable matter field in the bulk geometry, equivalently turning on a particular deformation to the TFD-state in the boundary QFT. Finally, we will also discuss Wormholes that emerge in the calculation of fine-grained entropies in Holography. These Wormholes are specific in the context of such fine-grained data in quantum gravity and are of the traversable-Wormhole type. To make this explicit, we will label each type of Wormhole in the subsequent sections and discussions, as they appear.

This review is divided into the following sections: In section $2$, we begin with a basic discussion of Euclidean instanton solutions in quantum mechanics and quantum field theory. These instantons, in certain ways, are similar to Euclidean Wormhole geometries which we review, in detail, later. In section $3$, we briefly introduce the basic statement of Holography, especially of AdS/CFT correspondence. Section $4$ is devoted to discussing Euclidean Wormhole geometries, their role in extracting the Page-curve and multi-partite entanglement structure. We provide a technical review on the construction of multi-boundary Wormhole geometries which play a foundational role in understanding the multi-partite entanglement structure in Holography. The next section is devoted to Wormholes in Lorentzian framework. In particular, we review the fate of energy-conditions for such geometries, explicit and varied constructions of traversable Wormholes and their physical significance, Wormholes on the Brane and a phenomenon of Regenesis. Finally, in section $6$ we conclude with a list of broad and open directions for future, including a list of puzzles that Wormholes raise. We have also included a technical and supplementary discussions in two appendices.

\section{Quantum Mechanics}\label{sec:instanton}

Let us begin the discussion with a simple model in Quantum Mechanics.\footnote{This is a standard discussion, available in many text books, {\it e.g.}~\cite{Marino:2015yie}.} Consider a particle in an unstable  potential, denoted by $V(q)$, where $q$ denotes the classical position co-ordinate of the corresponding particle. A prototypical classical action is given by
\begin{eqnarray}
S =  \int dt L [q(t), \dot{q}(t)] = \int dt \left(  \frac{1}{2} \left(\frac{dq}{dt}\right)^2 - V(q) \right) \ , \quad V(q) =  - \frac{q^2}{2} + \frac{ q^4}{4} + \frac{1}{4} \ ,
\end{eqnarray}
where, for convenience, we have chosen specific coefficients for each monomial in $q$ in the potential. Classically, there are two unstable maxima of the potential, which are obtained by solving $\left. \partial_q V(q) \right|_{q_0}= 0$ with $\left. \partial_q^2 V(q)\right|_{q_0} > 0$: $q_0 = \pm 1$. Choosing one minima will break the symmetry $q \to -q$, spontaneously. Note that, a linear combination of both the minima will manifestly restore the symmetry, although, such a configuration is not physically meaningful in the classical regime. 

To quantize the system, we consider the path integral:
\begin{eqnarray}
Z = \int \D q(t) e^{i S[q(t)]} \ , \label{pit}
\end{eqnarray}
where $\D q(t)$ is some integration measure that we need not specify at this point. As is well-known, the path-integral is not a well-defined object, since the integrand is a widely oscillating function. Instead, we perform a Wick rotation: $t \to - i \tau$, where $\tau$ is the Euclidean time, and define the Euclidean path integral as:
\begin{eqnarray}
Z = \int \D q(\tau) e^{- H[q(\tau)]} \ , \quad H = \int d\tau \left( \frac{1}{2} \left( \frac{dq}{d\tau} \right)^2 + V(q) \right) \ . \label{pitau}
\end{eqnarray}
It is straightforward to check that (\ref{pitau}) follows from (\ref{pit}), by simple a substitution $t \to - i\tau$.\footnote{At this point, we are not specifying the boundary conditions on the functions $q(\tau)$; in fact, we have not yet specified the range of $\tau$ yet. We can choose the integration range as $\tau \in [-\infty, + \infty]$. An alternative choice is to compactify the $\tau$-direction, such that $\tau \in [0, 2\pi]$. }

Now, vacuum correlation functions can be obtained by
\begin{eqnarray}
&& \left\langle q(\tau_1) q(\tau_2) \ldots q(\tau_n) \right \rangle = \frac{1}{Z} \int \D q(\tau) q(\tau_1) q(\tau_2) \ldots q(\tau_n) e^{-H[q(\tau)]} \ , \label{corrZ} \\
&& Z = \int \D  e^{-H[q(\tau)]} \ . \label{euclZ}
\end{eqnarray}
Thus, the quantum correlation functions can now be obtained by computing classical correlation functions of a one-dimensional statistical mechanical system with an Euclidean action $H[q]$. This classical statistical mechanical system is described by the corresponding Euler-Lagrange equation derived by extremizing the functional $H[q]$. The resulting equations of motion are:
\begin{eqnarray}
\frac{d^2 q}{d\tau^2} =   q^3 -  q \ . \label{eucleom}
\end{eqnarray}
The simplest solutions of the above equation of motion are the static ones, for which $q(\tau) = 0, \pm 1$. The corresponding on-shell energies of these solutions are, respectively: $H_{\rm on-shell} = 0, - \int d\tau (1/4)$. Note that, all three solutions yield finite action when integrated over a compact support on $\tau$, but the latter two diverge on an infinite/semi-infinite line.\footnote{If we compactify the $\tau$-direction, the corresponding partition function in (\ref{euclZ}) becomes the thermal partition function. On a finite strip of $\tau$, however, the Euclidean path integral can be interpreted as the wavefunction of {\it e.g.}~an excited state. In the limit of the infinite strip length, this corresponds to the ground state wavefunction, for a system with a mass gap.}

\begin{figure}
  \includegraphics[width=\linewidth]{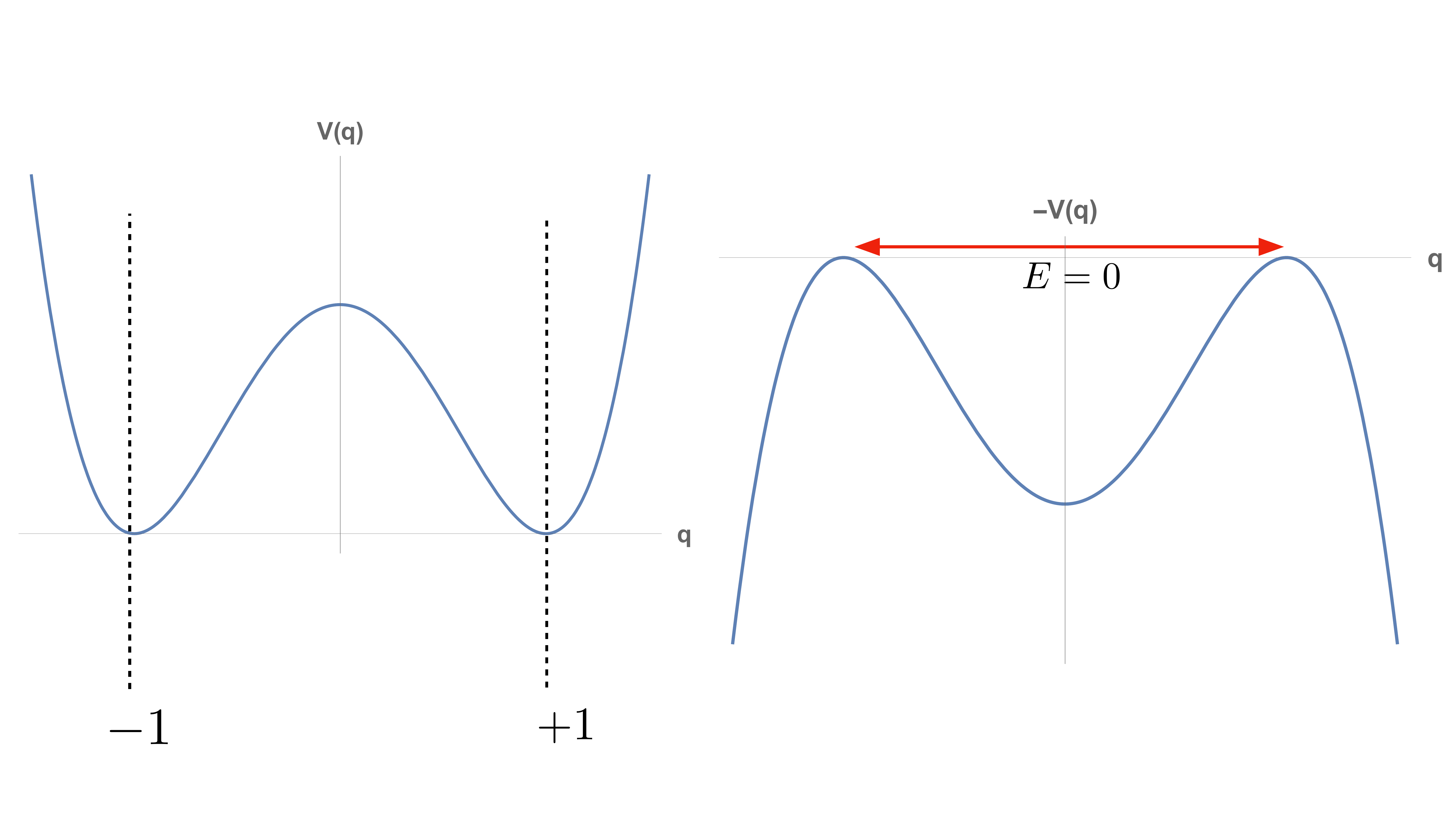}
  \caption{A schematic diagram for the instanton configuration. On the left, we have the original potential, and on the right we have the inverted potential. The instanton configuration of (\ref{soleuclidean}) begins at $\tau = -\infty$ from $q=-1$ and approaches $q=+1$ at $\tau = +\infty$. }
  \label{figinstanton}
\end{figure}

There is an obvious integral of motion for the equation in (\ref{eucleom}), corresponding to the symmetry under $\tau \to \tau + c$ of the Euclidean action in (\ref{pitau}), where $c$ is a constant. The integral of motion is given by
\begin{eqnarray}
\E = \frac{\partial H}{\partial \dot{q}} \dot{q} - H = \int d\tau \left[ \frac{1}{2} \left(\frac{dq}{d\tau}\right)^2 - V(q) \right] \ .
\end{eqnarray}
The solutions $q(\tau)= \pm 1$ correspond to $\E= (1/4)$. One can now solve for a more general $q(\tau)$ by setting $\E={\rm const}$, in (\ref{pitau}), which yields: 
\begin{eqnarray}
&& \frac{dq}{d\tau} = \pm \sqrt{ 2 V(q) + \E} \ , \\
&& \implies \quad q(\tau) = \pm   \tanh \left(   \sqrt{\frac{1}{2}} (\tau - \tau_0) \right)  \ , \quad {\rm with} \quad \E=0 \ , \label{soleuclidean}
\end{eqnarray}
where $\tau_0$ is an integration constant. This solution behaves as $ q \to 0 $ as $\tau \to \tau_0$ and $ q \to \pm 1$ as $\tau \to \pm \infty$. A pictorial representation is provided in figure \ref{figinstanton}.
In summary, we have three classical saddles: (i) $q(\tau)=0$, (ii) $q(\tau) = \pm 1$ and (iii) the interpolating solution in (\ref{soleuclidean}). Let us denote them by $q^{(1)}$, $q^{(2)}$ and $q^{(3)}$, respectively.

Suppose now, we want to evaluate the ``classical" Euclidean path integral in (\ref{pitau}), subject to some boundary condition, {\it i.e.}~we want to compute the contribution of the classical configuration satisfying the condition. Consider the boundary condition that $\lim_{\tau\to \pm \infty}q(\tau) = \pm 1 $. Without the knowledge of the interpolating solution in (\ref{soleuclidean}), {\it i.e.}~$q^{(3)}$, the corresponding path integral is obtained by a single saddle $q^{(1)}(\tau) = \pm 1$. This yields a path integral that consists of two identical contributions:
\begin{eqnarray}
Z = Z_{+\infty} + Z_{-\infty} = 2 \N \ , \label{zpm}
\end{eqnarray}
where the subscript $\pm \infty$ denotes the corresponding boundary conditions are imposed at $\tau = \pm \infty$. Here $\N$ is an unimportant numerical constant.

Subsequently, we can consider fluctuations around the classical saddle $q^{(2)}(\tau)$, let us denote them by $\delta q^{(2)}$. There are two copies of these fluctuations, associated to the two terms in (\ref{zpm}): $Z_{+\infty}[\delta q^{(2)}]$ and $Z_{-\infty}[\delta q^{(2)}]$. These path integrals define correlators of the type: $ \left\langle \delta q^{(2)} \ldots \delta q^{(2)} \right \rangle_{+\infty}$ and $\left \langle \delta q^{(2)} \ldots \delta q^{(2)}  \right \rangle_{-\infty}$, and no correlation between the $+\infty$ degrees of freedom and the $-\infty$ degrees of freedom exists. Also, the degrees of freedom near $q^{(1)}$ saddles (as well as $q^{(3)}$ ones) are completely absent.

On the other hand, subject to the same boundary condition $\lim_{\tau\to \pm \infty} q(\tau) =\pm 1 $, once we also include the interpolating saddles in (\ref{soleuclidean}), the corresponding path integral now takes the form:
\begin{eqnarray}
&& Z  =  Z_{+\infty} + Z_{-\infty} + Z_{\rm inter} \ , \label{zpminter} \\
&& Z_{\rm inter} =  {\rm exp}  \left[ -\frac{1}{2} \int_\infty^{\infty} d\tau \sech^4\left( \frac{\tau-\tau_0}{\sqrt{2}} \right) \right]  \ .
\end{eqnarray}
By construction, now the path integral yields a non-trivial correlator between the $+\infty$ degrees of freedom and the $-\infty$ degrees of freedom. Moreover, since the interpolating part also goes through the $q^{(1)}$ (near $\tau \to \tau_0$) saddle, the semi-classical degrees of freedom around this saddle are also coupled, in an indirect manner.

More quantitatively, suppose we semi-classical quantize using (\ref{zpm}). This manifestly implies the following:
\begin{eqnarray}
Z\left[ \delta q\right]  = Z_{+}\left[ \delta q^{(2)}, \J_+ \right]  + Z_{-} \left[ \delta q^{(2)}, \J_-\right] \ . \label{semiclpm}
\end{eqnarray}
Here we have dropped $\infty$ in the subscript, for simplicity. Furthermore, $\J_\pm$ represent the corresponding sources, conjugate to the fields $\delta q_\pm^{(2)}$. The above follows directly from the definition in (\ref{euclZ}), which, in this case, simply yields two unrelated semi-classical systems respectively localized at $\tau \to \pm \infty$. By construction, any correlator of the form $\left\langle \delta q_+^{(2)} \ldots \delta q_-^{(2)} \ldots \right \rangle = 0 $.\footnote{This is easily seen by {\it e.g.}~computing
\begin{eqnarray}
\frac{\delta^2}{\delta \J_+ \delta \J_-} Z = \frac{\delta}{\delta \J_+} \frac{\delta Z_-}{\delta \J_-} = 0 \ ,
\end{eqnarray}
since $Z_\mp$ is independent of $\J_\pm$. Here, $\J_\pm$ are the corresponding sources. 
} 

On the other hand, if we begin with (\ref{zpminter}), the corresponding expectation value is given by
\begin{eqnarray}
Z\left[ \delta q\right]  = Z_{+\infty}\left[ \delta q^{(2)}, \J_+ \right]  + Z_{-\infty} \left[ \delta q^{(2)}, \J_+ \right] + Z_{\rm inter} \left[ \delta q^{(3)}, \J_\pm \right] \ . \label{semiclinter}
\end{eqnarray}
Since $\delta q^{(3)}$ asymptotes to $\delta q_{\pm}^{(2)}$, in the semi-classical theory above, we obtain: $\left\langle \delta q_+^{(2)} \delta q_-^{(2)} \ldots \right \rangle \not =  0$. In this sense, including the instanton-saddle in (\ref{soleuclidean}) introduces additional correlations in the quantum system that no longer factorizes in terms of degrees of freedom near $+\infty$ and a $-\infty$. Note that, one way to obtain a non-vanishing correlator of the form $\left\langle \delta q_+^{(2)} \delta q_-^{(2)} \ldots \right \rangle $, one could begin with (\ref{semiclpm}), with a simple modification:
\begin{eqnarray}
Z\left[ \delta q\right]  = Z_{+}\left[ \delta q^{(2)}, \J_+ \right]  + Z_{-} \left[ \delta q^{(2)}, \J_-\right] \ , \quad \J_{\pm} = \frac{\J}{2} \ , \label{aveqm}
\end{eqnarray}
where $\J$ is an averaged coupling. It is suggestive that by averaging over the local semi-classical systems near $\pm \infty$, one can induce a non-trivial correlation function between the $+$ and the $-$ degrees of freedom, similar to what the interpolating solution does in (\ref{semiclinter}). This is not an equivalence, but a qualitative similarity. Later we will see this idea playing a sharper role in the Holographic context, in which Wormholes play a crucial role.

\section{Holography Basics}\label{sec:holobasic}

Before delving deeper into the physics of Wormholes, let us collect some basic facts and features of Holography that we will assume for the subsequent discussions in this review. The basic idea is that quantum gravity in an asymptotically anti de-Sitter (AdS) geometry is described by a quantum field theory (often a conformal field theory) defined on the conformal boundary of AdS. A basic idea came from holographic proposals in \cite{Stephens:1993an, Susskind:1994vu}, and within string theory it took a particularly sharp and precise form in terms of AdS/CFT, see {\it e.g.}~\cite{Maldacena:1997re, Gubser:1998bc, Witten:1998qj}. Over the years, these ideas have been generalized to a much wider class of examples and are sometimes used as a consistent definition of a theory of quantum gravity in AdS.

The best understood examples are in the class of SU$(N)$ gauge theories with an $N^2$ degrees of freedom at the conformal boundary of the AdS-geometry. The gravitational dual is described by Einstein-gravity in an AdS-geometry. We will mainly use this framework and our discussion will be confined within classical or semi-classical gravitational physics in AdS. A special case appears in AdS$_2$, where the dual QFT is a quantum mechanical system with $N$ Majorana Fermions that are interacting {\it via} random coupling, see {\it e.g.}~\cite{Sachdev_1993, Polchinski:2016xgd, Maldacena:2016hyu, Maldacena:2016upp}. We will also explicitly use this example. There are other dualities as well, for example, when the boundary QFT is an O$(N)$ vector model in which case the holographic dual is description is given in terms of infinite number of higher spin fields in the bulk, see {\it e.g.}~\cite{Klebanov:2002ja, Giombi:2009wh, Vasiliev:2003ev}. We will, however, not review the latter case in this article.

Qualitatively, the essential premise in the following classical gravity action:
\begin{eqnarray}
S_{\rm total} = S_{\rm gravity} + S_{\rm matter} \ , \label{actiongrav}
\end{eqnarray}
where $S_{\rm gravity}$ is the standard Einstein-gravity action, with a negative cosmological constant and $S_{\rm matter}$ is a generic matter contribution. Inverse Newton's constant plays the role for the gravitational action $S_{\rm gravity}$, and we work in the limit $G_N \to 0$. From the perspective of the boundary QFT, this sets the number of degrees of freedom $N^2 \sim G_N^{-1} \to \infty$. In this limit, gravity is purely classical\footnote{Note that, in general arbitrary higher derivative terms can be included in the gravitational action. However, these are parametrically suppressed by inverse string length or Planck length. Thus, as long as we discuss physics well below such scales, it is consistent to turn all of them off.} and by introducing a quantum matter field in the matter action $S_{\rm matter}$, we explore a semi-classical description.

The basic gravitational ingredient is an AdS$_{d+1}$-BH geometry, which extremizes the action (\ref{actiongrav}), in the absence of any matter source. The most familiar metric for this solution is given by
\begin{eqnarray}
ds^2 = \frac{L^2}{z^2} \left( - f(z) dt^2 + \frac{dz^2}{f(z)} + \sum_{i=1}^{d-1} dx_i^2 \right)  \ , \quad f(z) = 1 - \left( \frac{z_{\rm H}}{z}\right)^d \ , 
\end{eqnarray}
where $L$ is the curvature of the geometry, $z_{\rm H}$ is the location of an event-horizon. The dual QFT is defined on the ${\mathbb R}^{1,d}$ along $\{t, x_i\}$-directions. The presence of the horizon assigns a temperature to the QFT-state, with $T=d/(4\pi z_{\rm H})$. The conformal boundary is located at $z \to 0$; this radial coordinate is related to the corresponding energy-scale in the dual QFT.

Given the geometry, a generic bulk field corresponds to a gauge-invariant operator in the boundary. For example, the boundary QFT stress-tensor is dual to the metric, a conserved current is dual to a bulk gauge field, a scalar deformation of the boundary QFT is dual to a bulk scalar field, {\it etc}. Thus, one can use the gravitational description to compute correlation functions of such gauge-invariant operators in the boundary QFT.

Furthermore, a precise notion of quantum information in the boundary QFT is realized in the bulk description as well. For example, given a density matrix $\rho$ of a particular state in the boundary QFT, one can bi-partition the Hilbert space by looking at a spacelike sub-region $A$: $\H = \H_A \otimes \H_{\bar A}$, where $\bar{A}$ is the complement of the region $A$. Subsequently, one defines a reduced density matrix: $\rho_A = {\rm Tr}_{\bar A} \left( \rho \right) $, and a corresponding von Neumann entropy: $S_A = - {\rm Tr}_A \left( \rho_A \log \rho_A \right) $.\footnote{More generally, one defines a class of entropies, called Renyi entropy:
\begin{eqnarray}
S_A^{(n)} = \frac{1}{1-n} \log \left[ {\rm Tr}_A \left( \rho_A \right) \right] \ . 
\end{eqnarray}
}
These entropies encode quantum entanglement, and therefore quantum information, structure of the given QFT. Gravitationally, the von Neumann entropy can be calculated by the Ryu-Takayanagi prescription\cite{Ryu:2006bv, Ryu:2006ef}:
\begin{eqnarray}
S_A = \frac{{\rm Area}\left( \gamma_A \right) }{4 G_N} \ , \label{RT-formula}
\end{eqnarray}
where $\gamma_A$ is a co-dimension two minimal-hypersurface in the geometry, satisfying: (i) $\partial \gamma_A = \partial A$, (ii) $\gamma_A$ is homologous to $A$, (iii) $\gamma_A$ is defined on the same time-slice as $A$. While the above prescription holds for static states, it is further generalized to arbitrary time-dependent state in \cite{Hubeny:2007xt}. For further extensive review on this, we refer the reader to \cite{Rangamani:2016dms}.

Before leaving this section, let us note that it will be important to go beyond the classical limit and therefore consider a correction to (\ref{RT-formula}): a quantum corrected RT-formula. This can be obtained by computing entanglement between quantum fields in the bulk region, separated by the classical RT-surface. The relevant quantity to compute is the so-called generalized entanglement entropy, using the prescription of \cite{Engelhardt:2014gca}:
\begin{eqnarray}
S_A ={\rm min} \left[  \frac{{\rm Area}\left( \gamma_A \right) }{4 G_N} + S_{\rm bulk}  \right] \ ,
\end{eqnarray}
where $S_{\rm bulk}$ is the entanglement between quantum fields partitioned by the surface $\gamma_A$.

One needs to carry out an extremization of the above functional and subsequently choose the minimum of the extrema. For each given $\gamma_A$, $S_{\rm bulk}$ contributes to the generalized entropy functional and therefore alters the corresponding extrema. One needs to therefore analyze all possible $\gamma_A$, subject to the homology condition, and subsequently carry out the extremization. This is a technically difficult problem and only a few cases are hitherto analytically tractable. We will not make any explicit technical use of this functional, but it will play an important conceptual role.

\section{Euclidean Quantum Gravity}

We will now discuss instantons in Euclidean (Quantum) Gravity, specifically solutions that are categorized as wormholes. Before discussing that, it is incumbent that we define, at least operationally, the Euclidean Quantum Gravity action. As usual, the naive Lorentzian path integral is ill-defined and we can Wick rotate to define an Euclidean path integral accordingly. This is defined as\cite{gibbons1993euclidean}:
\begin{eqnarray}
&& Z = \int \D g e^{-S[g]} \ , \\
&& S[g] = - \frac{1}{16 \pi G} \int_\M d^D x \sqrt{g} \left( R - 2 \Lambda \right)  - \frac{1}{8\pi G} \int_{\partial \M}  K d^{D-1}\Sigma + C \ , \nonumber\\
\label{eqg}
\end{eqnarray}
where $G$ is the $D$-dimensional Newton's constant, $R$ is the Ricci-scalar, $\Lambda$ is the cosmological constant, $K$ is the trace of the second fundamental form on the boundary and $C$ is a constant that can be tuned to achieve a convenient on-shell configuration, {\it e.g.}~in flat space $S[g]=0$. The boundary term in (\ref{eqg}), known as the Gibbons-Hawking term, renders the variational problem well-defined and only contributes when $\partial \M \not = \emptyset$. Because this action is linear in the curvature which has no lower bound\footnote{For example, Ricci-scalar transforms non-trivially under a conformal transformation, under $\tilde{g}_{\mu\nu} = \Omega^2 g_{\mu\nu}$:
\begin{eqnarray}
\tilde{R} = \Omega^{-2} R - \left(D-4 \right) \left(D-1 \right) \Omega^{-4} \partial_\mu \partial^\mu \Omega - 2 \left( D-1\right) \Omega^{-3} g^{\mu\nu} \nabla_\mu \partial_\nu \Omega\ ,
\end{eqnarray}
 and by choosing a rapidly oscillating conformal factor, $\Omega$, the Ricci-scalar can be made as large as one wants.}, the Euclidean action is not bounded from below, unlike in ordinary QFT. One way to get around this issue is to define the Euclidean path integral by first separating the space of metrics into different conformal classes and integrating over a finite Ricci-scalar metric in each class, for more details see {\it e.g.}~\cite{Gibbons:1978ac}, \cite{Eguchi:1980jx}.

\begin{figure}
  \includegraphics[width=\linewidth]{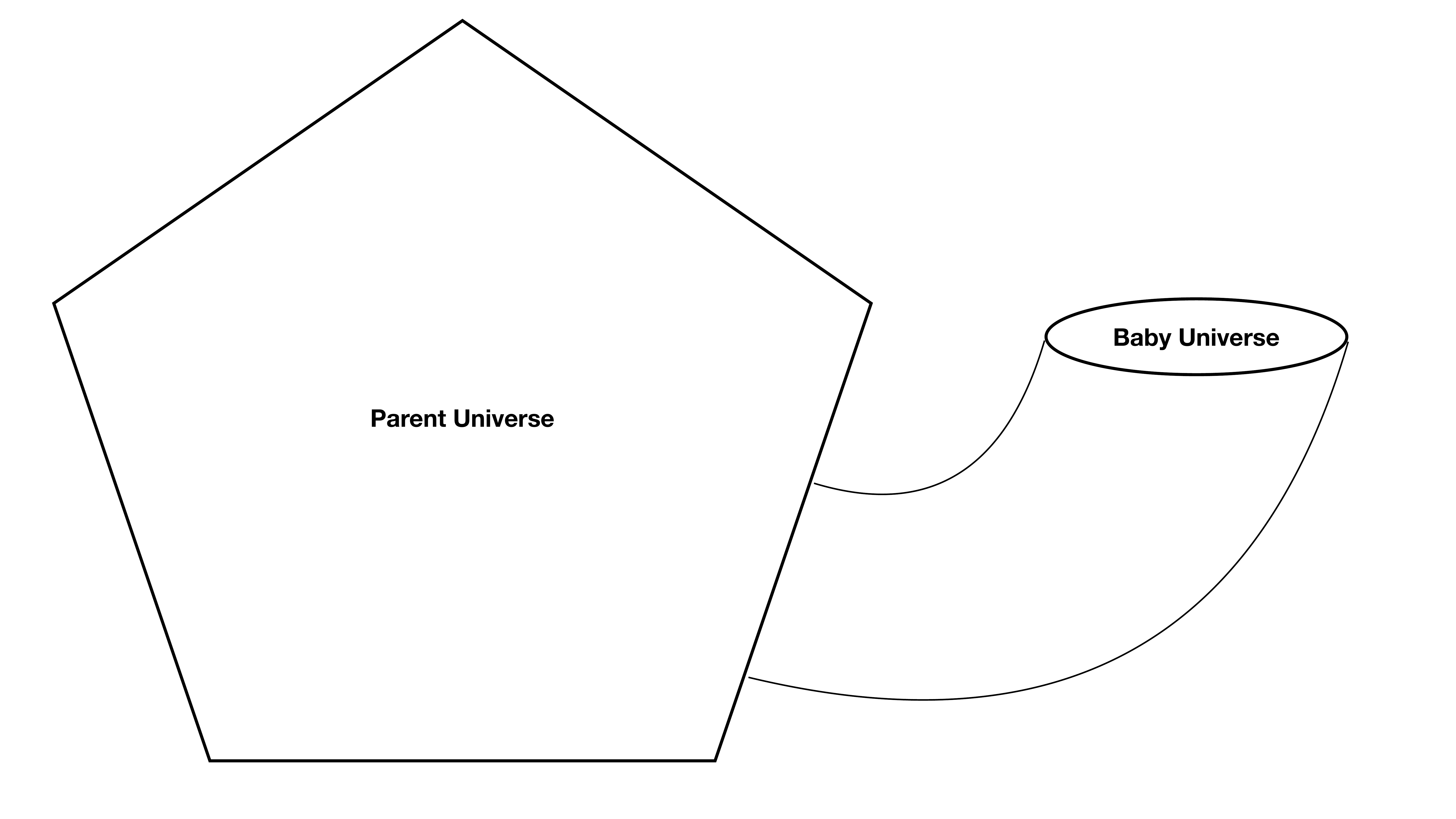}
  \caption{A schematic cartoon of a baby Universe popping out of a parent Universe. }
  \label{fig2}
\end{figure}

In this article, however, we will not delve into these issues. Rather, we will focus on the saddles of the Euclidean action, in particular the wormhole configurations. In a theory of quantum gravity, it is expected that topology-changing process take place, by simple quantum tunnelling. In particular, such topology-changing processes have been extensively investigated in the literature, within the context of potential loss of quantum coherence in quantum gravity, see {\it e.g.}~\cite{Lavrelashvili:1987jg, Hawking:1987mz, Giddings:1988cx, Coleman:1988cy, Giddings:1988wv}. The prototype of such topology changing process involves a Planck-scale baby Universe branching out from a parent Universe, see figure \ref{fig2}, for a representative processes.

In Lorentzian signature, the total spacetime is connected\footnote{It is important to note, however, that the Lorentzian continuation of the corresponding geometry are expected to be either complex or singular. We thank an anonymous Referee for emphasizing this point.} and therefore the corresponding Cauchy surface is also connected, and thus the baby Universe is not causally independent from the parent Universe. However, in Euclidean signature there is obstacle in having the baby Universe configuration. Therefore, one can now explore the solution space of Euclidean Gravity (equivalently, the saddles of the Euclidean Quantum Gravity path integral), for such configurations. From now on, we will refer to these configurations as Wormholes, motivated by the picture in figure \ref{fig2}.

The simplest of all is, of course, pure Einstein-gravity. Hawking showed in \cite{Hawking:1987mz} that there exists no such Wormhole saddle in pure gravity. However, Giddings-Strominger showed in \cite{Giddings:1987cg} that one-parameter family worth of Euclidean Wormhole solutions indeed exist if we add a generic massless-spinless axion field as a matter sector. We will momentarily review how this construction works, for now, let us assume they exist and ponder over the physical meaning of these configurations.

\begin{figure}
  \includegraphics[width=\linewidth]{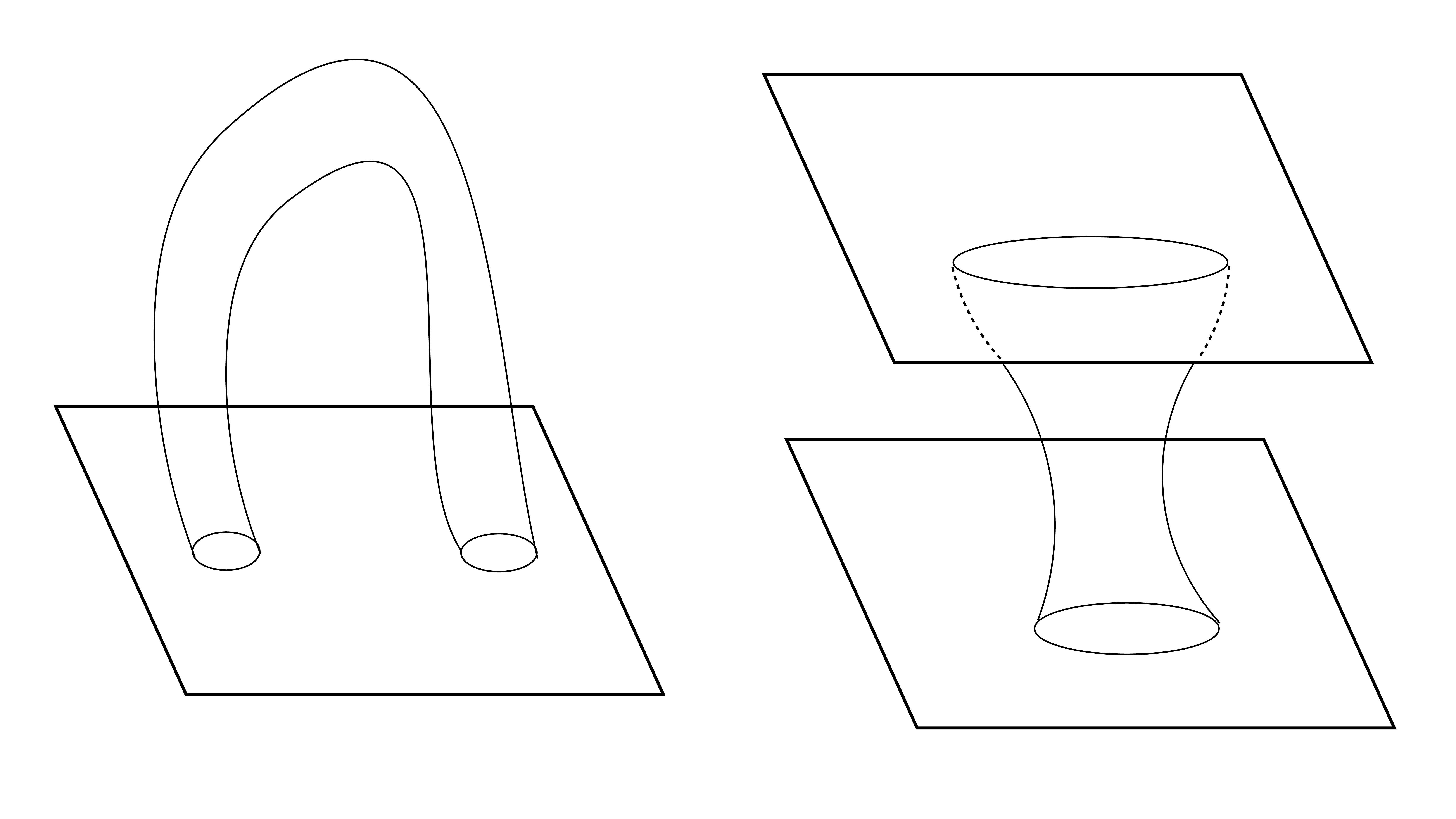}
  \caption{ A schematic cartoon of how Wormholes configurations may exist. On the left, the Wormhole begins and ends on the same asymptotic Universe, on the right, the Wormhole connects two asymptotic Universes. In the first case, an effective non-local description emerges within the same Universe, while in the latter, operators living on the two Universes couple through the Wormhole connecting them.}
  \label{fig3}
\end{figure}

From the definition of the Euclidean path integral, and the prescription of integrating over the conformal class of metrics (reviewed above), it is clear that the full quantum gravity calculation must take these Wormhole saddles into account. While it is not completely understood how this may work in detail, we can already make qualitative statements, following the suggestion by \cite{Preskill:1988na}. We draw the basic motivation from figure \ref{fig3}, in which the Euclidean Wormhole has two asymptotic regions, on which two classes of (gauge-invariant) operators are defined, denoted by $\O_I(x)$ and $\O_J(y)$. 
These two asymptotic regions are separated by a length $\ell_{\rm WH}$. Now, physics at length-scales $\gg \ell_{\rm WH}$ will receive a non-trivial contribution from the Wormholes, in that they will introduce an effective coupling of the form:
\begin{eqnarray}
S_{\rm WH} = - \frac{1}{2} \sum_{I, J} \int d^Dx d^D y \O_I(x) \O_J(y) \ , \label{nlocal}
\end{eqnarray}
which is an inherently non-local interaction. Thus, the Euclidean path integral receives a non-trivial contribution, which we rewrite as follows:
\begin{eqnarray}
Z_{\rm WH} & = & e^{- S_{\rm WH}}  =  {\rm exp} \left( - \frac{1}{2} \sum_{I, J} \int d^Dx d^D y C_{IJ}^{-1} \O_I(x) \O_J(y) \right) \nonumber\\
& = & \frac{\sqrt{{\rm det} C}}{\sqrt{2\pi}}\int [d\alpha] {\rm exp} \left(-\frac{1}{2} \alpha_I C_{IJ} \alpha_J\right) {\rm exp} \left( - \int d^Dx \sum_I \alpha_I \O_I(x) \right) \ , \nonumber\\ \label{alphalocal}
\end{eqnarray}
where $\{\alpha_I\}$ are auxiliary parameters, and $C_{IJ}$ is a symmetric matrix.\footnote{In deriving this identity, we have used the Gaussian integration result:
\begin{eqnarray}
\int d\alpha \, {\rm exp}\left( - \frac{1}{2} C \alpha^2\right)  {\rm exp}\left( - \alpha A\right)  = {\rm exp} \left( \frac{A^2}{2C}\right) \sqrt{\frac{2\pi}{C}} \ ,
\end{eqnarray}
where $A$ and $C$ are constants. The above result is valid for a single variable $\alpha$. This readily generalizes to an array of $\alpha_I$, in which case $C_{IJ}$ becomes a symmetric matrix and the number $C$ on the RHS, under the square root, is replaced by ${\rm det}[C]$. The $C^{-1}$ in the exponential argument is replaced by $C_{IJ}^{-1}$ matrix. 
} 

Let us pause and take stalk of how the physical interpretation has changed in going from (\ref{nlocal}) to (\ref{alphalocal}). Clearly, in (\ref{alphalocal}), (i) the resulting action is local, (ii) there are additional parameters $\{\alpha_I\}$ with each $\alpha_i \O_I$ term, which are then integrated out with a Gaussian distribution, (iii) the additional parameters $\{\alpha_I\}$ are spacetime independent and therefore does not break the Poincar\'{e} symmetry.\footnote{Note that this is not necessarily true for a global symmetry. For example, action in (\ref{nlocal}) is invariant under $O_I \to - \O_I$, however (\ref{alphalocal}) is not, unless we also demand that $\alpha_I \to - \alpha_I$ simultaneously.} The integral over a distribution of the couplings $\{\alpha_I\}$ suggests that there is an ``ensemble averaging" over a class of quantum systems, each member of which is characterized by a specific choice of $\alpha_I$.

\subsection{General Framework for Wormhole Solutions: Euclidean}

In this section we will review the basic ingredients required to construct a gravitational Wormhole solution. These are spacetime Wormholes that connect two asymptotic regions and introduce an effective coupling between them. In the dual CFT perspective, these geometries cause a factorization problem, which we will mention at the end of this section. To find Wormhole solutions, we need axion matter field:\footnote{See {\it e.g.}~the discussion in \cite{ArkaniHamed:2007js}. See also \cite{Hertog:2017owm, Katmadas:2018ksp, Hertog:2018kbz} for a further thorough analyses of the model. See \cite{Loges:2022nuw} for a recent account of stability of Wormholes in such models.}
\begin{eqnarray}
S = \frac{1}{2\kappa^2 } \int d^Dx \sqrt{g} \left(- R + \frac{1}{2} \partial_\mu \phi \partial^\mu \phi + \frac{1}{2(D-1)!} e^{\beta\phi} F_{\mu\nu \ldots } F^{\mu\nu\ldots} \right) \ ,
\end{eqnarray}
where $\phi$ is a scalar field, $\beta$ is a constant coupling and $F_{\mu\nu\ldots}$ is a $(D-1)$-form, which can be dualized to a scalar: 
\begin{eqnarray}
F_{\mu \ldots \lambda} = \epsilon_{\mu \ldots \nu\lambda} e^{-\beta \phi} \partial^\lambda A \ , 
\end{eqnarray}
where $A$ is the corresponding scalar. This is known as the axion-field and since it is defined as a dual to a $(D-1)$-form, the corresponding kinetic sign has a wrong signature. In general, from an Euclidean compactification of a supergravity (or stringy) system, one obtains the following action:
\begin{eqnarray}
S_{\rm A} = \frac{1}{2\kappa^2} \int d^D x \sqrt{g} \left[ - R + \frac{1}{2} G_{IJ} \partial_\mu \phi^I \partial^\mu \phi^J\right] \ , \label{phigen}
\end{eqnarray}
where $G_{IJ}$ is the metric in the space of scalars. There is no constraint on the sign of $G_{IJ}$, and this is crucial to allow us Wormhole solutions in the Euclidean signature, as we will see momentarily.

Consider that the Euclidean Wormhole connects two maximally symmetric geometries in $d=(D-1)$-dimensions, denoted by $\Sigma_d$. Clearly, $\Sigma_d$ can be either ${\mathbb R}^d$ ({\it i.e.}~an Euclidean plane), an ${\mathbb S}^d$ ({\it i.e.}~a round sphere) or a ${\mathbb H}^d$ ({\it i.e.}~a hyperbolic plane). Let us assume that the full Wormhole geometry obeys a ``left-right" symmetry, and hence it admits a natural slicing in terms of the transverse direction, which we denote by $r$. The corresponding geometric data  can be written in the following manner:
\begin{eqnarray}
&& ds^2 = N(r) dr^2 + a(r)^2 d\Sigma_d^2 \ , \label{an1} \\
&& \phi_I = \phi_I(r) \ . \label{an2}
\end{eqnarray}
Here we can further gauge-fix to choose $N(r)=1$. The geometry above can be interpreted as a homogeneous and isotropic Euclidean cosmology, with a Euclidean time $r$. Einstein equations now yield:
\begin{eqnarray}
&&\frac{a'^2}{a^2} - \frac{k}{a^2} - \frac{G_{IJ} (\phi^I)' (\phi^J)'}{2(D-1)(D-2)} - \frac{s}{L^2} = 0 \ , \label{flrw} \\
&& \left( a^{D-1} G_{IJ} (\phi^J)'\right) ' - \frac{1}{2} G_{JK,I} (\phi^J)' (\phi^K)' = 0 \ , \label{kg}
\end{eqnarray}
where, $k=1, 0, -1$ for $\Sigma_d \equiv {\mathbb S}^d, {\mathbb R}^d, {\mathbb H}^d$, respectively; the $' \equiv \frac{d}{dr}$, $s=0,1$ for asymptotically flat or anti de-Sitter geometry, and $L$ is the corresponding curvature scale.

One can rewrite the equations of motion, in the following form:
\begin{eqnarray}
&& a'^2 = k + \frac{s}{L^2} a^2 + \frac{C} {2(D-1)(D-2) a^{2D-4}} \quad \implies \quad a'^2 + V_{\rm eff}(a) = 0 \ ,  \\
&& \frac{C}{a^{2D-2}} = G_{IJ} (\phi^I)' (\phi^J)' \ , 
\end{eqnarray}
where $C$ is a constant of motion.\footnote{This integral of motion comes from the fact that the effective action, obtained by substituting the geometric ansatz in (\ref{an1}), and (\ref{an2}) in (\ref{phigen}), is independent of $r$, it only depends on functions of $r$. Therefore, there is a conserved Hamiltonian. } The nature of the solution now depends crucially on $C$; we sketch various possibilities below: \\
(i) $C>0$: In this case, $V_{\rm eff}(a) \to -\infty$, as $a\to 0$, and therefore $a' \to \infty$. This clearly is a singular limit and we do not expect any physical solution to exist in this case. \\
(ii) $C=0$: This can be exactly solved readily to obtain:
\begin{eqnarray}
a(r) = \pm \left( a_0 e^{-r \frac{\sqrt{s}}{L}} - \frac{k L^2}{2 a_0} e^{r \frac{\sqrt{s}}{L}} \right) \ ,  \label{mockwh}
\end{eqnarray}
where $a_0$ is an integration constant. Clearly, as $r \to \infty$, $a(r) \to \pm \infty$, which is consistent with having an asymptotically flat, as well as an AdS-geometry. For an asymptotically flat spacetime, we want $a(r) \to r^2$ as $r\to \infty$, while for an asymptotically AdS-geometry we need $a(r) \to e^{r/L}$ as $r\to \infty$.

However, if we want a Wormhole solution connecting the asymptotia, there must exist a point $r = r_*$ such that $a'(r_*) = 0$. We observe that setting $k + \frac{s}{L^2}a(r_*)^2 = 0$, this condition can be achieved. For asymptotically flat geometry, {\it i.e.}~with $s=0$, this occurs only for the trivial case of $k=0$; but for an AdS-geometry, {\it i.e.}~with $s=1$, this can be achieved for $k=-1$. On the other hand, if $k=0$, then an exact solution can also be found:
\begin{eqnarray}
a(r) = e^{ \pm r \frac{\sqrt{s}}{L}} \ , 
\end{eqnarray}
which also satisfies $a'(r_*)=0$ at $a(r_*)=0$. In this case, a Wormhole may be constructed by gluing the two branches of the solutions at $a(r_*)=0$, see the discussion in \cite{Marolf:2021kjc}. \\
(iii) $C<0$: This is the most interesting case. Let us write $C = - 2(D-1)(D-2) a_0^{2D-4} <0$, which yields an equation of the form:
\begin{eqnarray}
a'(r)^2  = k + \frac{s}{L^2} a(r)^2 - \frac{a_0^{2D-4}}{a(r)^{2D-4}} \ , \label{whreduced}
\end{eqnarray}
which can be solved in terms of hypergeometric functions. For our purposes, it is enough to notice that, as $r\to\infty$, we get:
\begin{eqnarray}
&& a(r) = \pm \sqrt{k} r \ , \quad s = 0 \quad \& \quad k = 0, 1 \ , \\
&& a(r) = e^{\pm \frac{r}{L}} \ , \quad s=1 \quad \& \quad k = 1, 0, -1 \ , 
\end{eqnarray}
where $s=0$ corresponds to asymptotically flat and $s=1$ corresponds to asymptotically AdS spacetime. 

Now, the condition $a'(r_*)=0$ can be satisfied for all cases, provided the RHS of (\ref{whreduced}) has a vanishing point. This appears to hold generically since the RHS has one positive term, one definitely negative term and one term that can take both positive as well as negative values.\footnote{It appears that for $k=0$ and $s=1$, an arbitrarily small value of $a_0$ can also satisfy the equation $a'(r_*)=0$, and thus a Wormhole geometry can be obtained for an arbitrarily small value of the matter field contribution\cite{Marolf:2021kjc}.} However, the following choices are not allowed: (i) $k=0$ and $s=0$, {\it i.e.}~between asymptotically flat geometry with planar boundaries, (ii) $k=-1$ and $s=0$, {\it i.e.}~between asymptotically flat geometry with hyperbolic boundaries.

Before concluding this section, a few comments are in order. Wormhole configurations lead to various conceptual issues, that become sharp contradictions in the context of AdS/CFT correspondence. In this context, the main conceptual problem Wormhole configurations lead to can be summarized in the following way: \\
(i) Wormholes can be arbitrarily separated, and therefore the bulk amplitudes will not factorize. In terms of the dual gauge theory this implies that the corresponding correlators do not satisfy cluster decomposition. \\
(ii) In standard AdS/CFT, the dual gauge theory is local and therefore it should obey cluster decomposition. \\
Clearly, (i) and (ii) are in direct contradiction with each other. 

This contradiction can be stated in a more quantitative way\cite{ArkaniHamed:2007js}: Consider the dual gauge theory on a compact Euclidean time, with a long period $\beta$. This corresponds to the low temperature regime of the theory. In this limit, a generic two-point function in the gauge theory, between two generic local gauge-invariant operators $\O_1$ and $\O_2$ are given by
\begin{eqnarray}
\left \langle \O_1 \O_2 \right \rangle_{\rm QFT} & = &  \sum_{i, j} e^{-\beta E_i} \left\langle E_i | \O_1 |E_j \right \rangle \left \langle E_j | \O_2 | E_i  \right \rangle \nonumber \\
& \approx & \left\langle 0 | \O_1 |0 \right \rangle \left \langle 0 | \O_2 | 0  \right \rangle + \O\left( e^{- \beta \Delta E}\right) \ ,\label{nowh}
\end{eqnarray}
where $\Delta E$ is the mass gap.\footnote{The ground state can also be degenerate, but with a finite degeneracy.} On the other hand, the Wormhole effective action will yield a contribution of the following form:
\begin{eqnarray}
\left \langle \O_1 \O_2 \right \rangle_{\rm WH} = \int [d\alpha ] e^{- \frac{1}{2} \alpha_I C_{IJ} \alpha_J} \left\langle 0 | \O_1 |0 \right \rangle_\alpha \left \langle 0 | \O_2 | 0  \right \rangle_\alpha + \O\left( e^{- \beta \Delta E_\alpha}\right) \ . \label{effwh}
\end{eqnarray}
Clearly, (\ref{nowh}) and (\ref{effwh}) will yield different answers for the same physical question. This is essentially the factorization problem.

\subsection{Euclidean Wormholes \& Black Hole Information}\label{sec:eadstfd}

In this section we will heavily discuss Euclidean Wormhole geometries and several physical questions in quantum aspects of gravity where these play a crucial role. We will discuss the relevance of these Wormhole geometries before we elaborate more on their technical constructions in later sections. The Euclidean Wormholes also yield a precise version of the factorization problem, which will momentarily become crucial in the context of the black hole information problem. For a working intuition, such Wormholes are essentially very similar to the Einstein-Rosen bridges which can also be made traversable by turning on an appropriate matter field. We will elaborate on the details of these constructions later in this review.

As the first application of Euclidean Wormholes, we will briefly review its role in addressing the Black Hole information paradox. To do so, let us recall the basic issues related to an information paradox in Black Hole physics. This is an incredibly rich and subtle subject and there is a wide variety of perspectives on this. We will not attempt to elaborate on them. Instead we refer the interested Reader to {\it e.g.}~\cite{Almheiri:2020cfm, Raju:2020smc} for a review on recent progress. Our discussion here will be based on the premise of \cite{Almheiri:2020cfm}.

The core issue is to understand precisely how the information inside of an event horizon comes out of it. This is a dynamical question that requires us to address this issue in a completely time-dependent framework. However, a version of the information paradox and what it implies to resolve the same can be provided in a completely static set up, following the pioneering work in \cite{Maldacena:2001kr}. Based on this, we will now review this idea.

\subsubsection{Eternal Black Holes \& Thermofield-Double State}\label{sec:eadstfd}

We begin with a discussion of the eternal AdS-BH and the corresponding thermo-field double (TFD) state in the dual CFT. Standard AdS/CFT states a duality between an asymptotically AdS gravitational background and a corresponding state in the boundary CFT. Specifically, an AdS-BH geometry is dual to a thermal state in the CFT. This description naturally generalizes for the corresponding thermo-field double (TFD) state as well.

Given two copies of the same quantum mechanical system, for which the individual Hilbert spaces, $\H_{\rm L,R}$, are spanned by the Hamiltonian eigenbasis $|n\rangle_{\rm L,R}$, the TFD state is defined as:
\begin{eqnarray}
| {\rm TFD}\rangle  = \frac{1}{\sqrt{Z}} \sum_{n} e^{-(\beta E_n)/2} | n \rangle_{\rm L} \otimes | n \rangle_{\rm R} \ , \label{tfd}
 \end{eqnarray}
where $E_n$ is the corresponding energy of the $|n\rangle_{\rm L,R}$ state and $\beta$ is a real-valued parameter.\footnote{More precisely, the TFD state is given by
\begin{eqnarray}
&& | {\rm TFD}\rangle  = \frac{1}{\sqrt{Z}} \sum_{n} e^{-(\beta E_n)/2} | n \rangle_{\rm L} \otimes | n^* \rangle_{\rm R} \ , \\
&& | n^* \rangle_{\rm R} = \Theta | n \rangle_{\rm R} \ ,
\end{eqnarray}
where $\Theta$ is an anti-unitary operator, such as the CPT-transformation. However, for our purposes, this aspect will have no consequences and hence we ignore this feature.} This state can be prepared as the ground state of a suitably chosen Hamiltonian, see {\it e.g.}~\cite{Maldacena:2018lmt} for an explicit Hamiltonian interaction in $(0+1)$-dimension and \cite{Cottrell:2018ash} for a generalization in higher dimensions. In particular, we will review the basic construction of \cite{Maldacena:2018lmt} later. 

It is now easy to check a few basic features of the TFD-state. These are:\\
(i) TFD is a pure state. This is easily seen, since $\rho_{\rm TFD}^2 = \rho_{\rm TFD}$, where $\rho_{\rm TFD} = | {\rm TFD}\rangle \langle {\rm TFD} |$. \\
(ii) Upon partial tracing over, {\it e.g.}~the left degrees of freedom, one is left with a reduced density matrix which is mixed: $\rho_\beta = \Tr_{\rm L}\left( \rho_{\rm TFD} \right) = e^{- \beta H_{\rm R}}$. Clearly, $\rho_\beta^2 \not = \rho_\beta$. \\
(iii) There is a choice in the total Hamiltonian with which the TFD-state can evolve. We can either choose to evolve the TFD state with $H= H_{\rm L} - H_{\rm R}$, or $\tilde{H}= H_{\rm L} + H_{\rm R}$. Under these two Hamiltonian time-evolution, the TFD state behaves qualitatively differently. For example:
\begin{eqnarray} \label{TFD_H}
&& | {\rm TFD}(t)\rangle  = e^{- i H t } | {\rm TFD}(0)\rangle  = \frac{1}{\sqrt{Z}} \sum_n e^{- \beta E_n/2} e^{-i (H_{\rm L} - H_{\rm R})t} | n \rangle_{\rm L} | n \rangle _{\rm R} = | {\rm TFD}(0)\rangle \ ,\nonumber\\
&& | {\rm TFD}(t)\rangle  = e^{- i \tilde{H} t } | {\rm TFD}(0)\rangle = \frac{1}{\sqrt{Z}} \sum_n e^{- \beta E_n/2} e^{-i (H_{\rm L} + H_{\rm R})t} | n \rangle_{\rm L} | n \rangle _{\rm R} \not= | {\rm TFD}(0)\rangle  \ .\nonumber\\
\end{eqnarray}
Thus, TFD is a ground state for $H$, but not for $\tilde{H}$. In later sections, we will make explicit use of the TFD-state.

The discussion above applies for any quantum mechanical system, including quantum field theories as pointed out by \cite{Takahashi:1996zn}. For a large $N$ Holographic CFT, specifically in the context of AdS/CFT, the CFT TFD-state is dual to the eternal black hole in AdS, as proposed in \cite{Maldacena:2001kr},\footnote{For a different perspective on the gravitational dual of the TFD-state, see \cite{Mathur:2014dia}.} see also \cite{Israel:1976ur} for earlier ideas along similar directions.

An information paradox can now be phrased in this framewowork\cite{Maldacena:2001kr}. Suppose, we perturb the TFD-state by inserting a hermitian operator $\O_{\rm R}$ in CFT$_{\rm R}$, supported by a real-valued perturbative coupling $\epsilon\ll 1$. In this deformed TFD-state, suppose we measure the expectation value of an operator $\O_{\rm L}$ in CFT$_{\rm L}$. This is given by
\begin{eqnarray}
\left\langle {\rm TFD} \right| \left(1 + \epsilon \O_{\rm R}  \right) \O_{\rm L} \left(1 + \epsilon \O_{\rm R}  \right)\left| {\rm TFD}\right \rangle = \left\langle \O_{\rm L} \right \rangle_{\rm TFD} + 2 \epsilon \left \langle \O_{\rm L} \O_{\rm R} \right \rangle_{\rm TFD} + \O(\epsilon^2) \ , \label{2ptTFD}
\end{eqnarray}
where we have used $\left[\O_{\rm L}, \O_{\rm R} \right] = 0 $, since there is no explicit coupling between the left and the right CFTs.

Let us pause for a few comments. Note that, we arrived at the above conclusion by considering how the TFD state is prepared. For example, in the Euclidean description this state can be formally prepared by 
\begin{eqnarray}
| {\rm TFD}\rangle & = & e^{- \frac{\beta}{4} (H_{\rm R} + H_{\rm L})}  \frac{1}{\sqrt{Z}} \sum_n | n \rangle_{\rm L} | n \rangle _{\rm R} \nonumber \\
& \to & e^{- \frac{\beta}{4} (H_{\rm R} + H_{\rm L}) + \epsilon \O_{\rm R}} \frac{1}{\sqrt{Z}} \sum_n | n \rangle_{\rm L} | n \rangle _{\rm R} = \left(1 + \epsilon \O_{\rm R}  \right) | {\rm TFD}\rangle \ ,
\end{eqnarray}
if the deformation $\O_{\rm R}$ commutes with $H_{\rm R}$, and at leading order in $\epsilon$. Since the Euclidean evolution is non-unitary, a non-trivial effect appears already at the leading order in $\epsilon$ in computing the one point function of an operator $\O_{\rm L}$ in this deformed TFD-state. This is captured by the RHS of equation (\ref{2ptTFD}). Alternatively, one can ask a similar question in the Lorentzian description. Suppose we choose the Hamiltonian $H=H_{\rm L}+ H_{\rm R}$ that determines the evolution of the TFD-state. Furthermore, we deform $H_{\rm R}$ by an operator $\O_{\rm R}$. In the Lorentzian description, this yields:
\begin{eqnarray}
| {\rm TFD}(\epsilon)\rangle \to e^{- i (H_{\rm R} + H_{\rm L} + \O_{\rm R}) \epsilon } | {\rm TFD}\rangle \approx \left(  1 - i \epsilon \O_{\rm R} \right)  | {\rm TFD}\rangle \ ,
\end{eqnarray}
at the leading order in $\epsilon$. Clearly, this is a unitary evolution that preserves the inner product $\langle {\rm TFD} | {\rm TFD}\rangle$. Thus, in the Lorentzian picture, at leading order one does not obtain a non-trivial correlator between $\O_{\rm L}$ and $\O_{\rm R}$.\footnote{Explicitly, one obtains the expectation value $\langle {\rm TFD} |[\O_{\rm L}, \O_{\rm R}] | {\rm TFD}\rangle$, which vanishes identically since left and right degrees of freedom are non-interacting. } In this description, one can either consider a higher point function between $\O_{\rm L}$ and $\O_{\rm R}$, {\it e.g.}~a four-point function, or consider a one-sided two-point function, {\it e.g.}~$\left\langle \O_{\rm R} \O_{\rm R} \right\rangle_{\rm TFD}$ or $\left\langle \O_{\rm L} \O_{\rm L} \right\rangle_{\rm TFD}$.

Let us go back to equation (\ref{2ptTFD}). The $\O(\epsilon)$-term in (\ref{2ptTFD}) can be obtained from Holography, using basic AdS/CFT dictionary, by computing geodesic distances between the points where $\O_{\rm R,L}$ are inserted. This geodesic distance grows linearly with time, as {\it e.g.}~one holds $\O_{\rm R}$ at a fixed time and sends $\O_{\rm L}$ to very late times $t \gg \beta$. Therefore we obtain: $\left \langle \O_{\rm L} \O_{\rm R} \right \rangle_{\rm TFD} \sim {\rm exp} \left[ - A t /\beta \right] $, where $A$ is an order one constant. Clearly, this correlator decays exponentially at late times and all initial data are lost to a final thermal state. From the dual CFT perspective, however, this is not expected to happen since we begin with a pure state and the evolution is unitary. Instead, the late time behaviour is expected to be: $\left \langle \O_{\rm L} \O_{\rm R} \right \rangle_{\rm TFD} \sim {\rm exp } \left[ - B S\right] $, where $B$ is an unimportant numerical constant and $S$ is the entropy of the thermal state to which the late time CFT state approaches. However, they are not arbitrarily close to each other, once $S$ is fixed. In particular, at times $t \gg \beta S$, the gravity calculation yields an answer infinitesimally close to the thermal answer, but the CFT calculation does not.

\begin{figure}
  \includegraphics[width=\linewidth]{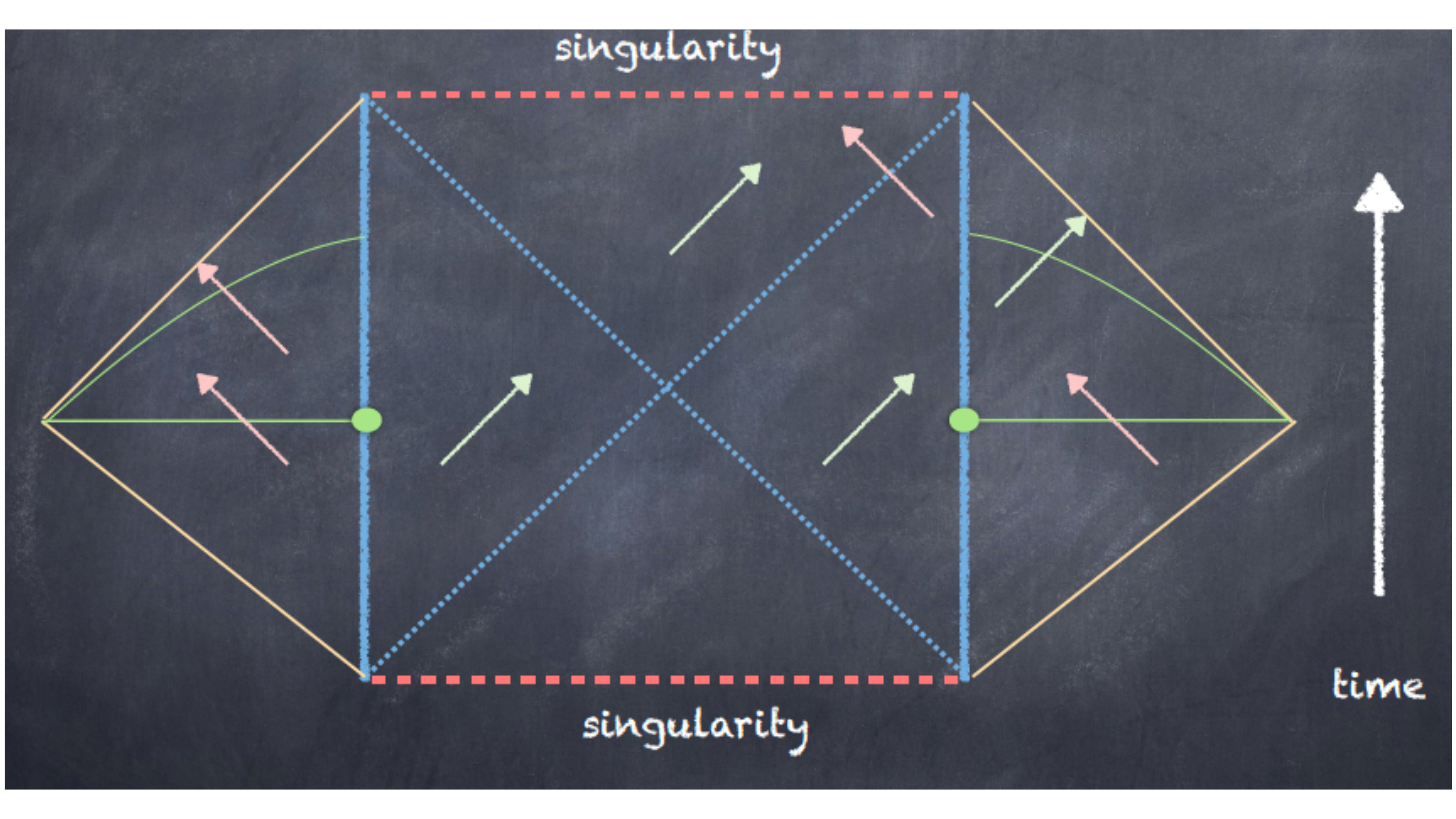}
  \caption{A pictorial representation of how an information paradox is realized within the framework of eternal black holes. The green and the red quanta are the Hawking pairs that can be detected by a measurement carried out at the boundary. These observations are made at regions denoted by green solid lines on the figures.}
  \label{eterentropy}
\end{figure}

An equivalent way to restate the same phenomenon is to notice the following, with reference to figure \ref{eterentropy}. This is essentially based on the work in \cite{Penington:2019kki, Almheiri:2019qdq}, see also \cite{Chen:2020hmv}. Suppose an observer, denoted by the green dot on figure \ref{eterentropy}, measures the Hawking quanta emitted from the eternal AdS black hole. To design an evaporating black hole in AdS, a transparent boundary condition is imposed on such quanta at the conformal boundary. This is denoted by gluing a Minkowski patch to the AdS-BH patch in the Penrose diagram in \ref{eterentropy}. These radiation quanta are denoted by the arrows in figure \ref{eterentropy}.  

At a constant time-slice, the observer is in contact with a constant time-slice of the Minkowski patch that acts as a thermal bath to the observer. Effectively, counting Hawking quanta for such an observer is equivalent to computing the corresponding entanglement entropy of the boundary sub-region, with the thermal bath. This is a well-defined calculation that can be addressed explicitly using the Ryu-Takayanagi formula, briefly reviewed in section \ref{sec:holobasic}. Qualitatively, this boils down to an intuitive understanding of what we have described above for the $2$-point correlator. Pictorially, at $t=0$, both the pink quanta are accounted for by the bath, and none of the green quanta are observed. At a later time, when the bath is denoted by the green curved lines in figure \ref{eterentropy}, only the pink quanta on the left and the green quanta on the right are detected. Therefore, entanglement entropy increases, as time increases. A detailed calculation yields an ever-increasing entanglement entropy without any bound. However, the eternal AdS-BH has a finite entropy $2S_{\rm BH}$, which should upper bound the entanglement entropy. This is a precise form of the information/entropy paradox in the eternal framework.

{\it A} resolution of the paradox above has been solved by generalizing the entanglement entropy prescription in (\ref{RT-formula}), to a quantum-corrected version of the Ryu-Takayanagi proposal. See \cite{Engelhardt:2014gca, Dong:2017xht}. Although the physical description is Lorentzian, all relevant calculations can be done in the Euclidean framework. In particular, the Euclidean Wormholes play a salient role in eventually upper bounding the growth of the entanglement entropy and thereby resolve the paradox. We will not discuss formal aspects of the quantum-corrected RT-prescription, instead we will demonstrate with an example how Euclidean Wormholes achieve this upper bound in the entanglement entropy computations.

This is a good place to take a few steps back and re-asses the current status of the field. An essential assumption in the above construction is that the Minkowski patch glued to the AdS-BH geometry is non-gravitational. In several known explicit examples, and also on general grounds, this fact alone leads to massive gravitons, see {\it e.g.}~\cite{Geng:2020qvw}, and subsequent studies in \cite{Geng:2020fxl, Geng:2021hlu}. It is also argued that a massive graviton facilitates a Hilbert space factorization, which does not happen in the limit of the vanishing graviton mass, and therefore one can obtain the physics of a Page curve only when the graviton is massive. The perspective in {\it e.g.}~\cite{Raju:2021lwh} (see also \cite{Laddha:2020kvp}) draws upon this technical point of gravitons acquiring a mass and advocates an alternative interpretation of the black hole information paradox and its possible resolution. For further discussion, we refer the interested Reader to \cite{Raju:2020smc}.

\subsubsection{Fine-grained Entropy \& Information Resolution}

The basic ideas were proposed and explored in detail in \cite{Penington:2019npb, Almheiri:2019psf, Almheiri:2019hni}. We will heavily draw on the subsequent works in \cite{Penington:2019kki, Almheiri:2019qdq}. The prototypical model in which these aspects are best demonstrated is the so-called Jackiw-Teitelboim (JT) gravity, coupled to an End-of-the-World (EOW) Brane. Spacetime is demanded to end on the EOW Brane, which will only provide certain boundary conditions on the bulk classical fields. The combined action is given by
\begin{eqnarray}
 &&S = S_{\rm JT} + S_{\rm Brane} \ , \label{2djtwh} \\
&& S_{\rm JT} = -\frac{S_0}{2\pi} \left( \frac{1}{2} \int_{\M} \sqrt{g} R + \int_{\partial\M} \sqrt{h} K\right) -  \left[ \frac{1}{2} \int_{\M} \sqrt{g} \phi \left(R + 2 \right) + \int_{\partial\M} \sqrt{h} \phi K \right] \ , \nonumber\\
&& S_{\rm Brane} = T \int ds \ ,
\end{eqnarray}
where $S_0$ is a constant that measures the extremal entropy of the extremal geometry, $\M$ is the manifold on which the JT-theory is defined, with a boundary $\partial\M$, $T$ is the tension of the Brane, $h$ is the induced metric at this boundary and $K$ is the corresponding extrinsic curvature. This JT-action can be obtained by dimensional reduction of a higher dimensional gravitational theory, near the extremal limit of a charged black hole, see {\it e.g.}~\cite{Nayak:2018qej}.

Note that the term proportional to $S_0$ is purely topological and therefore has no dynamics. The second term in the JT-action yields a rather simple dynamics. Variation of the scalar $\phi$ yields: $R+2 =0$, {\it i.e.}~two-dimensional geometries with a constant negative curvature. The complete set of equations obtained by varying the action in (\ref{2djtwh}) is given by
\begin{eqnarray}
&& R + 2 = 0 \ , \quad \left( R_{ab} + g_{ab} \nabla^2 - \nabla_a \nabla_b \right) \phi = 0  \ , \label{eom1} \\
&& {\rm with} \quad n^a \partial_a \phi = T \ , \quad K=0 \quad {\rm at \, \, EOW}  \ . \label{eom2}
\end{eqnarray}
The second line above simply provides a set of boundary conditions for the fields $\phi$ and the geometry $g_{ab}$, at the location of the EOW Brane. Furthermore, the asymptotic AdS boundary conditions are also needed:
\begin{eqnarray}
\left. ds^2 \right|_{\partial \M} = \frac{1}{\epsilon^2} d\tau^2 \ , \quad \left. \phi \right|_{\partial \M} = \frac{1}{\epsilon} \ , \quad \epsilon \to 0 \ . \label{adsboundary}
\end{eqnarray}

\begin{figure}
  \includegraphics[width=\linewidth]{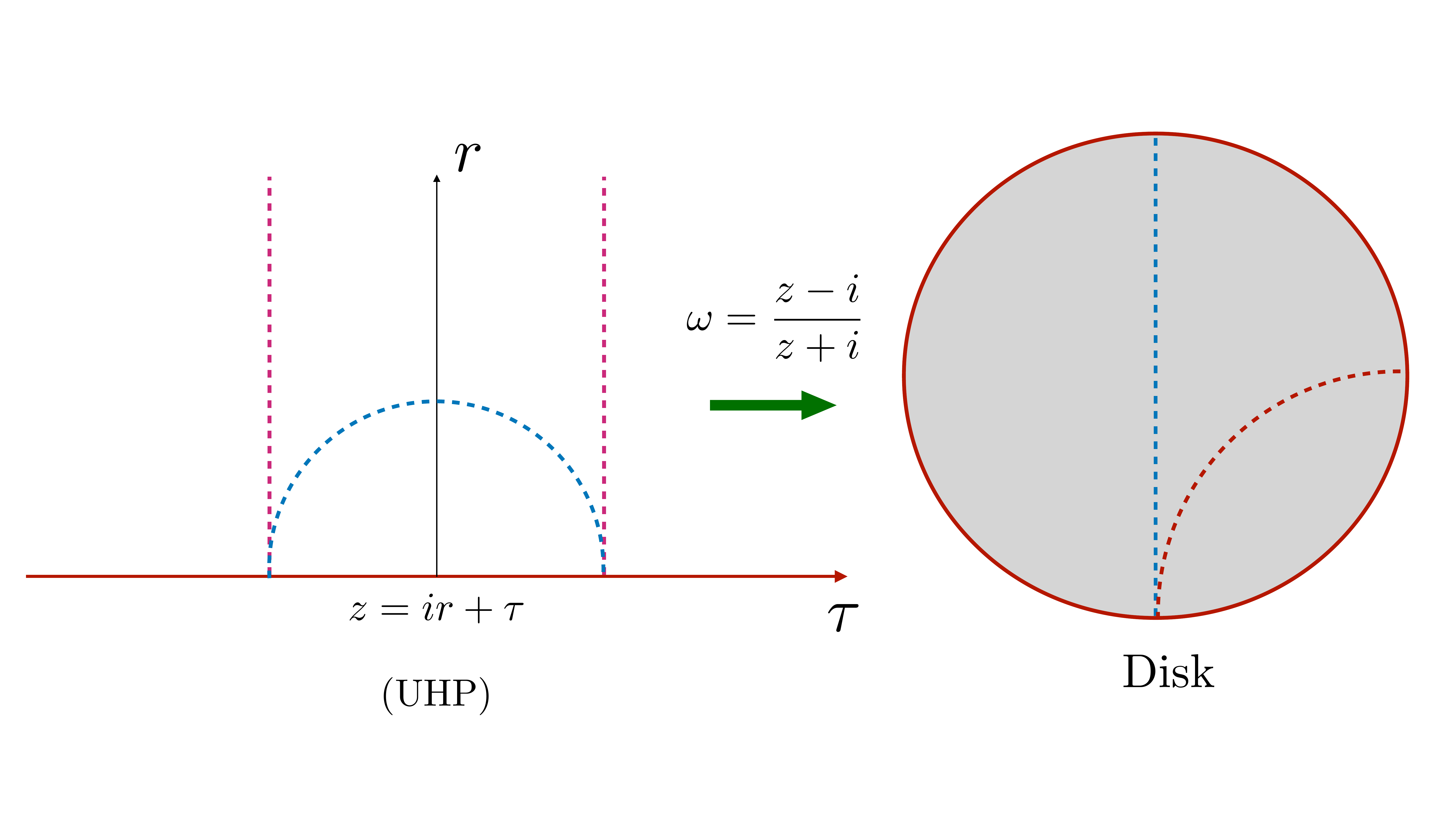}
  \caption{ The explicit map from the UHP to the unit Disk. With $z = ir +\tau$, the $\partial{\rm Disk}=S^1$ is described by $|\omega|^2=1$ $\iff$ $r=0$. There are two kinds of geodesics on the UHP, described by the dashed blue semi-circles and the dashed pink vertical lines. These map to the dashed blue and the dashed pink geodesics on the Disk.}
  \label{uhpdisk}
\end{figure}

It is simpler and more illuminating to understand the solutions pictorially. Without the Brane, {\it i.e.}~when $\mu=0$, only (\ref{adsboundary}) boundary conditions apply. The bulk equations of motion can be solved to obtain: 
\begin{eqnarray}
ds^2 = \frac{1}{r^2} \left( d\tau^2 + dr^2 \right) \ , \quad \phi = \frac{\tau^2+ r^2}{r}\ . \label{poinads2}
\end{eqnarray}
The scalar field can have a more general solution in the bulk, see {\it e.g.}\cite{Almheiri:2014cka, Maldacena:2016upp}. The metric describes the Upper Half Plane (UHP), and can be rewritten as $ds^2 = (1/r^2) dz d\bar{z}$, where $z = i r + \tau$. The UHP can subsequently be mapped to the unit disk, by the Cayley transformation: $ w = f(z) = (z-i)/(z+i)$. Therefore, the Euclidean AdS$_2$ geometry can be simply drawn as a disk, as shown in figure \ref{uhpdisk}, where the dual quantum system lives on the boundary-circle of the disk.\footnote{It is straightforward to see, using the Cayley map, that $r =0$ maps to $|\omega|^2=1$.}
For the sake of completeness, we have also explicitly presented how the geodesics on the UHP map to geodesics on the Disk. These geodesics are probes in the geometry. There are two kinds of geodesics in the complex $z$-plane: (i) semi-circles, described by $r^2 + \tau^2 = R_1^2$, where $R_1$ is a constant; (ii) vertical lines, described by $\tau=R_2$, where $R_2$ is also a constant. Under the Cayley map, these geodesics map to: 
\begin{eqnarray}
&& (i) \quad {\rm Re} [\omega] =  \frac{R_1^2 -1}{R_1^2+1+2 r} \ , \quad {\rm Im}[\omega] =  \frac{-2 \sqrt{R_1^2- r^2}}{R_1^2+1+2 r} \ , \\
&& (ii) \quad {\rm Re} [\omega]= \frac{r^2 + R_2^2 -1 }{ \left(r + 1 \right)^2 + R_2^2} \ , \quad  {\rm Im}[\omega] = - \frac{2 R_2}{\left(r + 1 \right)^2 + R_2^2 } \ . 
\end{eqnarray}
The corresponding geodesics are shown in figure \ref{uhpdisk}, for $R_1=1=R_2$. We do not want the EOW-branes to be described by $\tau= {\rm const}$, since these analytically continue to a space-like brane in the Lorentzian picture. 
\begin{figure}
  \includegraphics[width=\linewidth]{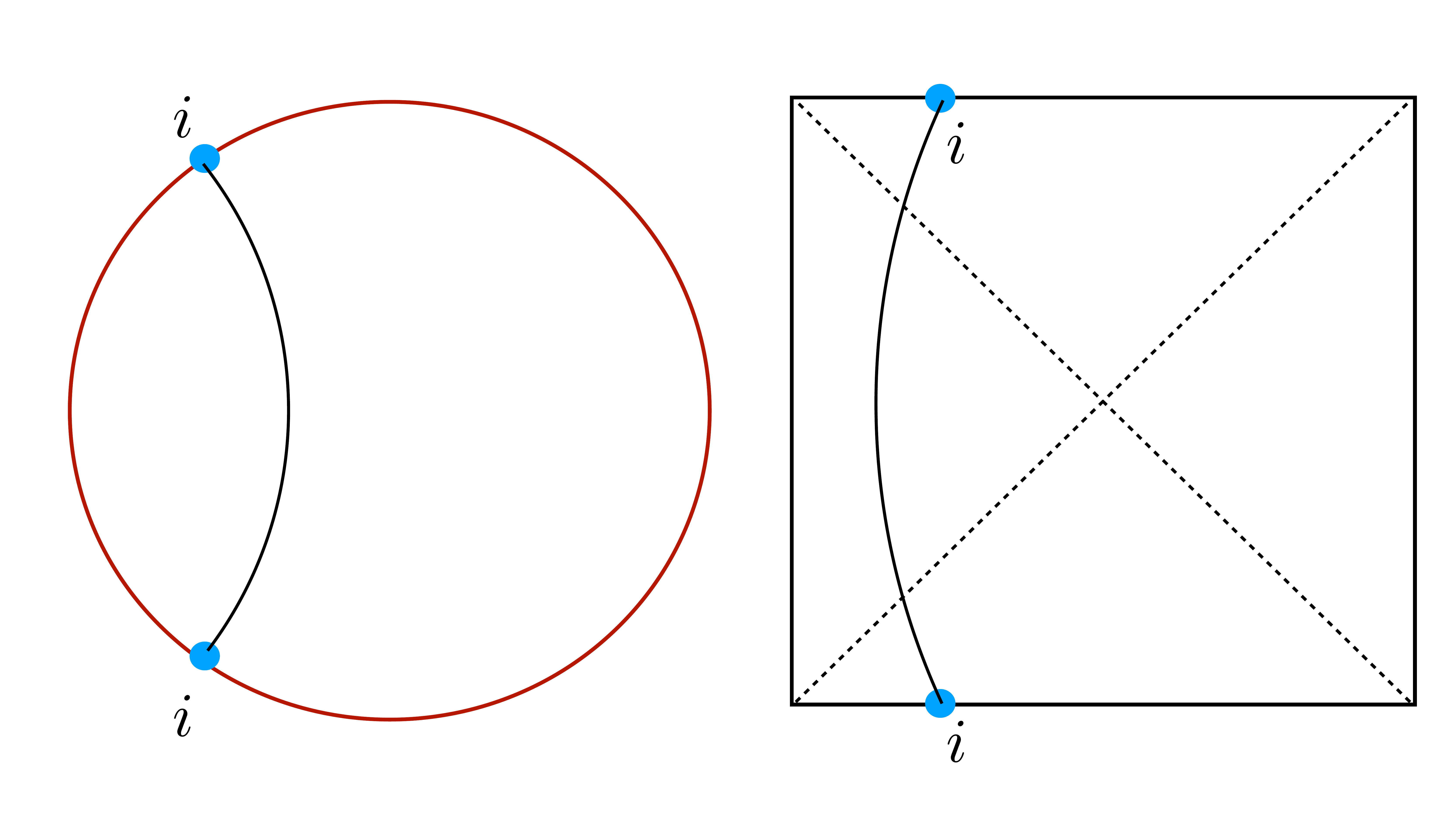}
  \caption{ The AdS$_2$ geometry with an EOW-brane, as the fully back-reacted solution. The brane is characterized by an index $i$ which captures the information of the microstates.}
  \label{eadslads}
\end{figure}

The EOW Brane itself is a back-reacting geodesic in the geometry which intersects with the boundary of the disk at two points in the geometry, as demonstrated in figure \ref{eadslads}. The EOW Brane is obtained as a solution to the equation $K=0$. In the coordinate patch of (\ref{poinads2}), this equation takes the form:
\begin{eqnarray}
1 + r'(\tau)^2 + r(\tau) r''(\tau) =0 \quad \implies \quad r(\tau) = \pm \left( C_1 - \left( \tau + C_2 \right)^2\right)^{1/2} \ ,
\end{eqnarray}
in the gauge where worldvolume coordinate is identified with $\tau$, and $C_{1,2}$ are two constants of integration corresponding to the location of the origin and the radius of the circle. The EOW-brane is described by circles intersecting the Disk, as demonstrated in figure \ref{eadslads}.

To model the basic features of Black Holes information paradox, let us consider the following ingredients:\\
(i) We can assign a quantum number, denoted by an index $i$, which captures all information about the quantum microstructure of the Black Hole. In string theory, it is standard to consider a stack of $k$ D-branes, on which open strings can end. The corresponding Chan-Paton factors can play the role of such quantum numbers.  \\
(ii) Assign a range $i = 1, 2, \ldots k$, with $k \sim e^{S_{\rm BH}}$, where $S_{\rm BH}$ is the Bekenstein-Hawking entropy of the Black Hole. This assignment clearly implies that the EOW branes cannot be treated in a probe limit. \\
(iii) Declare that the outgoing Hawking radiation quanta are entangled with the EOW brane degrees of freedom. 

With the above ingredients, let us consider the following state:
\begin{eqnarray}
| \Psi \rangle  = \frac{1}{\sqrt{k}} \sum_{i=1}^k | \psi_i \rangle _{\rm BH} \otimes | i \rangle_{\rm R}  \ , \label{state}
\end{eqnarray}
where $| \psi_i \rangle _{\rm BH}$ is the microstate of the Black Hole and $| i \rangle_{\rm R}$ is a reference state which will be accessed by an asymptotic observer. For our purposes, we assume $| i \rangle_{\rm R}$ form an orthogonal basis of the reference Hilbert space. The state $| \Psi \rangle$ is a maximally entangled state between the Black Hole degrees of freedom and the reference degrees of freedom. The asymptotic observer, who resides at the time-like conformal boundary of the AdS-geometry, can now calculate the entanglement entropy of the reference state.

\begin{figure}
  \includegraphics[width=\linewidth]{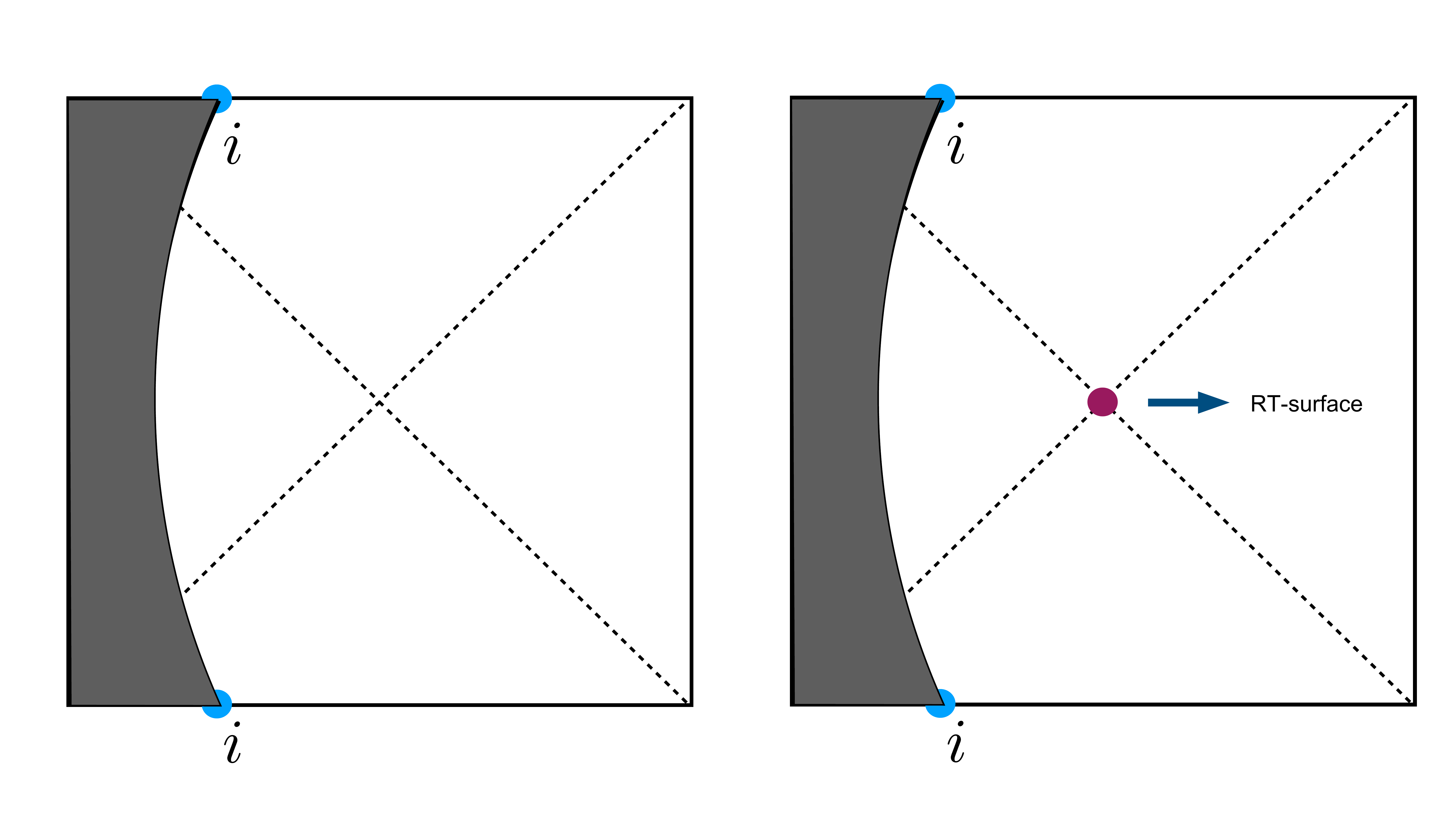}
  \caption{Two candidate RT-surfaces are shown here. On the left, the surface is an empty set, on the right it is a point, located at the bifurcation point. In both pictures, the dark shaded regions are excluded from the geometry.}
  \label{rtpos}
\end{figure}

This entanglement entropy calculation can be explicitly performed, by using the Holographic prescription of computing generalized entropy functional\cite{Engelhardt:2014gca}. The computation consists of two parts: The area of the Ryu-Takayanagi surface, and the bulk entanglement entropy between degrees of freedom across this surface. Since the RT-surface is a co-dimension two surface, in $2$d-geometry we have two options: (i) RT-surface is an empty set, (ii) RT-surface is just a point in the bulk. In the latter case of (ii), this bulk point is located at the bifurcation point of the Lorentzian Black Hole geometry, see figure \ref{rtpos}.
%

In the first case, the RT-surface area yields a vanishing result and the only contribution comes from the bulk entanglement between the EOW degrees of freedom and the reference state and thus behaves as $\sim\log k$. In the second case, one calculated the ``area" of the RT-point in the bulk, which is given by the dilaton field at the bifurcation point. This is easily seen from the JT-action which has an explicit $\phi R$-term, and the value of $\phi_{\rm bifurcation}$ sets the value of the corresponding Newton's constant. The resulting contribution equates the Black Hole entropy $S_{\rm BH}$. The so-called Island rule then dictates that the correct entanglement entropy is given by: ${\rm min}\left( \log k, S_{\rm BH}\right) $. Therefore, potentially, there can be a competition between the $\log k$ behaviour, which monotonically increases as a function of $k$, and $S_{\rm BH}$ which remains constant. If $k < e^{S_{\rm BH}}$, this transition never happens since $\log k$ is always the minimum of the two. This justifies the choice of $k > e^{S_{\rm BH}}$. 

\begin{figure}
  \includegraphics[width=\linewidth]{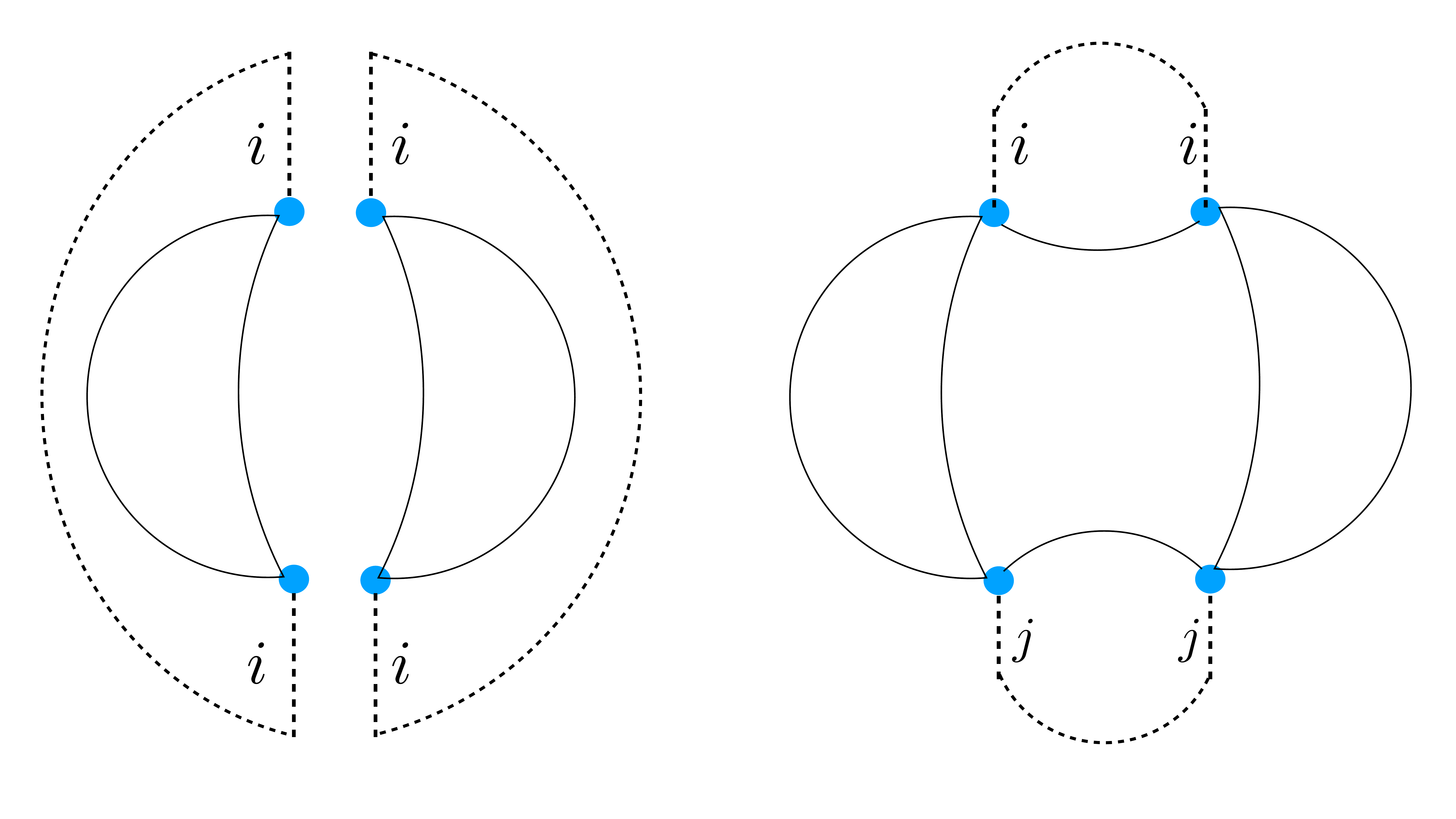}
  \caption{To calculate purity by setting $n=2$ (\ref{renyin}), two candidate geometries are shown. On the left, for the {\it disconnected} geometry the EOW-brane index $i$ is enforced on all dashed lines, which yields a single power of $k$, upon summation. On the right, for the {\it connected/Wormhole} geometry, two different EOW-brane indices can be assigned, which yields a factor of $k^2$ upon summation.}
  \label{eecomp}
\end{figure}

This transition can be derived from the gravitational path integral\cite{Penington:2019kki, Almheiri:2019qdq}, which we will briefly review in the rest of this section. Given the state in (\ref{state}), the reduced density matrix for the reference state is obtained by
\begin{eqnarray}
\rho_{\rm R} = {\rm Tr}_{\rm BH} \left( | \Psi \rangle \langle \Psi |\right) = \frac{1}{k}\sum_{i,j=1}^k | i \rangle \langle j |_{\rm R} \langle \psi_i | \psi_j \rangle_{\rm BH} \ ,
\end{eqnarray}
where, now, we will use AdS/CFT to compute the amplitude $\langle \psi_i | \psi_j \rangle_{\rm BH}$. This is obtained by calculating the on-shell gravity action, for the solution of (\ref{eom1}), (\ref{eom2}) and the boundary conditions in (\ref{adsboundary}). The required calculation is pictorially summarized in figure \ref{eecomp}, which yields: $\langle \psi_i | \psi_j \rangle_{\rm BH} = \delta_{ij} Z_1$, where $Z_1$ is the corresponding value of the on-shell Euclidean gravity action.\footnote{In principle, we can denote this on-shell action by $Z_i$, {\it i.e.}~the on-shell value can be indexed. However, this dressing will have no effect on our subsequent discussion, since we can redefine the reference state basis $|i \rangle$  by absorbing the overall normalization. Hence we ignore this possibility.} This yields:
\begin{eqnarray}
\rho_{\rm R} = \frac{Z_1}{k} \sum_{i=1}^k | i \rangle \langle i |_{\rm R}  \quad \implies \quad S_{\rm R} = \log k \ , \label{redrho}
\end{eqnarray}
where $S_{\rm R}$ is the von Neumann entropy (or the entanglement entropy) of the reference state. Subsequently, we can calculate the $n$-th Renyi entropies by evaluating ${\rm Tr}(\rho_{\rm R}^n)= \frac{1}{k^{n-1}}$. Note that, if we use $\rho_{\rm R} in (\ref{redrho})$, without making any reference to the bulk gravitational picture, then this is given by
\begin{eqnarray}
S_{\rm R}^{(n)} = \frac{1}{1-n} {\rm Tr} \left( \log \rho_{\rm R}^n \right) = \frac{n k}{n-1} \log k \ , \label{renyin}
\end{eqnarray}
in which we have only kept track of the $k$-dependent contributions. We will momentarily see that this is only a partial answer of what one obtains from the gravitational path integral.

On the other hand, from the bulk perspective one can also calculate ${\rm Tr}(\rho_{\rm R}^n)$. For $n=2$, we need to evaluate: $(1/k^2)\sum_{i,j} |\langle \psi_i | \psi_j\rangle|^2$. There are two classes of configurations that contribute to this calculation, summarized in figure \ref{eecomp}.
Restoring the factors of $Z_1$, this can be evaluated to yield:
\begin{eqnarray}
&& {\rm Tr}\left( \rho_{\rm R}^2 \right)  = \frac{1}{k^2 Z_1^2} \left [ {\rm disconnected} + {\rm connected} \right] \ , \label{purity2} \\
&& {\rm disconnected} = k Z_1^2 \ , \quad {\rm connected} = k^2 Z_2 \ ,
\end{eqnarray}
where $Z_2$ is the Euclidean on-shell contribution from the connected Wormhole saddle. The factors of $k$ arise by contracting the EOW indices, as explained in the figure \ref{eecomp}.

Let us denote the intersection point between $\partial({\rm EAdS})$ and the EOW Brane by $\sigma$, which is assigned an index $i$. In the disconnected diagram, in the summation above, all four $\sigma$ carry the same index; while, for the connected diagram, a pair of $\sigma$ carry the same quantum number. Thus, the former yields a factor of $k$ upon summing over the indices and the latter yields a factor of $k^2$ in the same process. 

In general, $Z_1$ and $Z_2$ can be calculated using the explicit solutions. For the JT-action, this is easily estimated from the purely topological terms in the action in (\ref{2djtwh}). The JT-path integral contributes $\sim e^{\chi S_0}$, where $\chi$ denoted the Euler characteristic of the corresponding bulk geometry and $S_0$ is the constant in front of the purely topological term in the JT-gravity. For both connected and the disconnected geometries, $\chi=1$, and the corresponding ratio $Z_2/ Z_1^2 \sim e^{-S_0}$. Therefore, the schematic form of (\ref{purity2}) is given by
\begin{eqnarray}
&& {\rm Tr}\left( \rho_{\rm R}^2 \right) = \frac{1}{k} + e^{-S_0} \ , \label{trrho2} \\
&& \implies \quad {\rm Tr}\left( \rho_{\rm R}^2 \right) \approx \frac{1}{k} \ , \quad k \ll e^{S_0} \ , \quad  {\rm and} \quad {\rm Tr}\left( \rho_{\rm R}^2 \right) \approx e^{-S_0} \ , \quad k \gg e^{S_0} \ . \nonumber\\ \label{trrho2limit}
\end{eqnarray}
The transition in the behaviour of purity, in the two regimes $k\gg e^{S_0}$ and $k\ll e^{S_0}$, captures unitarity of the underlying quantum dynamics of the Black Hole, albeit, in this case, for an eternal Black Hole. In particular, the Wormhole geometry dominates large $k$ regime and renders the fine-grained information plateau at a non-vanishing value. Note that, the result in (\ref{trrho2}) is in direct conflict with what one would obtain by squaring the reduced density matrix in (\ref{redrho}) -- the latter yields a contribution that one obtains only from the disconnected bulk geometry and this contribution becomes arbitrarily small for sufficiently large $k$.

\begin{figure}
  \includegraphics[width=\linewidth]{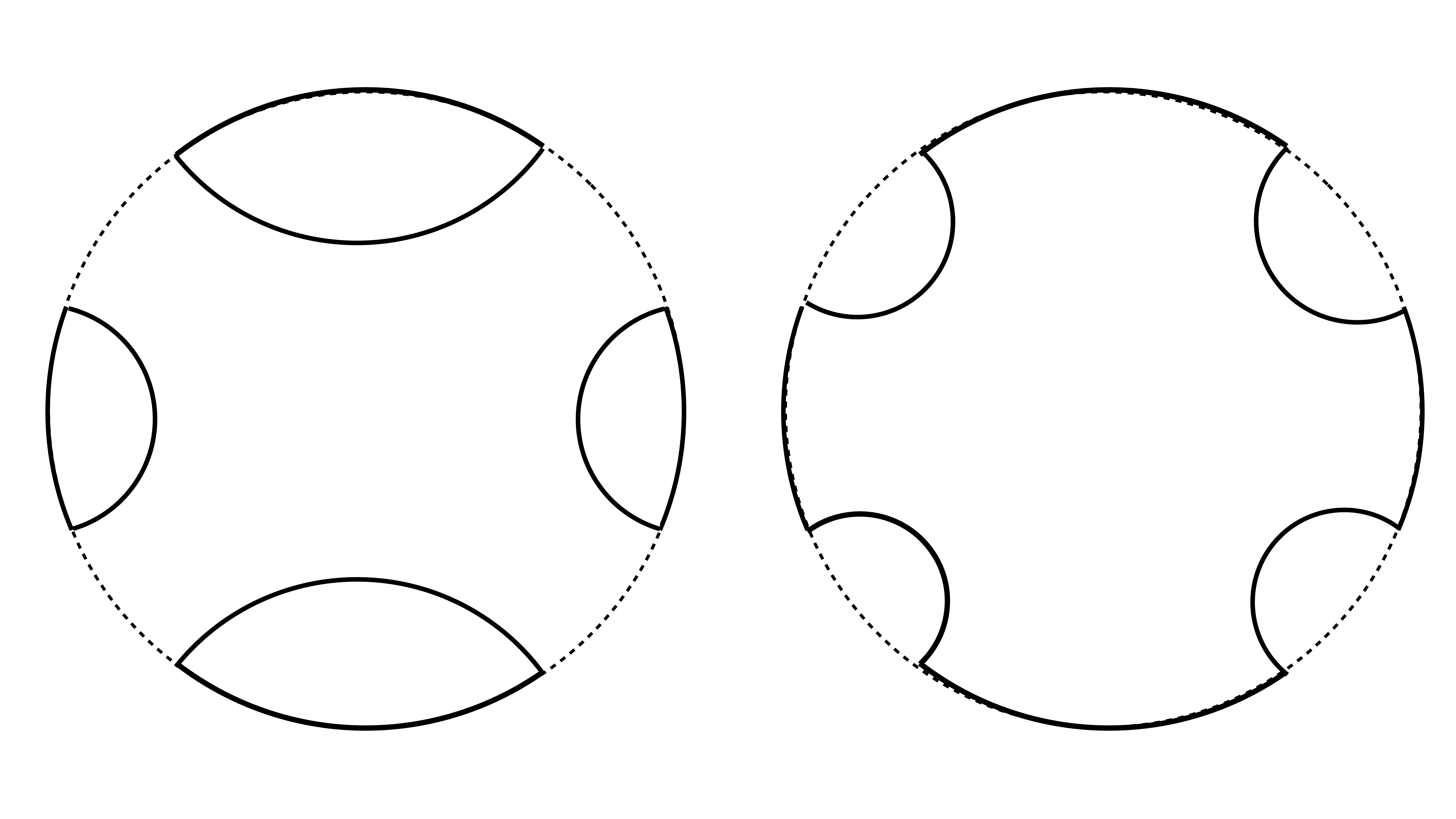}
  \caption{ Pictorial representation of two relevant saddles in the computation of ${\rm Tr}(\rho_{\rm R}^4)$. On the left, we have the completely disconnected geometry, and on the right, it is the completely connected one. There are additional saddles that also contribute to this computation, however, the above two are the two extreme examples. }
  \label{rhondiag}
\end{figure}

This feature holds generally, for ${\rm Tr}(\rho_{\rm R}^n)$. Now, all gravitational saddles with $n$-copies of the conformal boundary will contribute to this calculation. The qualitative similarity can be understood by considering the contributions coming from two extreme cases: One in which the $n$-boundaries are all disconnected in the bulk, and the other in which all $n$-boundaries are connected by a Euclidean Wormhole saddle. This is pictorially demonstrated in figure \ref{rhondiag}, for $n=4$.
The former diagram contributes $1/k^{n-1}$, while the latter contributes $Z_n/Z_1^n$, where $Z_n$ is the on-shell Euclidean action for the Euclidean Wormhole connecting $n$-boundaries -- a natural generalization of $Z_2$. Hence, $\Tr (\rho_{\rm R}^n)$ demarcates two distinct qualitative behaviour: for $k \ll e^{S_0}$, $\Tr (\rho_{\rm R}^n) \approx 1/k^{n-1}$ and for $k \gg e^{S_0}$, $\Tr (\rho_{\rm R}^n) \approx e^{-S_0}$. Subsequently, the von Neumann entropy is obtained from $\rho_{\rm R}^n$, using (\ref{renyin}) that computes $\Tr (\log \rho_{\rm R}^n)$, and then taking a limit $n\to 1$. In this limit, the $k \ll e^{S_0}$ and the $k \gg e^{S_0}$ behaviours are given by $\log k$ and $S_{\rm BH}$, respectively. Thus, one obtains the expected fine-grained entropy curve, as depicted in figure \ref{eecomp}.\footnote{Note that, in the limit, $k\gg 1$, $e^{S_0} \gg 1$ such that $k/e^{S_0}$ is held fixed, the entire summation over geometries can be carried out using a resolvent technique, since the sum only consists of planar diagrams. This result then can be analytically continued in $n$, which is non-trivial because of the presence of a branch cut. This discontinuity of the branch cut in the resolvent integral yields the correct behaviour for the von Neumann entropy. This is discussed in detail in \cite{Penington:2019kki}.}

Before concluding this section, let us allude to certain puzzling aspects that this computation raises. While it is remarkable that the Euclidean Wormhole solutions are necessary to reproduce fine-grained entropies of an underlying unitary theory, they also imply that, from the gravitational path integral:
\begin{eqnarray}
&& \left \langle \psi_i | \psi_j \right\rangle  = Z_1 \delta_{ij} \ , \label{inner1} \\
&& \left| \left \langle \psi_i | \psi_j \right\rangle \right|^2  = \delta_{ij} + \frac{Z_2}{Z_1^2} \ , \label{inner2}
\end{eqnarray}
The second line above cannot be obtained from the first line, with the usual notion of an inner product between states in a Hilbert space. The only possibilities that resolve this appear to be the following: (i) $Z_2=0$ identically for every gravitational theory, or, these saddles are always prohibited for some generic reason\footnote{One such potential reason could be that these Wormhole solutions are always unstable saddles, and therefore the true path integral will never include them, instead will be dominated by stable saddles.}, (ii) gravitational path integral computes an averaged quantity. While the first possibility is easy to understand, at present, it is not clear how to formulate a general gravitational argument for this. The second possibility is therefore worth visiting more carefully.

The latter requires us to replace the first line in (\ref{inner1}), by:
\begin{eqnarray}
 \left \langle \psi_i | \psi_j \right\rangle  = Z_1 \delta_{ij} + r_{ij} e^{-S_0/2} \ , \label{gravaverage}
\end{eqnarray}
where $r_{ij}$ are random (or, even pseudo-random) variables, with a vanishing mean and a non-vanishing standard deviation. At this point, we do not need to make any assumptions about the higher moments of the distribution function $P[r_{ij}]$ -- that, therefore, can be chosen to be a Gaussian. Suppose we declare that the gravity Euclidean path integral computes the following $\int [dr_{ij}] \left \langle \psi_i | \psi_j \right\rangle$, with an appropriate measure for $r_{ij}$ -- such as the one provided by a Gaussian distribution. This yields: $\int [dr_{ij}] \left \langle \psi_i | \psi_j \right\rangle \sim \delta_{ij}$, as $\int [dr_{ij}] r_{ij} =0$. However,
\begin{eqnarray}
\int [dr_{ij}] [dr_{ji}] \left| \left \langle \psi_i | \psi_j \right\rangle \right|^2 = \int [dr_{ij}] [dr_{ji}] \left( \delta_{ij} \delta_{ji } + r_{ij} r_{ji} e^{-S_0}\right) = 1 + \frac{Z_2} {Z_1^2} \ . 
\end{eqnarray}

Ordinarily, in the standard lore of AdS/CFT, or Holography, there is no such notion of averaging. For example, the celebrate duality between string theory in AdS$_5\times S^5$ (or, in type IIB supergravity) and $\N=4$ super Yang-Mills theory at the conformal boundary of the AdS$_5$ is a statement about the equivalence between the gravity path integral and the gauge theory path integral, without invoking any averaging. One may conclude therefore that this gravitational averaging is somehow a low-dimensional effect.

At this point, let us make a qualitative similarity-check between the discussion above and what we alluded to earlier in section \ref{sec:instanton}, especially in the light of equation (\ref{aveqm}). First, note that the ``connected" and the ``disconnected" geometries in equation (\ref{purity2}) are analogous to the $q=\pm 1$ and the instanton solution in (\ref{soleuclidean}), respectively. In particular, the Euclidean Wormhole geometries are similar to the instanton configurations in a classical scalar field theory. The corresponding semi-classical quantization couples the local fluctuation modes around both the ``disconnected" and the ``connected" configurations. Subsequently, the ``effective" partition function (or the path integral) in (\ref{aveqm}) can be written in terms an averaged source-terms, which couples the local fluctuation modes at $\pm \infty$ with each other. A qualitative similarity can be drawn between equation (\ref{aveqm}) and equation (\ref{gravaverage}), where an averaging is introduced. It is also important to note that, at this level, it is only an analogy and there are several quantitative differences. For example, instantons do not cause any factorization problem in QFT. The Euclidean Wormholes, in the purely gravitational picture, are also not puzzling. It is only through AdS/CFT, when the existence of the Wormholes imply a non-trivial correlation between two otherwise decoupled boundary theories, the issue of factorization problem emerges. The averaging prescription in (\ref{gravaverage}) is one way to reconcile with this issue. On perhaps a more technical point, while the effective averaging in (\ref{aveqm}) only involves a uniform distribution of the coupling constant, the averaging in (\ref{gravaverage}) requires an averaging over a Gaussian distribution. While it is not completely clear whether this analogy has predictive powers, we will simply note that this appears more than a coincidence. For example, there are analogue configurations in gravity, known as ``half-Wormholes"\cite{Saad:2021rcu}, to the half-instanton configurations in QFT. Such Wormholes do not require an averaging prescription, the analogous instantons in QFT do not couple the fluctuation modes at the two asymptotia.

Let us ponder over the issue of averaging, from a slightly different perspective. At a mathematical level, the notion in (\ref{gravaverage}) is very similar to the celebrated Eigenstate Thermalization Hypothesis (ETH) in the context of many-body quantum dynamics. The idea here is the following: Given a quantum mechanical system, with a local Hamiltonian and its corresponding eigenbasis $| E_i\rangle $, the expectation value of a simple, $q$-local self-adjoint operator, $\O_A$, is given by
\begin{eqnarray}
&& \left \langle E_i | \O_A |E_j \right\rangle  = f_A(E) \delta_{ij} + e^{-S(E)/2} G_A \left( E, \omega \right) R_{ij} \ ,\label{eth1} \\
&& E = \frac{1}{2}\left( E_i + E_j \right) \ , \quad \omega = E_i - E_j \ , 
\end{eqnarray}
where $f_A, G_A$ are smooth functions, $R_{ij}$ are pseudo-random variables, such that
\begin{eqnarray}
&& \overline{R_{ij} R_{ji}} = 1 \ , \\
&& \overline{R_{ij} R_{jk} R_{kl } R_{li}} = \delta_{ik} + \delta_{jl} + e^{-S(\tilde{E})} g_A\left(\tilde{E}, \omega_1, \omega_2, \omega_3\right)  \ , \\
&& \tilde{E} = \frac{E_i + E_j +E_k + E_l}{4} \ , \quad \omega_1 = E_i - E_j \ ,  \omega_2 = E_j- E_k \ , \omega_3 = E_k - E_l \ . \nonumber\\
\end{eqnarray}
In the above, $g_A$ is again a smooth function and encodes the non-Gaussianity of the pesudo-random variables $R_{ij}$. Furthermore:
\begin{eqnarray}
&& f_A(E) = \Tr \left( \rho \O_A \right)  \ , \quad {\rm with} \quad \rho= e^{-\beta H} \ , \\
&& {\rm such} \, \, \,   {\rm that} \quad E = \Tr \left(\rho H \right) \ , \label{betasol}
\end{eqnarray}
and $S(E)$ is the microcanonical entropy. Equation (\ref{betasol}) is used to find $\beta$, for a given $E$. It is important that one considers a sufficiently local operator $\O_A$ for ETH, the Hamiltonian $H$ is itself an exception to this since it contains all possible interactions of the system and in this sense not sufficiently local.\footnote{Upon evaluating the expectation value of the Hamiltonian, only the diagonal term in ETH remains, the off-diagonal terms are identically zero.}

The ETH is a cornerstone in understanding thermalization of a closed isolated system, in that it also defines a precise notion of thermalization in terms of the expectation value of a given operator. It is generally understood that systems which thermalize, satisfy the ETH. Note that, ETH is understood to hold for individual systems, even though there is a notion of an averaging over random (or, pseudo-random) variables. From this perspective, the gravitational averaging in (\ref{gravaverage}) is similar. To push these ideas further, note that non-Gaussianity in the ETH can be directly measured by computing three-and-higher-point functions, {\it e.g.}~$\left \langle \O_1 \O_2 \O_3 \right \rangle_{\rm vac}$, by inserting a complete set of energy eigenbasis between the operators and repeatedly using the ETH. This can be directly calculated by {\it e.g.}~$\Tr (\rho_{\rm R}^3)$, which evaluates the third moment of $r_{ij}$ in the gravitational picture, and so on. Note, however, that even an exact Gaussian distribution would have reproduced the expected fine-grained information.

\subsection{Euclidean Wormholes \& Entanglement}

In this section we will briefly review the role of Euclidean Wormholes in encoding the entanglement properties of the boundary CFT. As we have mentioned earlier, these Wormholes are generalizations of Einstein-Rosen bridges and can further be made traversable with appropriate ingredients.

Recall our discussion in section \ref{sec:eadstfd}. Such a TFD-state can be prepared in the lab by designing an appropriate Hamiltonian of which the TFD-state is a ground state. In general dimensions, for a quantum mechanical system that satisfied the ETH (see {\it e.g.}~equation (\ref{eth1}), where, now, one takes a Gaussian distribution for the $R_{ij}$-variables), the (aproximate) TFD Hamiltonian takes a particularly simple form:
\begin{eqnarray}
H_{\rm TFD} = H_{\rm L} + H_{\rm R} + \sum_k \alpha_k \A_k^\dagger \A_k \ , \quad \A_k = \O_{\rm L}^k - (\O_{\rm R}^k)^\dagger \ , 
\end{eqnarray}
where $H_{\rm L, R}$ are the original Hamiltonian for the left and the right theory, $\O_{\rm L, R}$ are local operators in the respective left and right systems. Note that, while the operators $\O$ are themselves local, one needs a non-local interaction in the Hamiltonian to directly couple the left and the right degrees of freedom. In general, it is also possible to construct a similar TFD Hamiltonian whose ground state is the TFD state itself, see \cite{Cottrell:2018ash} for more details.

Generalizing this idea further, let us note that given $n$-copies of a quantum mechanical system or a quantum mechanical system partitioned into $n$-subsystems. One can similarly construct an $n$-fold state of the combined system. So, the full system consists of a tensor product of $n$ Hilber spaces: $\H_1 \otimes \ldots \H_n$, where the indices are just book keeping parameters for $n$-copies of the same system; or, they keep track of $n$-partitions of a single system. The corresponding $n$-fold state can be written as:
\begin{eqnarray}
|n-{\rm fold}\rangle  = \sum_{i_1\ldots i_n} A_{i_1\ldots i_n} |n \rangle_1 \ldots | n \rangle_n \ , 
\end{eqnarray}
where $A_{i_1 \ldots i_n}$ are the corresponding coefficients. The entanglement structure of such a multi-partite system is richer than a simpler bi-partition of it.

To better understand the differences, let us look closer at some basic features of a bi-partition. Consider a bi-partition of a Hilbert space: $\H = \H_A \otimes \H_B$ and a pure, arbitrary state $|\psi \rangle_{AB} \in \H$. Suppose $\{| i\rangle_A \}$ and $\{|\mu\rangle_B\}$  are orthonormal basis for $\H_A$ and $\H_B$, respectively. So, a general state $|\psi \rangle_{AB}$ can be written as:
\begin{eqnarray}
|\psi \rangle_{AB} = \sum_{i} | i \rangle_A | \tilde{i} \rangle_B  \ , \quad | \tilde{i} \rangle_B = \sum_\mu a_{i, \mu} | \mu \rangle_B \ . \label{bipartite}
\end{eqnarray}
Evidently, $\{| \tilde{i} \rangle_B\}$ is not an orthonormal basis. Now, without any loss of generality, we can choose $\{| i \rangle_A\}$ such that the reduced density matrix $\rho_A =\Tr_B \rho_{AB}= {\rm diag} (p_1, \ldots p_D)$, where $D = {\rm dim}\H_A$. Starting with the state in (\ref{bipartite}), and computing the reduced density matrix will yield: $\langle \tilde{i} | \tilde{j}\rangle_B = p_i \delta_{ij}$. Therefore, $\{ \sqrt{p_i} | \tilde{i}\rangle_B\}$ is an orthonormal basis for $\H_B$. Thus, the original bipartite state in (\ref{bipartite}) can be written as:
\begin{eqnarray}
|\psi \rangle_{AB} = \sum_{i} p_i^{1/2} | i \rangle_A | i' \rangle_B \ , \label{schmidt}
\end{eqnarray}
where $\{|i \rangle_A\}$ and $\{|i' \rangle_A\}$ are orthonormal basis for $\H_A$ and $\H_B$, respectively. This is known as the Schmidt decomposition. With this decomposition, it is now natural to demarcate entangled states and separable states. For example, given a bi-partite state, if the number of non-zero eigenvalues of either reduced density matrix $\rho_A$ or $\rho_B$ is larger than one, the the corresponding bi-partite state is entangled. Similarly, if there is only one non-vanishing eigenvalue of the reduced density matrix, then the bi-partite state is separable. This is intuitive, since in the Schmidt decomposition, for an entangled state, we will have more than one terms in the RHS of (\ref{schmidt}), whereas for a separable state, there will be only one term.

One important difference between a bi-partite state and a multi-partite state is that a multi-partite state does not admit a Schmidt decomposition in general,\footnote{In genral, the Schmidt decomposition of a muti-partite state is possible only if by tracing out any subsytem one obtains a separable state.} and therefore it is subtle to define notions of entangled states and separable states. Let us explore a bit more with an instructive example. Perhaps the simplest yet richest example of a three-partite system is the so-called Greenberger-Horne-Zeilinger (GHZ) state, which can be written as:
\begin{eqnarray}
\left| {\rm GHZ}\right \rangle = \frac{1} {\sqrt{2} }\left( \left | 000 \right \rangle + \left | 111 \right  \rangle \right ) \ . 
\end{eqnarray}
If we trace out any one of the three sub-systems, the resulting reduced density matrix takes the form: $(1/2) \left( \left | 00 \right \rangle \left \langle 00 \right | + \left | 11 \right  \rangle \left \langle 11 \right | \right) $, which is a classical mixed density matrix. On the other hand, suppose we rewrite the basis of the third sub-system, by introducing $| 0 \rangle = \frac{1}{\sqrt{2}} \left( | + \rangle + | - \rangle \right) $ and $| 1 \rangle = \frac{1}{\sqrt{2}} \left( | + \rangle - | - \rangle \right)$, and subsequently measure by projecting on the $| +  \rangle$ or the $| - \rangle$ state. This will yield a reduced density matrix of the form: $\left( \left | 00 \right \rangle + \left | 11 \right \rangle\right) \left( \left \langle 00 \right | + \left \langle 11 \right | \right) $, which is a maximally entangled one. Thus, even for this simple three-partite state it is clear that the pairwise entanglement structure is somewhat subtle and richer.

\subsection{Multi-boundary Wormholes: Explicit Constructions}\label{EWormMulti}

Motivated by this, it is natural to explore multi-partite entanglement structure in AdS/CFT. The simplest is the TFD state itself, which is dual to the eternal Black Hole in AdS\cite{Maldacena:2001kr}. The eternal Black Hole geometry is a Lorentzian background, with two asymptotic conformal boundaries in which two copies of the same CFT are defined. For a multi-partite state, the natural candidate is a geometry with $n$ conformal boundaries, corresponding to a tensor product of $n$ copies of the CFT Hilbert space. Though, these geometries are inherently Lorentzian, to prepare such a state in the gravitational theory, one considers the standard Hartle-Hawking approach: Consider the Euclidean gravitational path integral and look for Wormhole saddle solutions with the desired boundary behaviour. Along a moment of time-symmetry, both the Lorentzian and the Euclidean contain a hypersurface of vanishing extrinsic curvature. The Hartle-Hawking state can be constructed by gluing the Euclidean section with the Lorentzian section along this hypersurface.

To be more precise with the explicit construction, let us consider AdS$_3$, in which the Wormholes can be explicitly constructed by extensive use of the symmetry of the system. Let us begin with an elementary review of AdS$_3$ and its isometries: It is well-known that an AdS$_3$ geometry can be described by a hypersurface embedded in a $(2+2)$-dimensional Minkowski geometry. The hyperboloid is defined by 
\begin{eqnarray}
-\ell^2 = -\bv^2 - \bu^2 + \bx^2 + \by^2 \ , 
\end{eqnarray}
embedded in ${\mathbb R}^{2,2}$, with a curvature set by $\ell$. There are various ways of defining a local co-ordinate on the AdS hyperboloid, let us focus on the Poincare patch below:
\begin{eqnarray}
\frac{ds^2}{\ell^2} = \frac{-dt^2 + dx^2 + dy^2}{y^2} \ .
\end{eqnarray}
The $t=0$ surface has an induced metric $ds^2 = (1/y^2)(dx^2 + dy^2)$, that describes a hyperbolic $2$-manifold, denoted by ${\mathbb H}^2$. The full AdS$_3$ has an ${\rm SO}(2,2)\equiv {\rm SL}(2, R) \times {\rm SL}(2, R)$ isometry, and by quotienting it with a discrete subgroup, $\Gamma$, one generates non-trivial but locally AdS$_3$ geometries. Typically, the sub-group is chosen to belong to a diagonal ${\rm SL}(2,R)$.\footnote{Such discrete groups are known as Fuchsian groups.} The resulting quotient ${\mathbb H}^2/\Gamma$ is a smooth Riemann surface with genus $g$ and boundaries $b$, denoted by $(g, b)$. Correspondingly, the action of $\Gamma$ on the entire AdS$_3$ geometry is realized by its action on the $t=0$ slice. This ensures that both the Lorentzian and the Euclidean geometries admit a similar quotienting method. In the case of Euclidean AdS$_3$ geometry, the isometry group is ${\rm SO}(3,1) \equiv {\rm SL}(2,C)$ and $\Gamma$ belongs to a diagonal ${\rm SL}(2,R)$. This ensures that the analytic continuation, after quotienting by $\Gamma$, remains real. In this section, we will review how multi-boundary (Wormhole) geometries can be constructed using this quotient procedure, following mainly \cite{Caceres:2019giy}.

Note that, each isometry is in one-to-one relation with a Killing vector in the geometry. The identification ``by an isometry", or ``by a Killing vector" implies that we identify all points in the orbit of the corresponding group action. For example, given a Killing vector $\xi$, we consider $\{e^{\alpha\xi}\}_{\alpha \in {\mathbb R}}$, as the corresponding one-parameter sub-group. Each Riemann surface $(g, b)$ --- which has $(6g +3b -6)$ parameters, will correspond to a particular Wormhole geometry. Among these, $(3g+2b-3)$ are lengths of minimal, non-intersecting, periodic geodesics and $3g+b-3$ are the so-called twist parameters\cite{Skenderis:2009ju}. Among these, the Black Hole horizons correspond to the length of the periodic geodesics, and the remaining parameters belong to the interior degrees of freedom\cite{Skenderis:2009ju}.

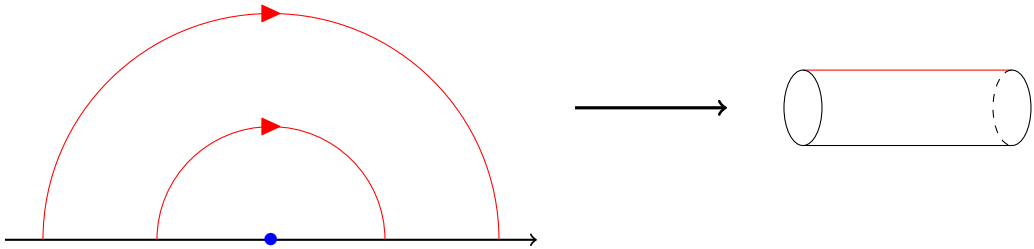
\begin{figure}
\centering
\begin{tikzpicture}
\draw[->,thick] (-3.5,0) to (3.5,0);
\draw[-,color=red] (-1.5,0) arc (180:0:1.5);
\draw[-,color=red] (3,0) arc (0:180:3);

\node[color=red,rotate=-90] at (0,1.5) {$\blacktriangle$};
\node[color=red,rotate=-90] at (0,3) {$\blacktriangle$};

\draw[->,very thick] (4,1.75) to (6,1.75);

\draw[-,color=red] (7,2.25) to (9.75,2.25);
\draw[-] (7,1.25) to (9.75,1.25);
\draw[-] (7,1.75) ellipse (0.25 and 0.5);
\draw[-] (9.75,2.25) arc (90:-90:0.25 and 0.5);
\draw[-,dashed] (9.75,2.25) arc (90:270:0.25 and 0.5);

\node[color=blue] at (0,0) {$\bullet$};
\end{tikzpicture}
\caption{The Riemann surface obtained by quotienting the upper half-plane by dilatation: $(t,x,y) \sim e^{2\pi \kappa} (t,x,y)$, as described in equation (\ref{dilAct}). The fundamental domain is bounded by the red semicircles, while the fixed point of the identification is the center of the semicircles. This is the $t = 0$ slice of the two-sided BTZ. This figure is taken from \cite{Caceres:2019giy}.}
\label{fig10}
\end{figure}

The structure of Riemann surfaces has been extensively used in \cite{Skenderis:2009ju, Balasubramanian:2014hda} to construct the $(g, b)$ Riemann surface, then lifting it to beyond the $t=0$-slice. The simplest example of this is the construction of a (two-sided) BTZ geometry, as described in figure \ref{fig10}. The construction methods in \cite{Skenderis:2009ju, Balasubramanian:2014hda} are patchy, piece-wise. Following \cite{Caceres:2019giy}, we will now describe a global approach of constructing the Wormhole geometries, by explicitly constructing the Killing vectors corresponding to that would yield a three-boundary geometry. 

Towards that, we begin with a suitable basis of Killing vectors on ${\mathbb H}^2$ and subsequently lift them to the full AdS$_3$-geometry (therefore, defined for all values of $t$). Now, geodesics on ${\mathbb H}^2$ are of two kinds: straight lines and semi-circles. To construct multi-boundary Wormholes, we will need to carry out identification of the semi-circular geodesics. The action of the global Killing vectors on these geodesics can be subsequently obtained. In particular, translations simply shift the semi-circles, dilatation scale up the radius as well as the center of the semi-circles and inversion simply flips the orientation of the semi-circle. We now have all the ingredients to construct explicitly the multi-boundary Wormhole geometries. For example, the BTZ-geometry is obtained by an identification shown in figure \ref{fig10}.

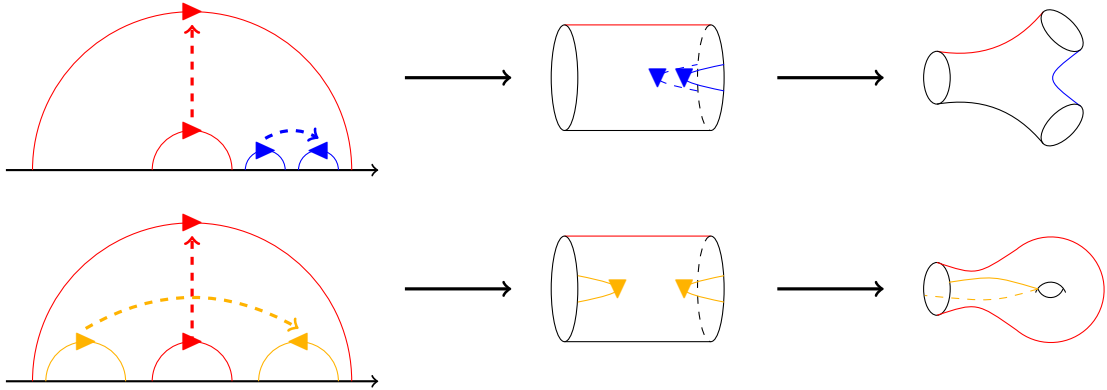
\begin{figure}
\centering
\begin{tikzpicture}[scale=0.7]
\draw[->,thick] (-3.5,0) to (3.5,0);
\draw[-,color=blue] (1,0) arc (180:0:0.75/2);
\draw[-,color=blue] (2.75,0) arc (0:180:0.75/2);

\draw[-,color=red] (0.75,0) arc (0:180:0.75);
\draw[-,color=red] (3,0) arc (0:180:3);

\node[color=blue,rotate=-90] at (1+0.75/2,0.75/2) {$\blacktriangle$};
\node[color=blue,rotate=90] at (2.75-0.75/2,0.75/2) {$\blacktriangle$};

\node[color=red,rotate=-90] at (0,0.75) {$\blacktriangle$};
\node[color=red,rotate=-90] at (0,3) {$\blacktriangle$};

\draw[->,very thick,dashed,red] (0,1) to (0,2.75);
\draw[->,very thick,dashed,blue] (1+0.75/2,0.6) to[bend left] (2.75-0.75/2,0.6);

\draw[->,very thick] (4,1.75) to (6,1.75);

\draw[-,color=red] (7,2.75) to (9.75,2.75);
\draw[-] (7,0.75) to (9.75,0.75);
\draw[-] (7,1.75) ellipse (0.25 and 1);
\draw[-] (9.75,2.75) arc (90:-90:0.25 and 1);
\draw[-,dashed] (9.75,2.75) arc (90:270:0.25 and 1);

\draw[-,dashed,blue] (9.5,1.5) .. controls (8.5,1.7) and (8.5,1.8) .. (9.5,2);
\draw[-,blue] (10,1.5) .. controls (9,1.7) and (9,1.8) .. (10,2);

\node[rotate=-180,color=blue] at (8.75,1.75) {$\blacktriangle$};
\node[rotate=-180,color=blue] at (9.25,1.75) {$\blacktriangle$};

\draw[->,very thick] (11,1.75) to (13,1.75);

\draw[-,color=red] (7+7,2.25) to[bend right] (7+9,3);
\draw[-,color=blue] (7+9.7,2.28) .. controls (7+9,3.49/2) .. (7+9.7,1.21);

\draw[-] (7+7,1.75) ellipse (0.25 and 0.5);
\draw[-] (7+7,1.25) to[bend left] (7+9,0.5);

\draw[-,rotate around={45:(7+9,2.5)}] (7+9.35,2.35) ellipse (0.25 and 0.5);
\draw[-,rotate around={135:(7+9,0.5)}] (7+9,0) ellipse (0.25 and 0.5);

\draw[->,thick] (-3.5,0-4) to (3.5,0-4);
\draw[-,color=red!30!yellow] (2.75,0-4) arc (0:180:0.75);
\draw[-,color=red!30!yellow] (-2.75,0-4) arc (180:0:0.75);

\draw[-,color=red] (0.75,0-4) arc (0:180:0.75);
\draw[-,color=red] (3,0-4) arc (0:180:3);

\node[color=red!30!yellow,rotate=90] at (2,0.75-4) {$\blacktriangle$};
\node[color=red!30!yellow,rotate=-90] at (-2,0.75-4) {$\blacktriangle$};

\node[color=red,rotate=-90] at (0.,0.75-4) {$\blacktriangle$};
\node[color=red,rotate=-90] at (0,3-4) {$\blacktriangle$};

\draw[->,very thick,dashed,red] (0,0.7-4) to (0,2.75-4);
\draw[->,very thick,dashed,red!30!yellow] (-2,1-4) to[bend left] (2,1-4);

\draw[->,very thick] (4,1.75-4) to (6,1.75-4);

\draw[-,color=red] (7,2.75-4) to (9.75,2.75-4);
\draw[-] (7,0.75-4) to (9.75,0.75-4);
\draw[-] (7,1.75-4) ellipse (0.25 and 1);
\draw[-] (9.75,2.75-4) arc (90:-90:0.25 and 1);
\draw[-,dashed] (9.75,2.75-4) arc (90:270:0.25 and 1);

\draw[-,red!30!yellow] (7.25,1.5-4) .. controls (8.25,1.7-4) and (8.25,1.8-4) .. (7.25,2-4);
\draw[-,red!30!yellow] (10,1.5-4) .. controls (9,1.7-4) and (9,1.8-4) .. (10,2-4);

\node[rotate=-180,red!30!yellow] at (8,1.75-4) {$\blacktriangle$};
\node[rotate=-180,red!30!yellow] at (9.25,1.75-4) {$\blacktriangle$};

\draw[->,very thick] (11,1.75-4) to (13,1.75-4);

\draw[-] (7+7,1.75-4) ellipse (0.25 and 0.5);
\draw[-,color=red] (7+7,2.25-4) .. controls (7+7.75,2-4) .. (7+8.5,2.5-4) arc(130:-130:1) .. controls (7+7.75,1.5-4) .. (7+7,1.25-4);

\draw[-] (7+8.9,1.75-4) arc (200:340:0.25);
\draw[-] (7+8.855,1.68-4) arc (160:20:0.3);

\draw[-,color=red!30!yellow] (7+7.225,1.875-4) .. controls (7+7.225/2+8.9/2,2-4) .. (7+8.9,1.75-4);
\draw[-,color=red!30!yellow,dashed] (7+8.9,1.75-4) .. controls (7+7.225/2+8.9/2,1.5-4) .. (7+6.775,1.625-4);
\end{tikzpicture}
\caption{The three-boundary and one-boundary, one-genus Riemann surfaces as quotients of the two-boundary Riemann surface. The three-boundary surface is obtained by ``pinching" one of the boundaries into two, while the one-boundary, one-genus surface is obtained by ``folding" one of the boundaries onto the other. This figure is taken from \cite{Caceres:2019giy}.}
\label{figs:pinchfold}
\end{figure}

To construct a multi-boundary geometry, let us first generalize the BTZ identification figure, for richer Riemann surfaces and in particular, a three-boundary Wormhole geometry. See figure \ref{figs:pinchfold}. 
The important point here is that, in order to create a Wormhole geometry with a non-trivial genus and a boundary, one needs an orientation-reversing isometry with which some quotienting needs to be carried out. This is explicitly presented in (\ref{orientrev}).

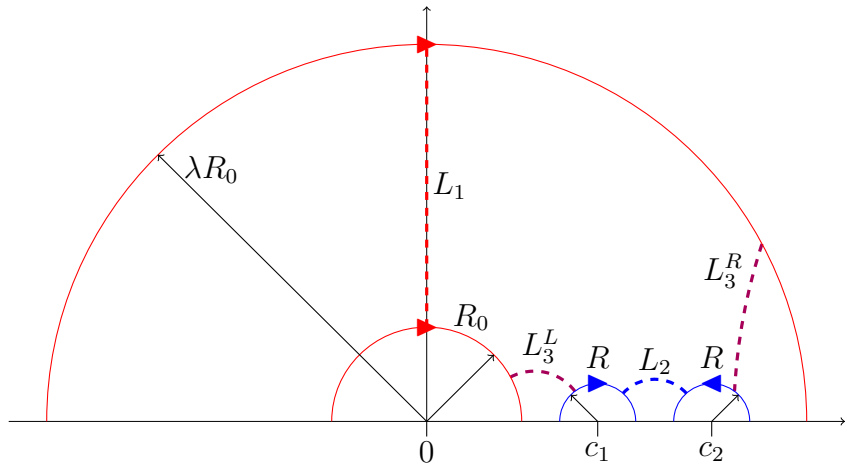
\begin{figure}
	\[
	\begin{tikzpicture}
	\draw[->,very thin] (-5.5,0) to (5.5,0);
	\draw[->,very thin] (0,0) to (0,5.5);
	
	\draw[-,dashed,very thick,color=red] (0,1.25) to (0,5);
	\draw[-,dashed,very thick,color=blue] (41/12,0.372678) arc (41.81:138.19:0.559017);
	\draw[-,dashed,very thick,color=red!65!blue] (39/20,2/5) arc (36.87:118.07:2/3);
	\draw[-,dashed,very thick,color=red!65!blue] (4*75/68,4*10/17) arc (162.02:176.99:7.62209);
	
	\draw[-,color=red] (5,0) arc (0:180:5);
	\draw[-,color=red] (1.25,0) arc (0:180:1.25);
	
	\draw[-,color=blue] (1.75,0) arc (180:0:0.5);
	\draw[-,color=blue] (3.25,0) arc (180:0:0.5);
	
	\node[color=red,rotate=-90] at (0,1.25) {$\blacktriangle$};
	\node[color=red,rotate=-90] at (0,5) {$\blacktriangle$};
	
	\node[color=blue,rotate=-90] at (2.25,0.5) {$\blacktriangle$};
	\node[color=blue,rotate=90] at (3.75,0.5) {$\blacktriangle$};
	
	\draw (0,0) to (0,-0.2);
	\draw (2.25,0) to (2.25,-0.2);
	\draw (3.75,0) to (3.75,-0.2);
	
	\node at (0,-0.4) {$0$};
	\node at (2.25,-0.4) {$c_1$};
	\node at (3.75,-0.4) {$c_2$};
	
	\draw[->] (0,0) to (1.25/1.414,1.25/1.414);
	\node at (1.25/1.414-0.3,1.25/1.414+0.5) {$R_0$};
	
	\draw[->] (0,0) to (-5/1.414,5/1.414);
	\node at (-5/1.414+0.7,5/1.414-0.2) {$\lambda R_0$};
	
	\draw[->] (2.25,0) to (2.25-0.5/1.414,0.5/1.414);
	\node at (2.25,0.5/1.414+0.5) {$R$};
	
	\draw[->] (3.75,0) to (3.75+0.5/1.414,0.5/1.414);
	\node at (3.75,0.5/1.414+0.5) {$R$};
	
	\node at (0.3,3.125) {$L_1$};
	\node at (3,0.8) {$L_2$};
	\node at (1.5,1) {$L_{3}^L$};
	\node at (3.9,2) {$L_3^R$};
	\end{tikzpicture}
	\]
	\caption{The fundamental domain of the three-boundary Riemann surface. The color-coded dashed lines $L_{1,2,3}$ are the minimal periodic geodesics, whose lengths are the three \textbf{physical parameters} of the system. The variables $\lambda$, $R_0$, $R$, $c_1$, and $c_2$ represent parameters for the picture. This figure is taken from \cite{Caceres:2019giy}.}
	\label{figs:3bdrypic}
\end{figure}

A more elaborate and precise pictorial representation is given in figure \ref{figs:3bdrypic}. 
The dashed geodesics in figure \ref{figs:3bdrypic}, after the identification, become periodic and subsequently imposing the minimal (within the homotopy class) condition, these become the Black Hole horizons. The corresponding isometry is presented in equation (\ref{orientrevfig12}). Furthermore, it can also be shown that the three minimal periodic geodesics yield three independent horizons. For more details, we refer the reader to \cite{Caceres:2019giy}.\footnote{As an example, the length of the minimal periodic geodesic for a two-sided BTZ geometry, from figure \ref{figs:3bdrypic}, is given by
\begin{eqnarray}
L_1 = \ell \int_{R_0}^{\lambda R_0} \frac{dy}{y} = \ell \log \lambda \ . 
\end{eqnarray}
Other periodic geodesic calculations are similar in nature, but algebraically more involved. }

It is natural to extend this construction to a Riemann surface $(g,b)$, which captures all physical parameters. We will now describe how this works for $(1,1)$-Wormhole. This has also been explicitly constructed in \cite{Aminneborg:1997pz, Aminneborg:1998si, Krasnov:2001va}, however the full moduli space was not provided. The $(1,1)$-Wormhole is also a two-step identification process, the logical steps are similar to the construction of the $(0,3)$-Wormhole. We demonstrate the details of this identification, at $t=0$ slice, in figure \ref{figs:11pic}.
\begin{figure}
	\[
	\begin{tikzpicture}
	\draw[->,very thin] (-5.5,0) to (5.5,0);
	\draw[->,very thin] (0,0) to (0,5.5);
	
	\draw[-,dashed,very thick,color=red] (0,1.25) to (0,5);

	\draw[-,dashed,very thick,color=red!30!yellow] (1.25/1.06,1.25/2.92) arc (120:45:0.6);
	\draw[-,dashed,very thick,color=red!30!yellow] (-3.75+0.5/1.414,0.5/1.414) arc (160:23.5:1.2);
	\draw[-,dashed,very thick,color=red!30!yellow] (-5/1.06,5/2.92) arc (30:16.75:6.5);
	\draw[-,dashed,very thick,color=red!30!yellow] (2.25+0.5/1.414,0.5/1.414) arc (160:87.5:2.125);

	\draw[-,color=red] (5,0) arc (0:180:5);
	\draw[-,color=red] (1.25,0) arc (0:180:1.25);
	
	\draw[-,color=red!30!yellow] (1.75,0) arc (180:0:0.5);
	\draw[-,color=red!30!yellow] (-3.25,0) arc (0:180:0.5);
	
	\node[color=red,rotate=-90] at (0,1.25) {$\blacktriangle$};
	\node[color=red,rotate=-90] at (0,5) {$\blacktriangle$};
	
	\node[color=red!30!yellow,rotate=-90] at (2.25,0.5) {$\blacktriangle$};
	\node[color=red!30!yellow,rotate=90] at (-3.75,0.5) {$\blacktriangle$};
	
	\draw (0,0) to (0,-0.2);
	\draw (2.25,0) to (2.25,-0.2);
	\draw (-3.75,0) to (-3.75,-0.2);
	
	\node at (0,-0.4) {$0$};
	\node at (2.25,-0.4) {$c_1$};
	\node at (-3.75,-0.4) {$c_2$};
	
	\draw[->] (0,0) to (1.25/1.414,1.25/1.414);
	\node at (1.25/1.414-0.3,1.25/1.414+0.5) {$R_0$};
	
	\draw[->] (0,0) to (-5/1.414,5/1.414);
	\node at (-5/1.414+0.7,5/1.414-0.2) {$\lambda R_0$};
	
	\draw[->] (2.25,0) to (2.25-0.5/1.414,0.5/1.414);
	\node at (2.25,0.5/1.414+0.5) {$R$};
	
	\draw[->] (-3.75,0) to (-3.75+0.5/1.414,0.5/1.414);
	\node at (-3.75,0.5/1.414+0.5) {$R$};
	
	\node at (0.3,3.125) {$L_1$};
	
	\node at (1.25/2.12+1.125-0.5/2.828,0.9) {$L_2^{(1)}$};
	\node at (-1.25/2.12-1.875+0.5/2.828,1.5) {$L_2^{(2)}$};
	\node at (-4.125,1.5) {$L_2^{(3)}$};
	\node at (4,2) {$L_2^{(4)}$};
	
	\end{tikzpicture}
	\]
	\caption{The fundamental domain of the $(1,1)$ Riemann surface. The color-coded dashed lines $L_{1,2}$ are non-intersecting periodic geodesics. In this system, two of the physical parameters are lengths of the minimal periodic geodesic, while the third physical parameter is a \textit{twist}. This picture is taken from \cite{Caceres:2019giy}.}
	\label{figs:11pic}
\end{figure}
Similarly, a $(1,2)$-Wormhole can also be generated by quotienting with the appropriate Killing vector, following a similar logical sequence. For further visual appeal, we have also demonstrated the details of this identification, at $t=0$-slice, in figure \ref{figs:12pic}.
\begin{figure}
\centering
\begin{tikzpicture}[scale=0.75]
\draw[->,thick] (-6.5,0) to (6.5,0);

\draw[-,color=red] (6,0) arc (0:180:6);
\draw[-,color=red] (1.25,0) arc (0:180:1.25);

\draw[-,color=blue] (1.75,0) arc (180:0:0.5);
\draw[-,color=blue] (4,0) arc (180:0:0.5);

\draw[-,color=red!30!yellow] (3,0) arc (180:0:0.375);
\draw[-,color=red!30!yellow] (5.125,0) arc (180:0:0.375);

\draw[-,color=red!30!yellow] (-1.75,0) arc (0:180:0.75);
\draw[-,color=red!30!yellow] (-5.5,0) arc (180:0:1);

\node[color=red,rotate=-90] at (0,1.25) {$\blacktriangle$};
\node[color=red,rotate=-90] at (0,6) {$\blacktriangle$};

\node[color=blue,rotate=-90] at (2.25,0.5) {$\blacktriangle$};
\node[color=blue,rotate=90] at (4.5,0.5) {$\blacktriangle$};

\node[color=red!30!yellow,rotate = -90] at (3.375,0.375) {$\blacktriangle$};
\node[color=red!30!yellow,rotate = -90] at (5.5,0.375) {$\blacktriangle$};

\node[color=red!30!yellow,rotate = 90] at (-2.5,0.75) {$\blacktriangle$};
\node[color=red!30!yellow,rotate = 90] at (-4.5,1) {$\blacktriangle$};

\draw[->,very thick,dashed,red] (0,1.5) to (0,5.75);
\node at (0.4,7.25/2) {$(1)$};

\draw[->,very thick,dashed,blue] (2.25,0.75) to[bend left] (4.5,0.75);
\node at (3.375,1.4) {$(2)$};

\draw[->,very thick,dashed,red!30!yellow] (3.375,0.625) to[bend right] (-2.5,1);
\node at (-2,1.6) {$(3)$};

\draw[->,very thick,dashed,red!30!yellow] (5.5,0.625) to[bend right] (-4.5,1.25);
\node at (-4,1.85) {$(4)$};
\end{tikzpicture}
\caption{The fundamental domain of the $(1,2)$ Riemann surface. Step (1) is quotienting by dilatation. Step (2) is quotienting by a pinching. Steps (3) and (4) are quotienting by foldings. This picture is taken from \cite{Caceres:2019giy}.}
\label{figs:12pic}
\end{figure}
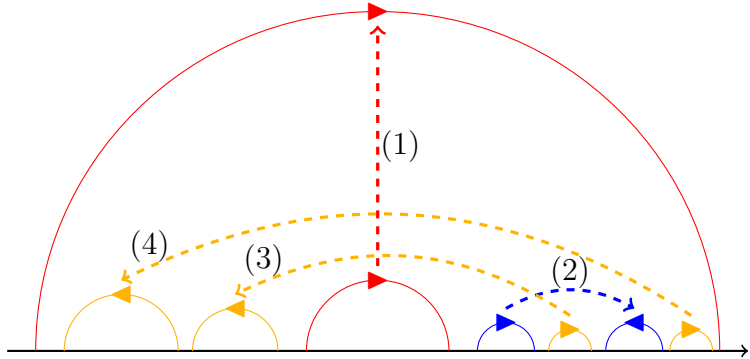

The astute reader will notice that the crucial difference between the figures in \ref{figs:11pic} and \ref{figs:12pic} is whether the identifying circles are on the same side or on the different sides of the center. Essentially, identifying circles on the same side of the center increases number of boundaries, whereas identifying two circles on the two sides of the center increases its genus. Thus, by repeating this process, one can generate an arbitrary $(g, b)$-Wormhole. This strategy of using explicit Killing vectors can also be used to generate $(g, b)$-Wormhole for a stationary geometry, {\it i.e.}~in rotating multi-boundary geometry.\footnote{Though, in practice, the tractable examples are $(0,3)$-Wormhole and $(1,1)$-Wormhole, as demonstrated in \cite{Caceres:2019giy}.} Hopefully, we have provided the reader with enough technology, as well as enough examples based on which a particular construction can be carried out. We will dwell no longer on the technical details of the construction, but move on to discussing general lessons about multi-partite entanglement based on the multi-boundary Wormhole geometries.

\subsection{Multi-partite Entanglement from Multi-boundary Wormholes}

We will now review entanglement properties based on the multi-boundary Wormhole geometries. These aspects have been explored in detail in \cite{Balasubramanian:2014hda} to which we refer the interested reader for more details. We present a broad-brush demonstration of how multi-boundary Wormhole geometries teach us qualitative lessons about multi-partite entanglement in the dual CFT. The primary tool for these computations is the Ryu-Takayanagi\cite{Ryu:2006bv, Ryu:2006ef} formula and its generalization to the Hubeny-Rangamani-Takayanagi formula\cite{Hubeny:2007xt}.

Consider the three-boundary example, where $L_{1,2,3}$ parametrize the lengths of three horizons denoted by $H_{1,2,3}$ and $B_{1,2,3}$ represent the three boundaries. As explained in \cite{Balasubramanian:2014hda}, the HRT minimal surfaces are given by unions of the horizon lengths. First, it is easy to see that there is a phase transition of the HRT-minimal surface at $L_3 = L_1 + L_2$. Consider, for example, $L_3 > L_1 + L_2$, the minimal surface for $B_1 \cup B_2$ is given by $H_1 \cup H_2$ and thus $S(B_3) = S(B_1 \cup B_2) \sim \left( L_1 + L_2\right) $. For $L_3< L_1 +L_2$, however, the minimal surface is given by $H_3$ and therefore $S(B_1\cup B_2) = S(B_3) \sim L_3$. in this limit, there is vanishing entanglement between $B_1$ and $B_2$, since the corresponding mutual information vanishes: $I(B_1 : B_2) = S(B_1) + S(B_2) - S(B_1 \cup B_2) = 0$. Thus, for $L_3 > L_1 + L_2$, the entanglement is purely bi-partite,\footnote{In terms of mutual information, one obtains: $I(B_1 : B_2)=0$, $I(B_1 : B_3)=2S(B_1)$ and $I(B_2 : B_3)=2S(B_2)$ in this regime.} while multi-partite entanglement exists for $L_3 < L_1+L_2$.

From a purely CFT-perspective, by tracing out the $B_3$-degrees of freedom, one obtains the following reduced density matrix:
\begin{eqnarray}
\rho_{12} = \sum_{iji'j'} \, \rho_{ii'jj'} |i \rangle_1 |j \rangle_2 \langle i' |_1 \langle j' |_2 \ , \quad \rho_{ii'jj'} = \sum_k A_{ijk} A_{i'j'k}^* \ .
\end{eqnarray}
It has been argued in \cite{Balasubramanian:2014hda} that in the limit $L_3\to \infty$, the reduced density matrix is obtained by a sphere partition function with four holes, in an anlaogue of the $t$-channel. The upshot is, in this limit, $\rho_{12}$ factorises: $\rho_{12} = \rho_1 \otimes \rho_2$. Thus, $B_1$ and $B_2$ are completely unentangled. The Holographic picture above further promotes this unentangled behaviour for any finite $L_3 > L_1 + L_2$. This is in stark contrast to a GHZ-state, in which no such regime exists such that the entanglement structure becomes purely bi-partite. Thus, although multi-boundary Wormhole geometries indeed encode multi-partite entanglement, they are unlike the GHZ-states.

\section{Lorentzian Wormholes}

In this section we will begin with a review of the Einstein-Rosen bridge and subsequently review how these Wormholes can be made traversable by introducing more degrees of freedom. We will now focus completely on the Lorentzian description, in terms of specific solutions of (semi)-classical gravity. In this context, ``semi-classical" would imply the existence of a matter stress-tensor which can arise from the expectation value of a quantum matter, in a suitable state. This stress-tensor is then fed into the classical Einstein-equations which are subsequently solved to obtain an on-shell geometric data. Such examples are far too many to enlist them here, rather we make explicit reference to some in later sections. 

The history of Lorentzian Wormholes is quite rich. Let us begin our discussion, stepping back almost a hundred years, with the discussion of the Einstein-Rosen bridge. Much of what we discuss here is standard knowledge, see {\it e.g.}~\cite{Visser:1995cc} on which we will heavily draw. We will also begin our discussion with an asymptotically flat space-time and subsequently discuss asymptotically AdS space-times.

\begin{figure}
  \includegraphics[width=\linewidth]{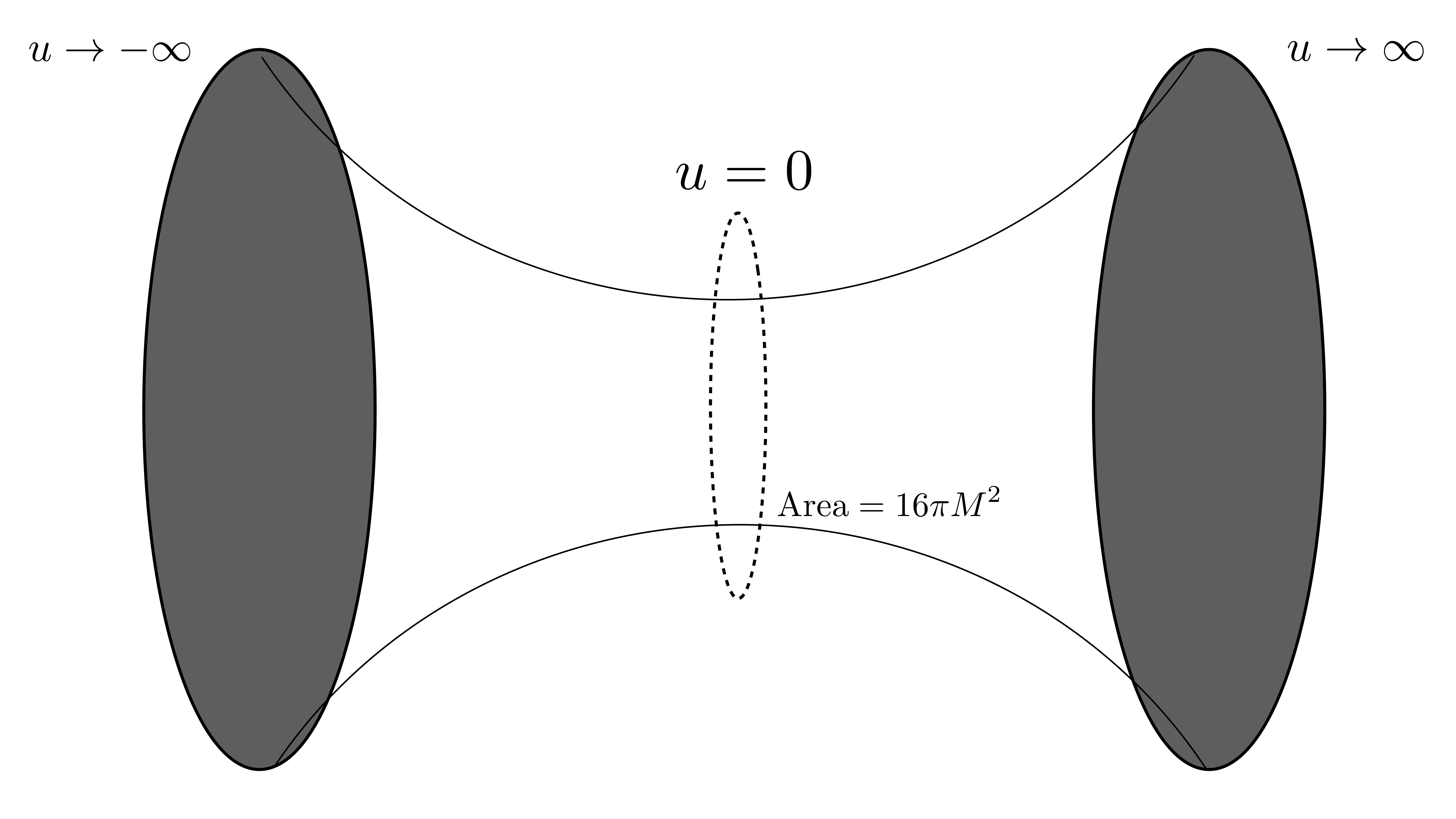}
  \caption{Einstein-Rosen bridge, as described by the metric in (\ref{schER}). The throat is located at $u=0$, where the area cross-section minimizes.}
  \label{er_bridge}
\end{figure}

Let us begin with Schwarzschild geometry in an asymptotically flat space-time in $(3+1)$-dimensions:
\begin{eqnarray}
ds^2 = \left( 1 - \frac{2M}{r} \right) dt^2 + \left( 1 - \frac{2M}{r}\right)^{-1} dr^2 + r^2 d\Omega^2  \ , \label{schflat}
\end{eqnarray}
where $M$ is the mass of the Black Hole and $d\Omega^2$ is the metric on the round unit-sphere. Here $r \in [0, \infty]$. Let us rewrite the geometry in (\ref{schflat}), by introducing a new co-ordinate: $u^2 = r - 2 M$, which yields:
\begin{eqnarray}
ds^2 = - \frac{u^2}{u^2+2M} dt^2 + 4 (u^2 + 2M) du^2 + (u^2+2M)^2 d\Omega^2 \ , \label{schER}
\end{eqnarray}
where $u \in [-\infty, +\infty]$. Clearly, the patch in (\ref{schER}) covers the $r \in [2M , \infty]$ of (\ref{schflat}) twice. Let us note the following salient features of the patch in (\ref{schER}): (i) There are two asymptotia, as $u \to \pm \infty$, (ii) The metric has a symmetry under $u \to - u$, (iii) Given any time-slice, the $u={\rm const}$ surface has an area $A(u) = 4\pi (u^2 + 2 M)^2$, which is minimized at $u=0$. This is pictorially summarized in figure \ref{er_bridge}.
This is what qualitatively defines a Wormhole, known as the Einstein-Rosen bridge. Of course, we have merely introduced a co-ordinate system but have not really done anything new to the Schwarzschild geometry.

Guided by the above features, consider a static, spherically symmetric geometry with an event-horizon:
\begin{eqnarray}
ds^2 = - e^{-\phi(r)} (1-b(r)) dt^2 + \frac{dr^2}{1- b(r)} + r^2 d\Omega^2 \ , \label{wormgen}
\end{eqnarray}
where the horizon is located at $b(r_H) = r_H$. We imagine that an appropriate matter field sources the geometry in (\ref{wormgen}). As before, we can introduce a new co-ordinate system $u^2 = r- r_H$, which yields all the Wormhole features listed above. As above, we have also introduced a new co-ordinate system that makes manifest two asymptotic region at $u \to \pm \infty$, which are connected by a throat at $u=0$. This, however, does not mean that an observer can travel from one asymptotic region to another, through the Wormhole.

Before discussing the traversability aspect of Wormholes, let us briefly review a curious feature of Einstein-Rosen bridges in AdS-BH geometries, which plays a ubiquitous role in many physical questions. In this part, we will closely follow \cite{Stanford:2014jda}. Consider an AdS$_{d+1}$-BH geometry:
\begin{eqnarray}
ds^2 = - f(r) dt^ 2+ \frac{dr^2}{f(r)} + r^ 2 d\Omega_{d-1}^2 \ , \quad f(r) = 1 + \frac{r^2}{\ell_{\rm AdS}^2} - \frac{M^{d-2}}{r^{d-2}} \ , \label{metadsbh}
\end{eqnarray}
where $\ell_{\rm AdS}$ is the AdS-curvature scale and $M$ is the mass of the black hole. Consider a bulk slice, denoted by $t=t(r)$, whose volume functional is given by
\begin{eqnarray}
{\rm Vol}_{d} = 2 {\rm Vol}_{S^{d-1}} \int_{r_{\rm min}}^{r_{\rm max}} d\lambda r^{d-2} \sqrt{f(r)^{-1}r'(\lambda)^2 - f(r) t'(\lambda)^2} \ , \quad ' = \frac{d}{d\lambda} \ . 
\end{eqnarray}
Here $r_{\rm min, max}$ are hitherto unspecified. 
\begin{figure}
  \includegraphics[width=\linewidth]{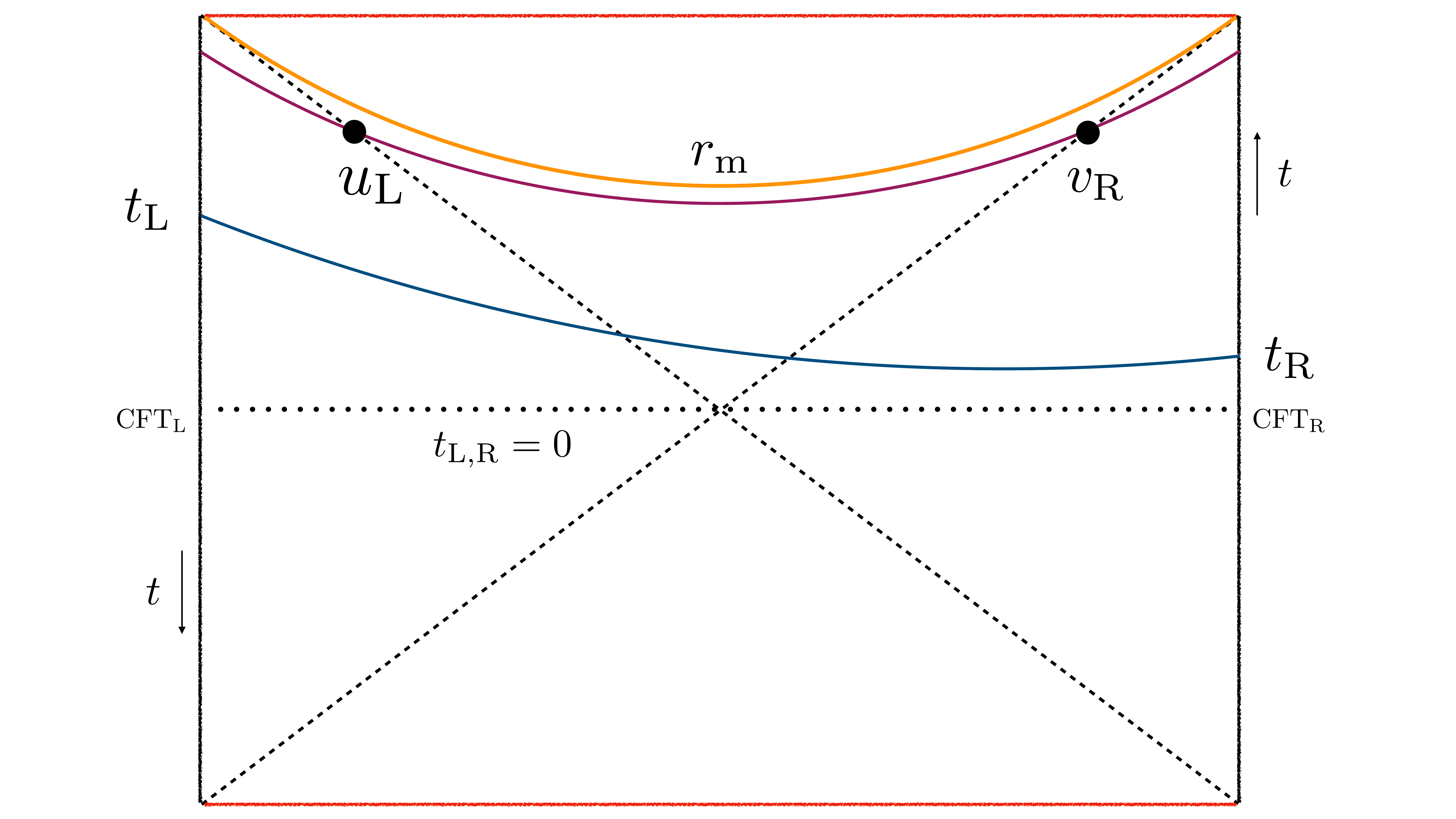}
  \caption{Various bulk slices of the eternal AdS-BH geometry. These slices are anchored at $t_{\rm R}$ on the right boundary and on $t_{\rm L}$ on the left boundary. For $E\ll 1$, the turning point is located near $r_{\rm H}$ and therefore the corresponding slice is close to the $t_{\rm L,R}=0$ one, as shown by the dotted black horizontal line. For $E\gg 1$, the turning point asymptotes to $r_{\rm m}$ and therefore corresponds to $t_{\rm L,R} \to \infty$, as shown by the orange slice. }
  \label{bsworm}
\end{figure}
The corresponding equations of motion are:
\begin{eqnarray}
t'(r)^2 = \frac{E^2}{f(r)^2} \frac{1}{f(r) r^{2(d-1)} + E^2} \ , \quad {\rm with} \quad r=\lambda \ ,
\end{eqnarray}
where $E$ is an integral of motion that results from the translational invariance of the volume functional ${\rm Vol}_d$, under $t \to t + {\rm const}$. The corresponding volume is obtained as:
\begin{eqnarray}
{\rm Vol}_d = \int_{r_*}^{r_{\rm max}} dr \frac{r^{2(d-1)}}{\sqrt{E^2 + r^{2(d-1)} f(r) }} \ , \label{volcal}
\end{eqnarray}
where $r_*$ is the turning point, defined by $\left. E^2 + r^{2(d-1)} f(r) \right|_{r_*}=0$,\footnote{This is also the location where $\frac{dt}{dr} \to \infty$.} and $r_{\rm H}$ is the horizon, defined by $f(r_{\rm H})=0$. Note further that the combination $r^{2(d-1)}|f(r)|$ has a maximum whose location is denoted by $r_{\rm m}$. Clearly, for $E^2 > r_{\rm m}^{2(d-1)}|f(r_{\rm m})|$, there is no solution for $r_*$. For very small values of $E$, $r_* \approx r_{\rm H}$, since it approximately satisfies $\left. r^{2(d-1)} f(r) \right|_{r_*} = \O(E^2) \ll 1$. On the other hand, for maximum allowed value of $E$, $r_* \to r_{\rm m}$. The former limit represents close to the $t_{\rm L,R}=0$ slices, while the latter limit yields $t_{\rm L,R} \to \infty$. Furthermore, in this limit, the integral in (\ref{volcal}) has a logarithmic divergence, as a function of $E$, and therefore most of its contribution comes from the $r_*\approx r_{\rm m}$ region. This is pictorially demonstrated in figure \ref{bsworm}.

Now, the integral in (\ref{volcal}) can be split into two parts: one between $r_*$ to $r_{\rm H}$ and the other between $r_{\rm H}$ to $\infty$. The integral contribution coming from the latter part yields a divergent answer due to an infinite volume of the AdS-space. Thus, upon a suitable renormalization, this piece does not have an interesting contribution. The former region of $[r_*, r_{\rm H}]$ is interestingly non-trivial. To quantitatively establish this, let us introduce the Kruskal co-ordinates of \cite{Stanford:2014jda} and rewrite the metric in (\ref{metadsbh}) as:
\begin{eqnarray}
&& ds^2 = - \frac{4 f(r) }{f'(r_{\rm H})^2} {\rm exp}\left[- f'(r_{\rm H}) r_t(r) \right] du dv + r^2 d\Omega_{d-1}^2 \ , \\
&& uv = {\rm exp}\left[ - f'(r_{\rm H}) r_t(r)\right]  \ , \quad \frac{u}{v} = {\rm exp}\left[ - f'(r_{\rm H}) t \right]  \ , \quad dr_t = \frac{dr}{f(r)} \ . 
\end{eqnarray}
The coordinate transformation above can be used to obtain $u_{\rm L}$ as a function of $E$, see figure \ref{bsworm}. This yields\cite{Stanford:2014jda}:
\begin{eqnarray}
\log u_{\rm L} = \log u_* + \frac{f'(r_{\rm H})}{2} \int_{r_*}^{r_{\rm H}} dr \frac{\sqrt{E^2 + r^{2(d-1)} f(r)} - E}{\sqrt{E^2 + r^{2(d-1)} f(r) }} \ ,
\end{eqnarray}
which also has a similar log-divergence as $E \gg 1$. An identical formula holds for $\log v_{\rm R}$. The denominator inside the integrand above yields a similar logarithmic divergence in $E$. Therefore, ${\rm Vol}_d$ is proportional to $\log u_{\rm L}$ and $\log v_{\rm R}$. The proportionality constant can also be found out, since all transformations are explicitly known. This yields: ${\rm Vol}_d \sim |t_{\rm L}+ t_{\rm R}|$.

This is a very intriguing feature. The eternal AdS-BH state is dual to a TFD-state in the boundary CFT. Any local perturbation will thermalize quickly with characteristic quasi-normal modes (on the one-sided black hole patch), or stop evolving beyond the scrambling time. The classical Einstein-Rosen bridge, however, will grow linearly in time, forever. In the dual CFT-description this must correspond to some property of a large $N$ quantum mechanical system that grows for a parametrically long time, even after thermalization. In recent years, this aspect has become a field in its own right and the essential ideas revolve around a notion called computational complexity for generic quantum mechanical systems, that exhibit such growth at a much longer time-scales. We will not delve into the details of this idea, instead we will refer the interested Reader to the early ideas in \cite{Susskind:2014rva, Brown:2015bva, Brown:2015lvg}, and for a recent review in \cite{Chen:2021lnq} and references therein.

There are physical constraints on how difficult it is to traverse a Wormhole of the kind in (\ref{wormgen}). There are two ways to estimate this: (i) Assume that such Wormholes are solutions of Einstein-gravity with some stress-tensor and subsequently investigate constraints on the stress-tensor, (ii) Assume a Wormhole geometry and analyze constraints on geodesics of probe particles in the same. We will discuss both approaches here.

\subsection{Lorentzian Wormholes: Energy Constraints}

\subsubsection{Energy Conditions on the Matter}

Let us summarize the main point of this section, since the subsequent discussion contains technical details. The main result is that the existence of Wormhole solutions (of Einstein-gravity ) in an asymptotically AdS-space does not violate {\it e.g.}~the Null Energy Condition for the matter sector, while for asymptotically flat space NEC is violated.

We will closely follow the discussion in \cite{Visser:1995cc}. Let us begin with a static and spherically symmetric space-time of the following form:
\begin{eqnarray}
&& ds^2 = - e^{2\phi_{\pm}(r)} dt^2 + \left( 1 - b_{\pm}(r)\right)^{-1} dr^2 + r^2 d \Omega^2 \ , \label{wormgen} \\
&& {\rm with} \quad  \lim_{r\to\infty} b_{\pm}(r) = b_{\pm} \ , \quad \lim_{r\to\infty} \phi_{\pm}(r) = \phi_{\pm} \ .
\end{eqnarray}
Here $b_\pm$ and $\phi_\pm$ are asymptotic data at the two asymptotic regions. These are connected by a throat region in the interior. Note that this Wormhole connects two Universes, but it does not have an event-horizon. The proper radial distance can be defined as $\ell(r) = \pm \int_{r_0}^r dr/\sqrt{1-b_\pm(r)/r}$. The Wormhole throat is characterized by a minimum of $r(\ell)$, which is denoted by $r_0$. Thus, the full geometry can be viewed as two branches of the metric data, denoted by the subscripts $\pm$, glued at $r=r_0$.

The minimality condition at $r(\ell) = r_0$ implies:
\begin{eqnarray}
&& \frac{dr}{d\ell} = \pm \sqrt{1- \frac{b(r_0)}{r_0}} =0 \ , \label{cond1} \\
&& \frac{d^2 r}{d\ell^2} = \frac{1}{2r_0} \left( \frac{b(r_0)}{r_0} - b'(r_0) \right) > 0  \ . \label{cond2}
\end{eqnarray}
Suppose now, that the Wormhole in (\ref{wormgen}) solves Einstein-equations: $G_{\mu\nu} = (8\pi G_N) T_{\mu\nu}$, where $T_{\mu\nu} = {\rm diag}\left( \rho, p_r, \vec{p}\right)$. Here $\rho$ is the energy-density, $p_r$ is the radial-pressure and $\vec{p}$ is the transverse pressure (along the $S^2$). Corresponding equations of motion yield:
\begin{eqnarray}
&& b' = 8 \pi G_N \rho r^2 \ , \quad \phi' = \frac{b + 8 \pi G_N r^3 p_r}{2r^2 \left(1 - \frac{b}{r} \right)} \ , \label{wormflat1} \\
&& - p_r' = (\rho+p_r) \phi' - 2 \frac{ p - p_r}{r} \ , \label{wormflat2}
\end{eqnarray}
Combining the Einstein equations, one obtains:
\begin{eqnarray}
8\pi G_N \left(\rho +  p_r\right) = - \frac{e^{2\phi}}{r} \left[ e^{-2\phi} \left( 1 - \frac{b}{r} \right)\right] ' \ .
\end{eqnarray}
Using conditions (\ref{cond1}) and (\ref{cond2}), at the throat, one obtains:
\begin{eqnarray}
&& \left. e^{-2\phi} \left( 1 - \frac{b}{r} \right) \right|_{r_0} = 0 \ , \\
&& \forall \, r \ge r_0 \ , \quad e^{-2\phi} \left( 1 - \frac{b}{r} \right) \ge 0 \ . 
\end{eqnarray}
Therefore, one arrives at the following conclusion:
\begin{eqnarray}
\exists \, \, \delta r_0 \, \, \, {\rm such \, \, \, that} \, \, \, \forall r \in [r_0, r_0 \pm \delta r_0] \quad \rho + p_r < 0 \ . \label{necviolate}
\end{eqnarray}
The condition in (\ref{necviolate}) implies that the stress-tensor $T_{\mu\nu}$ violates the so-called Null Energy Condition (NEC), at least within the range $[r_0 - \delta r_0, r_0 + \delta r_0]$.\footnote{In fact, in this case, the stress-tensor violates all other classical energy conditions: Weak Energy Condition, Strong Energy Condition and Dominant Energy Condition.} The upshot is: in order to realize a Wormhole solution of classical gravity, one must introduce some {\it exotic} matter to support the same. Note that, we have not yet addressed possible constraints that may appear from demanding that a signal passes through such a Wormhole, to reach from one asymptotic region to another.

The equations of motion in (\ref{wormflat1})-(\ref{wormflat2}) generalize easily in an asymptotically AdS-background and, in units of $8\pi G_N=1$, these are given by
\begin{eqnarray}
&& b' = r^2 \left(\rho+ \Lambda e^{2\phi} \right)  \ , \\
&& \phi' = \frac{1}{2r^2(1-b/r)} \left[b + r^3 \left( p_r - \frac{\Lambda}{1-b}\right)  \right]  \ , \\
&& - \left( p_r - \frac{\Lambda}{1-b}\right) ' = \phi' \left( \rho + \Lambda e^{2\phi} + p_r - \frac{\Lambda}{1-b}\right)  - \frac{2}{r} \left( p - \Lambda r^2 - p_r + \frac{\Lambda}{1-b}\right) \ . \nonumber \\ 
\end{eqnarray}
Proceeding as above, in this case one obtains:
\begin{eqnarray}
\left(\rho +  p_r\right) = - \frac{e^{2\phi}}{r} \left[ e^{-2\phi} \left( 1 - \frac{b}{r} \right)\right] ' - \Lambda \left( e^{2\phi(r)} - \frac{1}{1-b(r)} \right) \ .
\end{eqnarray}
While the first term on the RHS is always negative, the second term can have both signs. It is therefore still possible to fine-tune the matter sector in such a way that $(\rho+p_r)$ remains positive and NEC is satisfied. We will see in the next section that by analyzing null geodesics, we can rule out this possibility completely.\footnote{It is worth noting that various energy conditions in GR are important in proving theorems. At the same time, explicit exception-examples are also known corresponding to such energy conditions. For example, classical scalar fields violate strong energy condition, Casimir energies violate NECs and can violate a weaker integrated energy condition, {\it etc}. See \cite{Visser:1995cc} for a detailed account on many such examples.}

\subsubsection{Energy Conditions \& Geodesics} \label{ECG}

We have already observed that it is non-trivial to have a static Wormhole solution in asymptotically flat-space, within classical gravity, and one is forced to consider some {\it exotic} matter source to support it. However, in asymptotically AdS-geometry, no such issue arises. In this section, the main result that we will review is that irrespective of the asymptotics, constraints on the probe signal necessarily imposes a violation of the NEC. In fact, it necessitates a stronger violation of the ANEC condition. The rest of the section is devoted to a review of the technical details leading to this conclusion.

Let us further assume that such Wormholes are also traversable, {\it i.e.}~we can send null rays from one asymptotic region to the other one through this Wormhole. Consider a bundle of radial null geodesics that enter one mouth of the Wormhole and exit the other. Intuitively, the area cross-section of this null congruence should be initially decreasing as it approaches the Wormhole throat region, subsequently, it should then be increasing as the second mouth is approached. For a quantitative idea, let us begin with the Raychaudhuri equation:
\begin{eqnarray}
\frac{d\Theta}{d\lambda} = - R_{\mu\nu} k^\mu k^\nu - 2 \sigma^2 - \frac{1}{2}\Theta^2 + 2 \omega^2 \ , \label{RCeqn}
\end{eqnarray}
where $\Theta$ is the expansion of the area of the null congruence, $\sigma$ is the shear and $\omega$ is the vorticity, $k^\mu$ is a null vector and $\lambda$ is the affine parameter. For radial null geodesics, both $\sigma=0=\omega$.

For radial null geodesics, therefore, at the throat $d\Theta/d\lambda=0$ implies $R_{\mu\nu} k^\mu k^\nu = - 1/2 \Theta^2 \le 0$. This statement can be easily translated to a statement about the stress-tensor in the Einstein equations by:
\begin{eqnarray}
\left( R_{\mu\nu} - \frac{1}{2} g_{\mu\nu} R = 8 \pi G_N T_{\mu\nu} \right) k^\mu k^\nu \quad \implies \quad R_{\mu\nu} k^\mu k^\nu = (8\pi G_N) T_{\mu\nu} k^\mu k^\nu \ . \label{eomEin}
\end{eqnarray}
Therefore, at the throat, we must have: $T_{\mu\nu} k^\mu k^\nu \le 0$, {\it i.e.}~a violation of the NEC. Thus, we arrive at a qualitatively similar conclusion as before.

A more general constraint can be obtained by integrating the Raychaudhuri equation in (\ref{RCeqn}), for radial null inextensible geodesics. This yields:
\begin{eqnarray}
\int_{\lambda_1}^{\lambda_2} R_{\mu\nu} k^\mu k^\nu d\lambda = - \frac{1}{2} \int_{\lambda_1}^{\lambda_2} \Theta^2 d\lambda - \left. \Theta \right|_{\lambda_1}^{\lambda_2} \ ,
\end{eqnarray}
where, for inextensible null geodesics, we can send $\lambda_1 \to - \infty $ and $\lambda_2 \to \infty$. For such geodesics, it is expected that $\Theta(-\infty) =0 = \Theta(+\infty)$\footnote{Since these are inextensible, the expansion must vanish at the end points. Otherwise, the geodesics will be expanding into something, and therefore be extensible.}. Therefore, we obtain: $\int_{-\infty}^{\infty} R_{\mu\nu} k^\mu k^\nu d\lambda = \int_{-\infty}^{\infty} T_{\mu\nu} k^\mu k^\nu d\lambda \le 0$. Thus, while the existence of a minimum of the expansion of the radial null congruence implies a violation of the NEC, geodesic completeness of the Wormhole geometry further implies the violation of the ANEC.

The ANEC violation can be further quantified\cite{Visser:1995cc} in terms of two basic scales in the configuration: The length of the throat region and the location of the throat. Starting with the metric in (\ref{wormgen}), the ANEC integral is obtained to be:
\begin{eqnarray}
I_{\rm ANEC} = - \frac{1}{4\pi G_N} \int_{r_0}^\infty \left[ e^{-\phi_+(r)} \sqrt{1-\frac{b_+(r)}{r}} + e^{-\phi_-(r)} \sqrt{1-\frac{b_-(r)}{r}} \right] \frac{dr}{r^2} \ . \label{anecint}
\end{eqnarray}
Suppose that the ANEC violation occurs within a region $[r_0, R_{\pm}]$ of the geometry. For, $r> R_{\pm}$, the static, spherically symmetric space-time simply takes a Schwarzschild form, due to Birkhoff's theorem: $b_{\pm}(r) = 2 M_{\pm}$, $e^{\phi_{\pm}(r)} = e^{\phi_{\pm}(\infty)} \sqrt{1 - 2 M_{\pm}/r}$. The integral in (\ref{anecint}) splits into two parts: one within $[r_0, R_{\pm}]$ and the other within $[R_{\pm}, \infty]$. The latter range always yields an overall negative contribution to the ANEC integral. Consequently, the ANEC integral in (\ref{anecint}) is upper bounded by the Schwarzschild-region contribution. This yields:
\begin{eqnarray}
I_{\rm ANEC}  < - \frac{1}{4\pi G_N} \left[ \frac{e^{-\phi_+(\infty)}}{R_+} +  \frac{e^{-\phi_-(\infty)}}{R_-}\right]  \ . 
\end{eqnarray}
Here, $R_\pm$ are scales associated with the length of the throat.

Similarly a lower bound for $I_{\rm ANEC}$ can be obtained in terms of the location of the throat. Recall that, by construction $\phi_\pm(r)$ is finite-valued within $[r_0, \infty]$. Hence, $\exists$ $\phi_\pm^{\rm min} = {\rm min} \{\phi_\pm(r)\}$ in the same range of $r \in [r_0, \infty]$. Thus, a lower bound for the ANEC integral in (\ref{anecint}) is obtained by replacing the integrands by $e^{-\phi_\pm^{\rm min}}$:
\begin{eqnarray}
I_{\rm ANEC} > - \frac{1}{4\pi G_N} \frac{1}{r_0} \left[ e^{-\phi_+^{\rm min}} + e^{-\phi_-^{\rm min}} \right] \ .
\end{eqnarray}
Finally, note that, the same analyses will through for asymptotically AdS-geometry. The only use of the Einstein-equation has been made in (\ref{eomEin}), which yields the same answer upon contracting with null vectors. 


\subsection{Traversable Wormholes}

From what we have discussed so far, a violation of ANEC is unavoidable to obtain a Wormhole solution. Generally, this is a serious obstacle since ANEC-violating non-pathological examples are unknown. Furthermore, such a potential violation raises the possibility of violating topology-censorship theorems in GR. Moreover, ANEC is expected to hold for QFTs in general.

\begin{figure}
  \includegraphics[width=\linewidth]{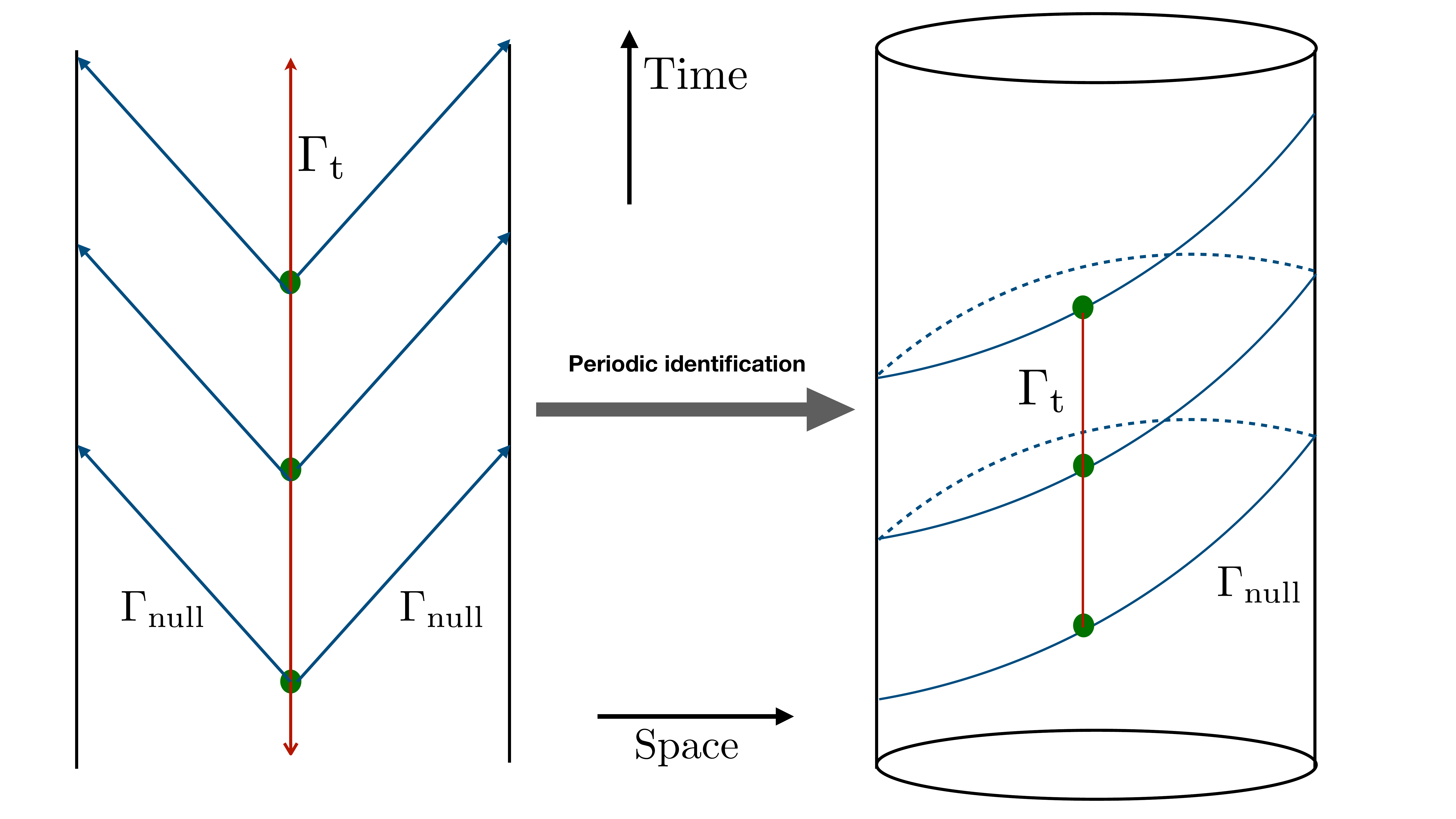}
  \caption{ A schematic pictorial representation of non-achronal null geodesics. On the left, the three green dots on the vertical line are connected by the red time-like paths (denoted by $\Gamma_{\rm t}$), but they are not connected by null geodesics (denoted by $\Gamma_{\rm null}$) emanating from any of these points. If the spatial direction is compactified, these green dots can now be connected by null geodesics $\Gamma_{\rm null}$, as shown on the right. However, the time-like red vertical path remains the shorter of the two paths. Thus, $\Gamma_{\rm null}$ are not achronal. }
  \label{achronal}
\end{figure}

Interestingly, there is a technical loophole. There exists a corner in which ANEC violation is admissible, keeping all the above constraints unperturbed\cite{Graham:2007va}. In summary, instead of ANEC one needs to consider SCAANEC (Self-consistent Achronal ANEC). It is instructive to ponder over what each letter signifies in this acronym. First of all, ``Self-consistent" simply means that the geometric data is on-shell, and in particular, the metric and various fields satisfy Einstein equations. ``Achronal" implies that no two points on the null geodesic can be connected by a time-like geodesic. Therefore, the corresponding null geodesic is the fastest path between the two said points. Examples of non-achronal null geodesics are provided in figure \ref{achronal}.
Finally, ANEC is what we have already encountered: this is simply an averaged version of the classical NEC. This averaging is expected in the quantum regime, since small quantum violations can occur in the classical energy constraints.

\begin{figure}
  \includegraphics[width=\linewidth]{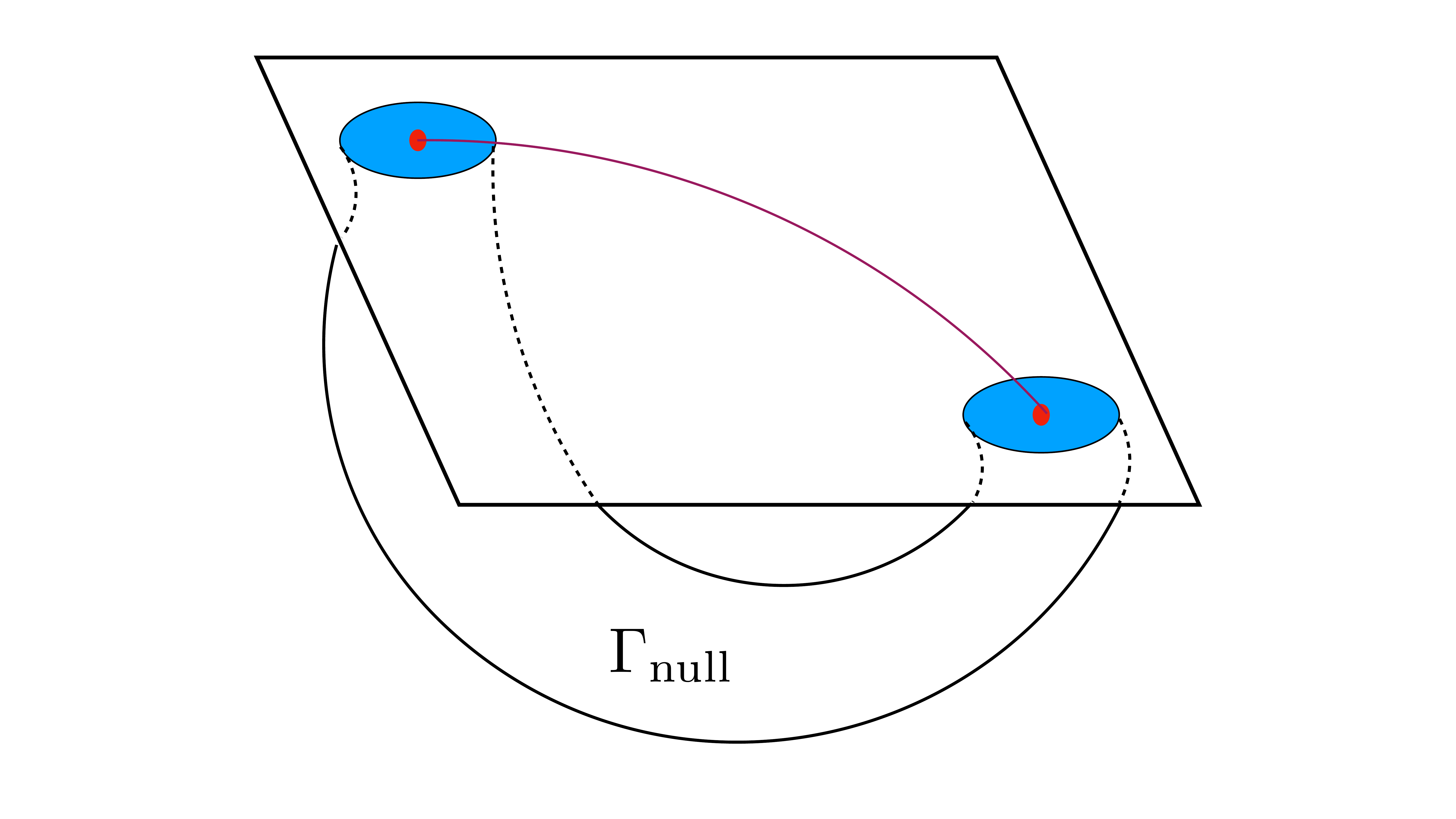}
  \caption{A schematic diagram demonstrating an ANEC-violating but SCAANEC-preserving  traversable Wormhole configuration. The red points are connected by the Wormholes, as well as by boundary time-like paths and the latter is shorter than the former. }
  \label{trworm}
\end{figure}

The upshot is: traversable Wormholes can indeed be found as solutions to Einstein-gravity with ANEC-violating, but SCAANEC-preserving matter field. This is qualitatively represented in figure \ref{trworm}.
 We will now review an explicit such construction, within the context of AdS/CFT and discuss some of its basic ramifications.

\subsubsection{Traversable Wormholes: a UV-complete Example}\label{GJW}

Recent years have seen truly interesting and fresh perspectives in which Wormholes have played a key role. The explicit example that led to precise and quantitative understanding of traversability and motivated subsequent works in this direction is the Gao-Jafferis-Wall traversable Wormhole construction in \cite{Gao:2016bin}.\footnote{It is worth mentioning that Wormhole solutions in gravity are usually constructed with exotic matter sources, {\it e.g.}~in \cite{Morris:1988cz, Morris:1988tu, Visser:1989kh, Visser:1989kg, Poisson:1995sv, Barcelo:2000zf, Visser:2003yf}, or in higher-derivative theories of gravity, {\it e.g.}~in \cite{Bhawal:1992sz, Thibeault:2005ha, Arias:2010xg, Chernicoff:2020tvr}.} In this section, we will heavily draw on \cite{Gao:2016bin} and review the GJW Wormhole.

Before discussing some of the necessary and basic details, let us ssummarize the construction. Consider an eternal Black Hole in AdS, {\it e.g.}~the one already discussed in section \ref{sec:eadstfd}. This is dual to the thermofield-double state in the CFT$_{\rm L} \times$CFT$_{\rm R}$ system. Let us now deform the TFD state, by turning on a relevant deformation at time $t=t_0$, of the following form:
\begin{eqnarray} \label{wh_cft_coupling}
\delta S = \int d^d x h(t, x) \O_{\rm R}(t, x) \O_{\rm L}(-t, x ) \ ,
\end{eqnarray}
where $\O_{\rm L,R}$ are defined in CFT$_{\rm L,R}$. It is clear from (\ref{wh_cft_coupling}) that this coupling breaks locality explicitly. In the dual gravitational picture, turning on (\ref{wh_cft_coupling}) turns on a scalar field with an appropriate boundary condition, which in the bulk of the eternal Black Hole, back-reacts and correspondingly deforms the geometry. It turns out that one can tune this system such that a traversable Wormhole geometry is created by this scalar back-reaction.

\begin{figure}
  \includegraphics[width=\linewidth]{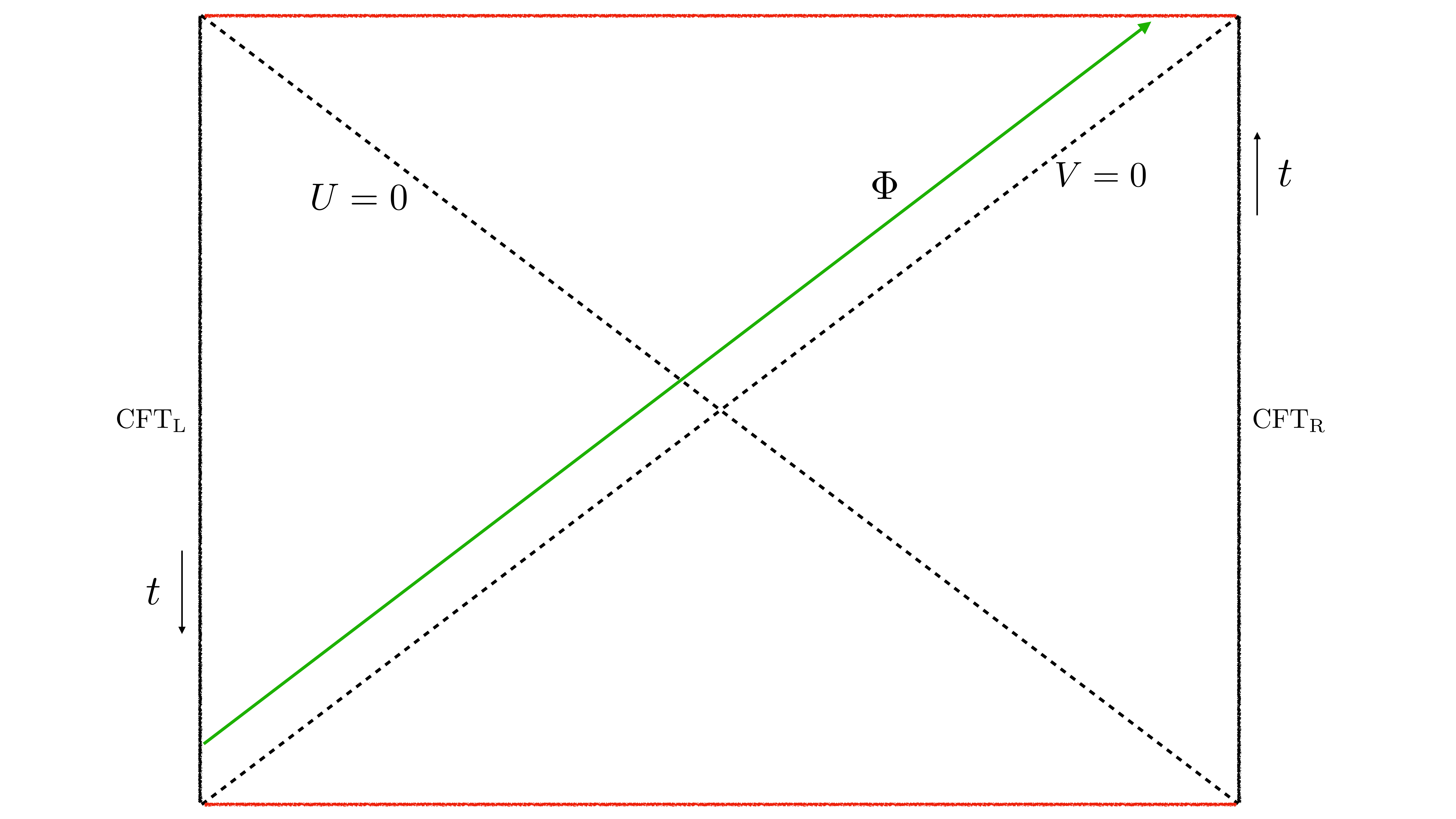}
  \caption{Penrose diagram for the eternal Black Hole in AdS. The red horizontal lines represent the singularity. The green arrow represents a signal that originates at the left boundary and propagates towards the right boundary.}
  \label{wh_no_traverse}
\end{figure}

Pictorially, consider figure \ref{wh_no_traverse}.
A null ray along $V=0$ will begin at $t=-\infty$ on the left boundary and reach at $t=+\infty$ at the right boundary. Any other null ray that begins at a finite time at the left boundary will end up in the singularity. Thus, no signal can propagate from the left to the right.

As was demonstrated earlier, a necessary condition for creating a traversable Wormhole configuration is to arrange for a stress-tensor such that $\int dU T_{UU} < 0$, where $T_{UU}$ corresponds to a matter field that propagates along the $U=0$ line in figure \ref{wh_no_traverse}. Let us consider throwing in a small matter field with a non-vanishing $T_{UU} \sim \O(\epsilon)$, where $\epsilon$ bookeeps the smallness of the matter-field. From Einstein equations, one can easily obtain the linearized back-reaction $\delta g_{\mu\nu} \sim \O(\epsilon)$ as a result:
\begin{eqnarray}\label{eineq_back}
\frac{d-2}{4} \left[ \left( \frac{d-3}{r_{\rm H}^2 } + \frac{d-1}{L^2} \right) \left( h_{UU} + \partial_U \left( U \delta g_{UU} \right) \right) - 2 r_{\rm H}^2  \partial_U^2 \delta g_{\phi\phi}\right] = 8\pi G_N T_{UU} \ , 
\end{eqnarray}
which is the only non-trivial component of the equations. Here $r_{\rm H}$ denotes the location of the horizon in the unperturbed geometry. Also, $h_{\phi\phi}$ is the azimuthal component of the metric correction.

Since the perturbation is assumed to be small, in a long time limit (after the scrambling time), we are expected to obtain a stationary state. This implies that, at sufficiently long time-scale, $\partial_U$ should be identified with the asymptotic Killing vector $U \partial_U = \partial_t$. Stationarity implies that $U \partial_U T_{UU} \to 0 $, and therefore, $T_{UU} \sim U^{-2}$ or faster as $U \to \infty$. This, in turn, implies that every term in the LHS of (\ref{eineq_back}) should also have a similar fall-off behaviour. Integrating the two sides of (\ref{eineq_back}), and using the above behaviour to drop out certain boundary terms, we get:
\begin{eqnarray} \label{anecviolate}
0 > 8 \pi G_N \int _{-\infty}^\infty dU T_{UU} = \frac{d-2}{4} \left[ \frac{d-3}{r_{\rm H}^2}  + \frac{d-1}{L^2} \right] \int_{-\infty}^\infty dU \delta g_{UU} \ . 
\end{eqnarray}
Thus, the ANEC violating matter-field will induce a particular signature in the linearized metric correction, integrated over the null line $V={\rm constant}$.

Before proceeding further, let us present an explicit example to demonstrate that explicit analytical solutions can be obtained. Consider the AdS-Schwarzschild geometry in $(2+1)$-dimensions:
\begin{eqnarray}
ds^2 = - \frac{4}{\left( 1 + UV\right)^2} dU dV + \frac{\left( 1 - UV\right)^2}{\left( 1 + UV \right)^2} dx^2  \ ,
\end{eqnarray}
where we have written the metric in the Eddington-Finkelstein co-ordinates. Suppose now, a matter is thrown along the $U=0$ line into the geometry. This will back-react and the back-reacted geometry, at the linearized order, can be obtained as:
\begin{eqnarray}
ds^2 = - \frac{4}{\left( 1 + UV\right)^2} dU dV + \frac{\left( 1 - UV\right)^2}{\left( 1 + UV \right)^2} dx^2  + 4 \delta(U) \delta g_{UU} dU^2 \ ,
\end{eqnarray}
where $\delta(U)$ is the delta-function and $\delta g_{UU}$ is obtained from linearized Einstein equations. Clearly, this structure transcends dimensionality of the system.

\begin{figure}
  \includegraphics[width=\linewidth]{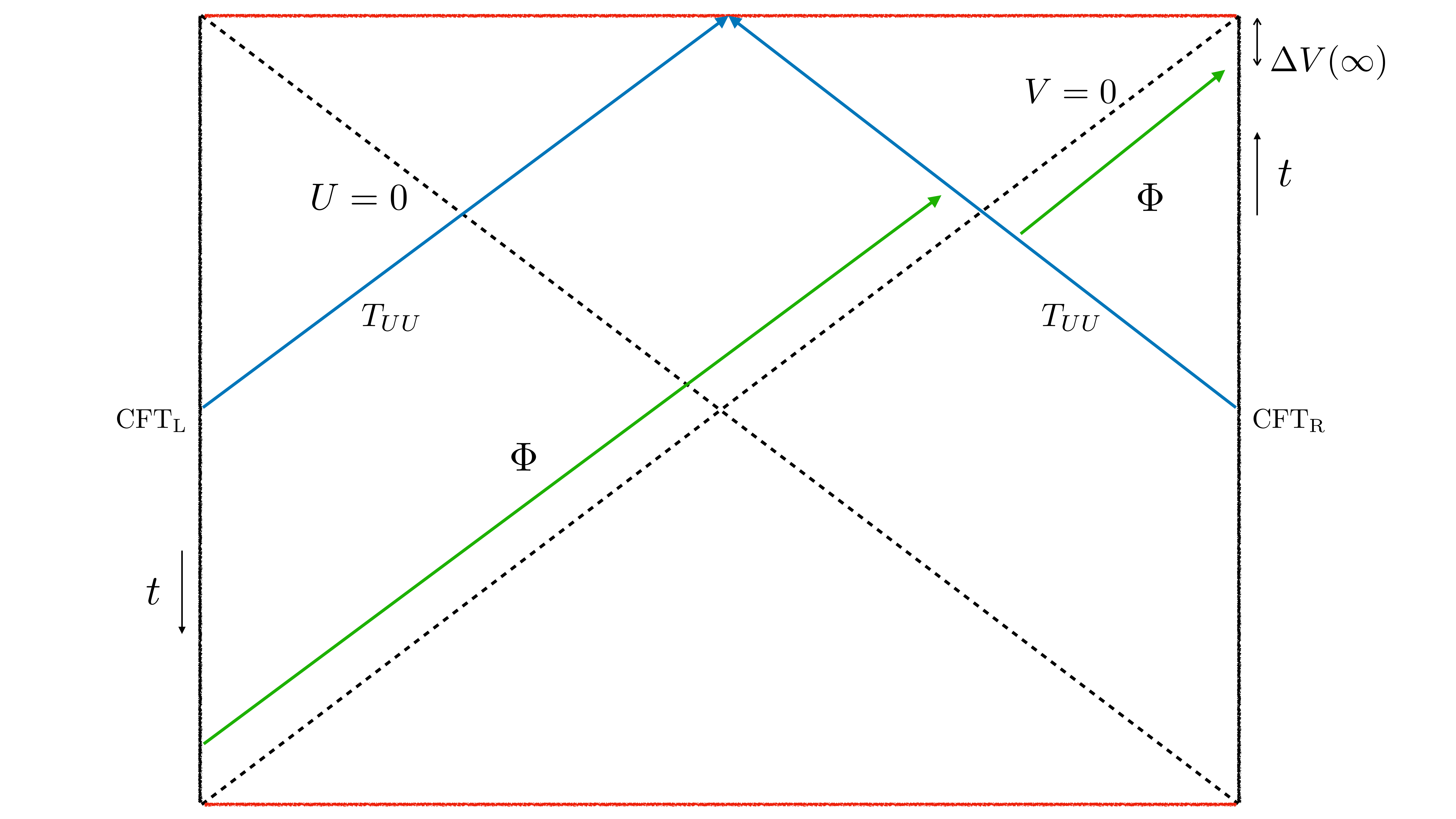}
  \caption{A schematic construction of the traversable wormhole. The blue lines represent the stress-tensor inserted on both boundaries at some time $t=t_0$, the green arrow represent a signal that is emitted at a sufficiently past on the left boundary and reaches, due to the shift caused by the $T_{UU}$ back-reaction, the right boundary at a sufficient future. }
  \label{wh_traverse}
\end{figure}

Let us now look at the null rays in this back-reacted geometry. First, we focus only on $x={\rm constant}$ null geodesics. These rays are obtained by solving:
\begin{eqnarray}\label{null_changed}
ds^2 = 2 g_{UV} dU dV + \delta g_{UU} dU^2  = 0 \ .
\end{eqnarray}
Clearly, the leading order solutions to (\ref{null_changed}) are simply $U={\rm const.}$ and $V={\rm const.}$ lines. The back-reaction will now deform these rays, specifically, the $V={\rm const.}$ lines. The shift, at order $\O(\epsilon)$, to these null rays are obtained by integrating (\ref{null_changed}):
\begin{eqnarray}
\Delta V(U) = - \int_{-\infty}^U \frac{\delta g_{UU} dU}{2 g_{UV}} = - \frac{1}{2 g_{UV}} \int_{-\infty}^U \delta g_{UU} dU \ . 
\end{eqnarray}
This finally yields: 
\begin{eqnarray}
\Delta V(+\infty) =  - \frac{1}{2 g_{UV}} \int_{-\infty}^\infty \delta g_{UU} dU \sim - \frac{1}{2 g_{UV}} \int_{-\infty}^\infty  T_{UU} dU < 0 \ ,
\end{eqnarray}
where, in the last line, we have used $g_{UV}<0$ and that $T_{UU}$ violates ANEC. This is pictorially demonstrated in figure \ref{wh_traverse}.
Therefore, with a sufficient back-reaction, a signal originating at the left boundary can now reach the right boundary.

We need to still specify the most crucial aspect of the construction: How one arranges $\left \langle \int dU T_{UU}  \right \rangle < 0 $. The perturbative deformation above is performed around the TFD-state, and suppose the corresponding deformation operator is denoted by $\O$. Then, schematically, we obtain a deformed state: $| {\rm TFD} \rangle \to (1 + i \epsilon \O) | {\rm TFD} \rangle $. The expectation value of the ANEC operator in this deformed state is obtained to be:
\begin{eqnarray}\label{anectfd}
\left \langle \int dU T_{UU}  \right \rangle  \approx i \epsilon \left \langle   \left[ \int dU T_{UU} , \O \right] \right \rangle_{\rm TFD} \ .  
\end{eqnarray}
At this level, it appears that by tuning the sign of $\epsilon$ one can easily arrange for an ANEC-violating configuration. However, this assumes that the RHS of (\ref{anectfd}) is non-vanishing.

We will now review why the last assumption above is non-trivial. First, recall that time runs upwards/downwards on the right/left boundary\footnote{Also true for the right/left Rindler wedges in figure \ref{wh_no_traverse}.} of the eternal Black Hole in AdS, see {\it e.g.}~figure \ref{wh_no_traverse}. The corresponding Killing symmetry, $i \partial_t$, is generated by the Hamiltonian $H = H_{\rm L}-H_{\rm R}$, see {\it e.g.}~the discussion around equation (\ref{TFD_H}). Also, recall that on $V=0$ slice, $\partial_t = U \partial_U$, as we have already used in arriving the result in (\ref{anecviolate}). Let us consider how the ANEC operator transforms under $\partial_t$.

Clearly, as $U \to \lambda U$, $T_{UU} \to \lambda^{-2} T_{UU}$ and therefore $\int dU T_{UU} \to \lambda^{-1} \int dU T_{UU}$. Therefore, the ANEC operator is an eigenoperator of the $U$-dilatation operator, with an eigenvalue $-1$. Now consider:
\begin{eqnarray}
\left[ H_{\rm L} - H_{\rm R} , \int dU T_{UU}\right]  = \left[ - i \partial_t , \int dU T_{UU} \right ] &=& - i \left[ U \partial_U, \int dU T_{UU}\right]  \nonumber \\
& = & i \int dU T_{UU} \ ,
\end{eqnarray}
where, in the last line, we have used: $\left[ U \partial_U, \int dU T_{UU}\right] = - \int dU T_{UU}  $.\footnote{This simply follows from the eigenvalue of $\int dU T_{UU} $ operator, when acted upon by the $U\partial_U$ dilatation. This relation holds for any state at the scale of scrambling-time or higher when $U\partial_U | {\rm state} \rangle=0$, for any state. } With this, let us calculate:
\begin{eqnarray}
\left( H_{\rm L} - H_{\rm R}\right)  \int dU T_{UU} | {\rm TFD} \rangle & = & \left[\left( H_{\rm L} - H_{\rm R}\right), \int dU T_{UU}  \right]  | {\rm TFD} \rangle \nonumber\\
& = & i \int dU T_{UU} | {\rm TFD} \rangle \ .
\end{eqnarray}
Note that: $H_{\rm L} - H_{\rm R}$ is hermitian and therefore its spectrum only contains real eigenvalues. The equation above, therefore, can be satisfied only iff $\int dU T_{UU} | {\rm TFD} \rangle = 0$. This implies, using (\ref{anectfd}): $\left \langle \int dU T_{UU}  \right \rangle = 0$.

Physically, this identity follows from a cancellation of the ANEC integral between the contribution from $U>0$ branch and the contribution from $U<0$ branch. This can be avoided by turning on the deformation in (\ref{wh_cft_coupling}) at time $t=t_0$. As is explicitly shown in \cite{Gao:2016bin}, such a deformation indeed induces an ANEC-violating matter-sector in the bulk. This is checked by computing the energy-momentum $2$-point function and extracting the expectation value of the ANEC-operator from this. Therefore, indeed, \cite{Gao:2016bin} constructs a traversable Wormhole with the desired qualities. We will not discuss more technical details related to this, and refer the interested reader to \cite{Gao:2016bin} instead. Moreover, another review article in this series will also discuss these aspects in more detail. We will now review other examples of traversable Wormholes in a similar context.

\subsubsection{Traversable Wormholes: Eternal \& Others}

In the previous section, we have demonstrated how a traversable Wormhole can be constructed by turning on a non-local coupling between the left and the right boundaries of the TFD-state and thereby inducing an ANEC-violating matter source. In this section, we will review some recent progress in generalizing this in various interesting directions. We will begin with the eternal Wormhole construction in \cite{Maldacena:2018lmt}, in which the left-right Hamiltonian interaction is always turned on. See also {\it e.g.}~\cite{Garcia-Garcia:2019poj, Garcia-Garcia:2020ttf} for further studies on such constructions. One begins with an action similar to the one in (\ref{2djtwh}) in Lorentzian signature, with $S_{\rm Brane}$ replaced by a generic matter action $S_{\rm matter}[\chi, g]$, where $\chi$ denotes the matter-field. Subsequently, the boundary conditions are exactly the same as in (\ref{adsboundary}), except now the metric is chosen in the global Lorentzian AdS: $ds^2 = 1/(\sin^2\sigma)(-dt^2 + d\sigma^2)$.

As described in \cite{Maldacena:2018lmt}, the $++$-component of Einstein equation, in the conformal gauge, yields:
\begin{eqnarray}
&& - \partial_+ \left(\sin^2 \sigma \partial_+\phi \right) = T_{++} \sin^2 \sigma \nonumber\\
&& \implies \quad - \left. \left(\sin^2 \sigma \partial_+\phi \right) \right|_{-\infty}^{\infty} = \int T_{++} \sin^2\sigma dx^+ = \int_{-\infty}^{\infty} dX^+ T_{X^+ X^+} \ , \nonumber\\ \label{eomanec}
\end{eqnarray}
the RHS of which is precisely the ANEC-integral introduced in section \ref{ECG}. The LHS, on the other hand, is negative for a growing dilaton near the boundary. Thus, to have a non-trivial solution to (\ref{eomanec}), we must introduce matter field violating ANEC.

Similar to the GJW Wormhole discussed in the previous section, this can be done by introducing an explicit interaction at the boundary of the form $\int dt \O_{\rm L}(t) \O_{\rm R}(t)$, where $\O_{\rm L,R}$ are operators of the left and the right boundary systems and $t$ is the time co-ordinate at the boundary. Note that the corresponding coupling is time-independent, and hence this represents an eternal set up. In \cite{Maldacena:2018lmt}, it is explicitly shown that such an interaction yields an ANEC-violating matter field with free, massless bulk fermions. Equivalently, from the perspective of the boundary theory, this corresponds to coupling two copies of the so-called SYK model, with a relevant deformation. For more details on the phase structure of this system, we refer the interested reader to \cite{Maldacena:2018lmt}. Note that, in dimensions higher than two, constructing eternal Wormholes appear subtle. In fact, there are no-go results that prohibit such constructions, provided Poincar\'{e} symmetry is preserved\cite{Freivogel:2019lej}.

We now shift gear towards a Wormhole construction in arbitrary dimensions. Clearly, GJW-construction applies to general dimensions, however, the underlying system is described either by a dual CFT {\it e.g.}~${\cal N}=4$ SYM theory in $(3+1)$-dimensions, or some scalar fields back-reacting on the eternal black hole geometry. It turns out that traversable Wormholes can also be constructed out of perhaps more earthy systems {\it e.g.}~the standard model and in asymptotically flat space-time. From an observational perspective, this is highly relevant for constructing states in the lab, or on a quantum computer, which are dual to a traversable Wormhole geometry. See {\it e.g.}~\cite{Susskind:2017nto, Brown:2019hmk, Nezami:2021yaq} for very interesting and related discussions. We will now review the pioneering construction in \cite{Maldacena:2018gjk}.

The basic ingredients are a U$(1)$ gauge field and a charged Dirac fermion, coupled to Einstein-gravity in $(3+1)$-dimensions:
\begin{eqnarray}\label{stdmodel_wh}
S = \frac{1}{16 \pi G_N} \int d^4 x \sqrt{-g} R - \frac{1}{g_{\rm ym}^2} \int d^4 x \sqrt{-g} F^2 + \int d^4x \sqrt{-g} i \bar{\psi} \left( \slashed{\nabla} - i \slashed{ A} \right) \psi \ . \nonumber\\
\end{eqnarray}
Note that, one can also begin with a negative cosmological constant in the action and essentially follow all the subsequent steps that we will review below. For more details on this case, see {\it e.g.}~\cite{Bintanja:2021xfs}. The strategy of constructing a traversable Wormhole, starting from the action in (\ref{stdmodel_wh}) comprises of the following steps: (i) Ignore the Dirac field and construct a magnetically charged black hole solution. This solution has a near-horizon AdS$_2\times S^2$ throat, whose throat size is determined by the U$(1)$-flux of the solution. If we begin with two oppositely charged magnetic black holes, then their respective AdS$_2$ throats can be glued to each other. This throat approximates the Wormhole region. (ii) Now introduce the Dirac fermions such that they contribute a negative energy and back-react on the geometry. This step is essential to make the Wormhole traversable. The negative energy is supplied by the lowest Landau levels of massless fermions in the magnetic field. (iii) Finally, this entire structure can be stabilized, at least to a desired extent, by rotating the entire system.

While the general strategy is intuitively clear, an explicit construction can be tedious and is perhaps best obtained by smoothly gluing various approximate geometric patches. We will now briefly discuss various scales associated with this construction and its subsequent generalizations. The near-horizon magnetically charged Reissner-Nordstrom solution is described by the metric:
\begin{eqnarray}
ds^2 = r_{\rm h}^2 \left( - \left( r^2 - 1 \right) dt^2 + \frac{dr^2}{r^2 - 1} + d\Omega_2^2\right) \ , \quad r_{\rm h} = \sqrt{\pi G_N} \frac{Q}{g_{\rm ym}}  \ ,
\end{eqnarray}
where $r_{\rm h}$ is the horizon radius, $Q$ is the charge of the black hole, and the AdS$_2$ throat has been written in the Rindler patch denoted by $\{t, r\}$. In the traversable Wormhole, the AdS$_2$ throat will vary in size and this can be captured by introducing simple variations to the metric functions above, {\it e.g.}~$\left( r^2 - 1 \right)  \to \left(1 +  r^2 +\gamma \right)$, $r_{\rm h}^2 d\Omega^2 \to r_{\rm h}^2 (1 + \varphi) d\Omega^2$. Magnetic field, looping through the $S^1\times {\mathbb R}$, creates localized Landau levels and yields $Q$ number of effective $(1+1)$-dimensional massless fermions and a negative Casimir energy, which makes the Wormhole traversable. In \cite{Maldacena:2018gjk}, it was shown that the energy difference between such a Wormhole and two disconnected extremal black holes is given by
\begin{eqnarray} \label{wh_extremal_diff}
\Delta E = \frac{r_{\rm h}^3}{G_N \ell^2} - \frac{N_f Q}{6} \left( \frac{\pi}{\pi \ell + d_{\rm out}} - \frac{1}{4\ell }\right) \ , 
\end{eqnarray}
where $\ell$ sets the length of the Wormhole throat, $d_{\rm out}$ sets the separation of the Wormhole mouths, $N_f$ is the number of fermions and $Q$ is the magnetic charge. Given (\ref{wh_extremal_diff}), it is straightforward to extremize the energy-difference. This becomes particularly simple in two limits: $d_{\rm out} \ll \ell$ and $d_{\rm out} \gg \ell$, both of which can be readily analytically solved to obtain a Wormhole biding energy: $E_{\rm bind} \sim - N_f Q/\ell$.

The two mouths will gravitationally attract each other. However, if $d_{\rm out}$ is large enough, instability due to this interaction can still yield a sufficiently stable traversable Wormhole geometry for long enough. In \cite{Maldacena:2018gjk}, this merger time is estimated as $\sim d_{\rm out}^{3/2}$, while the transit time through the Wormhole is $\sim d_{\rm out}$. Furthermore, additional ingredients can further increase stability of this construction, {\it e.g.}~simply having the two mouths rotate around each other.

The construction above can be further generalized to a multi-boundary traversable Wormhole solution, as is described in \cite{Emparan:2020ldj}. Qualitatively, given the two-boundary traversable Wormhole geometry, one introduces a {\it small} Wormhole mouth. There are numerous possibilities, discussed in detail in \cite{Emparan:2020ldj}, in which this opens up a new throat and connects to a third boundary. Perhaps the simplest is to consider a three-flavour fermion model, with three distinct U$(1)$-gauge fields. Each flavour couples to one of the U$(1)$'s. Suppose the {\it large} Wormhole mouths have charges $\left( Q_1, q_2, 0\right) $ and $\left(-Q_1, 0, q_3 \right) $ and the {\it small} mouth has the charge $\left(0, -q_2, -q_3 \right) $. In the limit $|Q_1| \gg |q_2|, |q_3|$, the configuration is mechanically stable. The fermions, correspondingly, yield the required negative energy. The basic point is: with a standard model like ingredient, such multi-boundary traversable Wormholes can be created, for long enough such that traversal affects are accessible. For a more detailed account, we refer the reader to \cite{Emparan:2020ldj}.

Alternatively, one can also begin with the multi-boundary Wormhole geometry of section \ref{EWormMulti} and add a double-trace deformation, as discussed in section \ref{GJW}. This non-local boundary interaction provides us with the required ANEC-violating matter field in the bulk. This matter field eventually renders the Wormhole traversable. This approach has been explored in detail in \cite{AlBalushi:2020kso}, to which we refer the reader for a detailed discussion on this. For more explicit examples of traversable Wormholes, see {\it e.g.}~\cite{Fu:2018oaq, Caceres:2018ehr, Marolf:2019ojx, Fu:2019vco}, for dynamical production of such geometries see {\it e.g.}~\cite{Horowitz:2019hgb, Maldacena:2019ufo}. From the boundary CFT perspective, such geometric solutions are understood as quantum teleportation. Several interesting directions have been explored with this theme, see {\it e.g.}~\cite{Maldacena:2017axo, Freivogel:2019whb, Berenstein:2019yfv, Fallows:2020ugr, Balasubramanian:2020ffd, Maldacena:2020sxe, Chen:2020tes}.

\subsubsection{Wormholes on the Brane}

Let us slightly shift gear and discuss possible Wormhole geometries outside Einstein-gravity. These geometries emerge from specific probe D-brane configurations in a given supergravity background, see {\it e.g.}~\cite{Karch:2007pd, Albash:2007bq, Alam:2012fw, Kim:2011qh, Sonner:2012if, Kundu:2013eba, Banerjee:2015cvy, Kundu:2015qda, Banerjee:2016qeu} and \cite{Kundu:2018sof, Kundu:2019ull} for a review. In this section we will briefly discuss them.

The basic premise is set with probe D-branes in a given supergravity geometry, following the pioneering work of \cite{Karch:2002sh}. This adds low-energy open string degrees of freedom in a geometry obtained from the low energy limit of the closed string spectrum. In the dual CFT, for example for the case of $\N=4$ SYM, this introduces an $\N=2$ hypermultiplet that transforms in the fundamental representation of the gauge group. Schematically, the corresponding path-integral takes the form: $Z_{\rm sugra} Z_{\rm probe} = Z_{\rm Adjoint} Z_{\rm fundamental}$, where the LHS is defined in the gravitational description. The probe action is given by the Dirac-Born-Infeld (DBI) action:
\begin{eqnarray}
S_{\rm DBI} = - N_f \tau_p \int d^{p+1} \xi e^{-\Phi} \sqrt{-{\rm det} \left[ \varphi^* \left[ G + B\right] + (2\pi \alpha') F\right] } \ , \label{openstring}
\end{eqnarray}
where $N_f$ denotes the number of $(p+1)$-dimensional probes, $\tau_p$ is the tension of each such probe, $\{\Phi, G, B\}$ represent of the supergravity data, $\alpha'$ is the tension of the open string and $F$ is gauge field on the D-brane and $\varphi^*$ denotes the pull-back map.

Suppose the probe embedding is characterized by the classical on-shell data $\{\theta_i, F_{ab}\}$ where $a, b$ run over the worldvolume indices of the brane and $i$ runs over the transverse directions. Consider now the corresponding fluctuation modes: $\theta_i + \delta \theta_i$ and $F_{ab} + \F_{ab}$. At the quadratic order, the fluctuations actions are obtained to be:
\begin{eqnarray}
&& \S_{\rm scalar} = - \frac{\kappa}{2} \int d^{p+1} \xi \left(\frac{{\rm det} G}{{\rm det} \S} \right)^{1/4} \sqrt{-{\rm det} \S} \S^{ab} \partial_a \delta \theta_i \partial_b \delta \theta_i + \ldots \ , \\
&& \S_{\rm vector} = - \frac{\kappa}{2} \int d^{p+1} \xi \left(\frac{{\rm det} G}{{\rm det} \S} \right)^{1/4} \sqrt{-{\rm det} \S} \S^{ab} \S^{cd} \F_{ac}\F_{bd} + \ldots  \ , \\
&& \S = \varphi^*  G - \left(F \left( \varphi^* G\right)^{-1} F \right) \ . 
\end{eqnarray}
The emergent metric $\S$ is known as the Open String Metric (OSM).

A particularly interesting configuration consist of a probe D-brane, embedded in an exactly AdS-background, with a worldvolume constant electric field: $F = E dt\wedge dx + \ldots$, where the $\ldots$ represent other components that need to be turned on to satisfy the equations of motion resulting from (\ref{openstring}). The supergravity background here is the vacuum solution (AdS), which is dual to the ground state of the SU$(N)$ adjoint gauge theory. On the other hand, the open string degrees of freedom propagates in a geometry with an event-horizon and therefore the dual fundamental matter sector is in a thermal state. There is a heat-flow due to this. However, in the probe limit, this flow is parametrically suppressed. For a detailed review on such constructions, see {\it e.g.}~\cite{Kundu:2018sof, Kundu:2019ull}.

Here, we present the following observation. As discussed in detail in \cite{Banerjee:2016qeu}, the scalar and vector (as well as fermion) modes propagate in a geometry with a horizon. Correspondingly, there is a two-sided Kruskal extension and a geometric notion of the TFD-state that we have discussed earlier, for standard black hole geometries. Subsequently, as we have discussed before, an Einstein-Rosen bridge connects the two sides of this maximally extended geometry. The OSM-geometry or the conformal OSM-geometry in which such semi-classical modes propagate have no {\it a priori} reason to obey a null Ricci condition: $n^\mu R_{\mu\nu} n^\nu > 0$, where $n^\mu$ is the null vector and $R_{\mu\nu}$ is the Ricci-tensor. This condition is important for traversability of the corresponding Einstein-Rosen bridge. It is straightforward to check, using the results in \cite{Banerjee:2016qeu} that indeed $n^\mu R_{\mu\nu} n^\nu \ge 0$ appears to hold generally. A particularly interesting case is an AdS$_4$-background, where the OSM-metric yields $n^\mu R_{\mu\nu} n^\nu=0$. This yields, using (\ref{RCeqn}): $d\Theta/d\lambda = - (1/2) \Theta^2$. In this case, even an infinitesimal violation of the null Ricci-condition can make the Einstein-Rosen bridge traversable.

At present, there are no explicit traversable Wormhole construction on such branes. Long, traversable Wormholes may be constructible making use of the fermionic fluctuations (as part of the effective supersymmetric ) on the D-brane, see {\it e.g.}~\cite{Marolf:2003vf}. We note that the possibility of a traversable Wormholes on the Brane has been implicitly hinted in \cite{Sonner:2013mba, Jensen:2014bpa, Jensen:2014lua}, which predates the GJW-Wormhole construction. In these cases, a Holographic dual description of a single EPR pair have been constructed, in terms of an open string embedded in an AdS-geometry. An Einstein-Rosen bridge forms, due to acceleration of the end-points of the open string. At a qualitative level, this is similar to the framework above, since a constant electric field on a D-brane will indeed accelerate open string end points. The non-traversable Wormholes in these cases have been viewed as explicit examples of the ER=EPR idea of \cite{Maldacena:2013xja}. A slightly different manifestation of essentially the same basic physics also draws a connection to the physics of Out-of-Time-Order correlators (OTOCs) on such worldsheets, see {\it e.g.}~\cite{deBoer:2017xdk, Murata:2017rbp, Banerjee:2018twd, Banerjee:2018kwy, Banerjee:2019vff}. This brings us to a very interesting connection between OTOCs, quantum regenesis and traversable Wormhole\cite{Gao:2018yzk}, which we will briefly review now. 


\subsubsection{Traversable Wormholes \& Regenesis}

Traversable Wormholes can be understood as quantum teleportation in the dual QFT. We will not review this aspect in detail, however, in this subsection we will briefly touch upon a related aspect that bridges various topical ideas of current research. This section will be based heavily on \cite{Gao:2018yzk} to which we refer the interested reader for a detailed account. The basic physics follows from the entanglement between the left and the right degrees of freedom, as well as the notion of information scrambling in a quantum system. In what follows, we will only review the basic and salient technical details of this argument.

One key ingredient is the TFD-state, introduced in (\ref{tfd}). Consider turning on a sufficiently local operator, denoted by $J^{\rm R}$, with a coupling $\phi^{\rm R}$, on the right boundary at time $t = - t_{\rm s}$. The corresponding expectation value $\left\langle {\rm TFD} | J^{\rm R}(t) | {\rm TFD} \right \rangle \not = 0$. However, since $\left [ J^{\rm R}, J^{\rm L}\right ] = 0$, for a sufficiently local operator $J^{\rm L}$ on the left boundary, $\left\langle {\rm TFD} | J^{\rm L}(t) | {\rm TFD} \right \rangle  = 0$. Suppose now a deformation is introduced to the total Hamiltonian, similar to the GJW-interaction:
\begin{eqnarray}
H = H^{\rm L} + H^{\rm R} - \frac{g}{k} \sum_{i=1}^k \O_i^{\rm L} \O_i^{\rm R} \delta(t=0) \ , \label{inttfdnl}
\end{eqnarray}
where $\{ \O_i^{\rm R,L}\}$ are sufficiently local operators on the right and the left boundaries. Upon inserting a signal on the right boundary at $-t= t_{\rm s}>t_*$, where $t_*$ is the corresponding scrambling time, a signal re-appears on the left boundary due to the explicit interaction in (\ref{inttfdnl})\cite{Maldacena:2017axo}. The scrambling time $t_*$ is defined as the time-scale when $\left\langle \left[ \O_1(t_1), \O_2(t_2)\right]^2 \right \rangle$ becomes of $\O(1)$. Here $\O_{1,2}$ are sufficiently local operators and the expectation value is computed in a given state. This $\left\langle \left[ \O_1(t_1), \O_2(t_2)\right]^2 \right \rangle$ expectation value is given by a Time-Order correlator and an Out-of-Time-Order correlator (OTOC); the limit $\left\langle \left[ \O_1(t_1), \O_2(t_2)\right]^2 \right \rangle\sim \O(1)$ is equivalent to setting ${\rm OTOC} \sim \O(0)$.

Let us pause for a brief discussion on the scrambling time. Given a local QFT, Lorentzian causality implies that any two space-like separated operators, outside each others light-cones, always commute.\footnote{Note that, even in the non-relativistic limit, for quantum systems with sufficiently local interactions, there is an emergent light-cone that is defined in terms of the so-called Lieb-Robinson bound\cite{Lieb:1972wy}.} Conversely, any commutator is non-trivial only inside the light-cone. For two given operators $\O_1(t_1)$, $\O_2(t_2)$, the commutator-squared evaluates to:
\begin{eqnarray}\label{commexpect}
\left[ \O_1(t_1), \O_2(t_2)\right] ^ 2 &=&  \O_1(t_1)\O_2(t_2) \O_1(t_1) \O_2(t_2) +  \O_2(t_2) \O_1(t_1) \O_2(t_2) \O_1(t_1)  \nonumber\\
& -&  \left[\O_1(t_1) \O_2(t_2) \O_2(t_2) \O_1(t_1) + \O_2(t_2) \O_1(t_1) \O_1(t_1) \O_2(t_2)  \right] \ . \nonumber\\
\end{eqnarray}
In the above expressions, we have suppressed the spatial directions since our goal is to make statements related to time-scales. One can consider the operators are spatially located at given points, but are not exactly co-incident. The latter condition eliminates contact singularities.

Now consider a thermal state with a density matrix $\rho = e^{-\beta H}$. In this case, the second line of the RHS of equation (\ref{commexpect}) can be written as TOCs.\footnote{For example:
\begin{eqnarray}
\left\langle \O_1(t_1) \O_2(t_2) \O_2(t_2) \O_1(t_1) \right \rangle_{\beta} = {\rm Tr} \left( \rho \O_1(t_1) \O_2(t_2) \O_2(t_2) \O_1(t_1) \right) \sim {\rm Tr} \left( \rho \O_1(t_1) \O_1(t_1) \O_2(t_2) \O_2(t_2)  \right)\ , \nonumber\\
\end{eqnarray}
where we have used $\rho = e^{-\beta H}$, $\O(t) = e^{i Ht} \O(0) e^{-iHt}$ and $\O(t- i\beta) \sim \O(t)$. The last condition assumes a periodicity condition on the operator $\O$, with a period $\beta$.}The other two terms can be written as OTOCs. At early times, the operators $\O_{1,2}$ are expected to commute yielding a vanishing expectation value of the commutator-squared. At sufficiently late time, however, when information of the operator $\O_1$ becomes accessible to $\O_2$, and vice-versa, the commutator-squared should approach an $\O(1)$ number. This can be restated in terms of the OTOC: at early times, the OTOC is expected to be an $\O(1)$ contribution, and at late times the OTOC is expected to approach zero. Scrambling time is defined as: ${\rm OTOC}(t_*) \sim 0$. Often, an exponential decay of the OTOCs is observed for thermal states, which is subsequently identified as a signature of an ``early-time quantum chaos".\footnote{Note that, this ``early-time chaos" applies till $t_*\sim \log N$, for a system with $N\gg 1$ degrees of freedom. The more conventional notion of ``quantum chaos" is defined in terms of level-repulsion of the corresponding quantum spectrum. This effect is visible at time-scales $\sim e^{N} \gg \log N$.}

Although the above ideas are cleanest for a thermal state, we can take this as a working definition of scrambling time for a wider class of systems. The essential physical understanding is that at the order of scrambling time, an initial local information of the system becomes accessible at the scale of the system size. This happens due to the Hamiltonian evolution, which couples various local physics at subsequent larger time-scales. In particular, for systems with a large number of degrees of freedom, scrambling time is parametrically larger than a local dissipation time-scale.

With this slight detour, let us come back to the main result of \cite{Gao:2018yzk}, which they referred to as ``regenesis" in the context of generic quantum chaotic systems. We need two key features of the TFD-state: Given a hermitian operator $J_{\rm R}$, we get:
\begin{eqnarray}
&& J_{\rm R} \left | {\rm TFD} \right\rangle = J_{\rm L} \left( \frac{i\beta}{2} \right)  \left | {\rm TFD} \right\rangle \ , \\
&& {\rm and} \quad e^{-i H_{\rm L} t}  \left | {\rm TFD} \right\rangle = e^{i H_{\rm R} t}  \left | {\rm TFD} \right\rangle \ , 
\end{eqnarray}
from which it follows that:
\begin{eqnarray}
J_{\rm R}(t) \left | {\rm TFD} \right\rangle = J_{\rm L} \left( \frac{i\beta}{2}  - t \right)  \left | {\rm TFD} \right\rangle \ . \label{LRrel}
\end{eqnarray}
The last relation above entangles states on the right boundary at time $t$, with states on the left boundary at time $-t$. This basic property becomes very important in what follows. For example, a two-point function of $J_{\rm L,R}$ in the TFD-state is given by
\begin{eqnarray}
\left\langle {\rm TFD}| J_{\rm L}(t) J_{\rm R}(-t')|{\rm TFD} \right\rangle  & = & \left\langle {\rm TFD}| J_{\rm L}(t) J_{\rm L} \left(t'+ \frac{i\beta}{2} \right)|{\rm TFD} \right\rangle  \nonumber\\
& = & \left\langle  J_{\rm L}(t) J_{\rm L} \left(t'+ \frac{i\beta}{2} \right)  \right\rangle_\beta \sim e^{- \frac{2\pi}{\beta}|t-t'|} \ , \label{LRcorr}
\end{eqnarray}
where the second line above follows from standard expectation of thermal two-point correlators, in the limit $|t-t' | \gg \beta/(2\pi)$. This two-point function is non-vanishing at when $t\sim t'$.

Given the TFD-state, consider perturbing the CFT$_R$ with $\int d^dx \phi_{\rm R} J_{\rm R}$. At the linear order, this deformation induces a non-vanishing expectation value for the one-point function $\left\langle J_{\rm R}\right\rangle \not = 0$. When $g=0$ in the interacting Hamiltonian in (\ref{inttfdnl}), we still have $\left\langle J_{\rm L}\right\rangle = 0$. Consider now turning on $g\not =0$, centered around $t=0$. Consider further that the source $\phi_{\rm R}$ is supported at a small time-window around $t = - t_*$. This ensures that the non-vanishing one point function $\left\langle J_{\rm R}\right\rangle$ has long decayed before $g$ could be turned on. Now we obtain a non-vanishing one point function for $\left\langle J_{\rm L}\right\rangle$, given by
\begin{eqnarray}
&& \left\langle J_{\rm L}(t) \right\rangle_g = \left\langle \tilde{\rm TFD} |J_{\rm L}(t) |\tilde{\rm TFD}\right \rangle  \ , \label{JLonepoint}\\
&&\left  |\tilde{\rm TFD}\right \rangle = e^{i g V} e^{i \int d^dx \phi_{\rm R} J_{\rm R}} \left  | {\rm TFD}\right \rangle \ . \label{tildetfd}
\end{eqnarray}
Let us spell this out. From the definition of $\left  |\tilde{\rm TFD}\right \rangle$, it is clear that for $g=0$, $\left\langle J_{\rm L}\right\rangle = 0$ since $J_{\rm L}$ and $J_{\rm R}$ commute. For small values of $t$, we still have $\left[ V, J_{\rm L}\right] \approx 0$, since $J_{\rm L}$ and all individual terms in $V$ are sufficiently local operators. At sufficiently later times, $t \ge t_*$, the effect of $J_{\rm L}$ spreads across the system size and therefore $\left[ V, J_{\rm L}\right] \not= 0$, which yields: $\left\langle J_{\rm L}\right\rangle_g \not= 0$. In light of (\ref{LRcorr}), any expectation value is peaked around $t_{\rm L} \sim - t_{\rm R}$.

Consider the limit $g\ll 1$, which means we can expand the exponential in (\ref{tildetfd}) in powers of $g$. At $n^{\rm th}$-order, there are terms of the following form: 
\begin{eqnarray}
\left\langle J_{\rm L}(t) \right\rangle_g^{(n)} = \left\langle {\rm TFD}| e^{-ig V} J_{\rm L}(t_{\rm L}) \left( \O_{\rm L}(0)\right)^n  J_{\rm L}(t_{\rm R}+ i\beta/2) \left( \O_{\rm L}(i\beta/2)\right)^n |{\rm TFD} \right\rangle \ ,
\end{eqnarray}
using the relation in (\ref{LRrel}). The RHS above is a manifest OTOC. For $t_{\rm L} \sim t_{\rm R} \ge t_*$, $\left\langle J_{\rm L}(t) \right\rangle_g^{(n)}  \to 0$ by definition. The remaining contribution in (\ref{JLonepoint}) is obtained from terms of the following form:
\begin{eqnarray}
\left\langle {\rm TFD}| e^{-i g V} J_{\rm L}(t_{\rm L}) J_{\rm R}(- t_{\rm R})|{\rm TFD} \right\rangle \approx \left\langle e^{-i g V} \right \rangle \left \langle J_{\rm L}(t_{\rm L}) J_{\rm R}(- t_{\rm R}) \right \rangle \ ,
\end{eqnarray}
which finally yields:
\begin{eqnarray}
\left\langle J_{\rm L}(t_{\rm L}) \right\rangle_g \approx C \phi_{\rm R} (-t_{\rm R}) \ , \quad t_{\rm R} \ge t_* \ , 
\end{eqnarray}
for sufficiently slowly varying field $\phi_{\rm R}$. Here, $C$ is a constant that depends on $g$. The bottom-line is: Physical information contained in $\phi_{\rm R}$, inserted at $- t_{\rm R}$ is contained in the one-point function of $J_{\rm L}$ at $t_{\rm L} \sim t_{\rm R}$. Thus, a local information at sufficiently early times on the right-boundary emerges as a local information at a sufficiently late time on the left boundary. This happens, as emphasized in \cite{Gao:2018yzk}, because of the entanglement structure of the TFD-state, and the direct coupling of the left and the right boundary degrees of freedom.

\section{Conclusion \& Outlook}

In this review, we attempted to catalogue a set of recent ideas, with simple and analytically tractable examples to elucidate them further, based on explicit Wormhole geometries in both Euclidean as well as Lorentzian frameworks. To re-emphasize, Wormholes play an integral role in extracting quantum dynamics of Black Holes, understanding properties of quantum entanglement in quantum gravity and Holography.\footnote{Let us emphasize that our discussion has not been, by any stretch of imagination, inclusive of all ideas. For example, see {\it e.g.}~\cite{Lavrelashvili:1987jg, Hawking:1987mz, Giddings:1988wv, Coleman:1988cy, Giddings:1987cg, Giddings:2020yes, Marolf:2020xie} for the so-called ``Wormhole calculus", which we have left untouched.} These features are most transparent in low-dimensional systems: $2$-dimensional and $3$-dimensional gravity.

Perhaps more interestingly, Wormholes raise rather intriguing puzzles as well. The basic conundrum arises from the perspective in \cite{Preskill:1988na, Coleman:1988cy}, in which one has to either sacrifice macroscopic locality at some level, or re-interpret the defining path-integral of the system. In particular, \cite{Preskill:1988na} advocated a path-integral interpretation that involves an averaging over the parameter space of the system, and restore locality. On one hand, such an averaging procedure conflicts with an intrinsic unitary UV-complete description of {\it e.g.}~in string theory \cite{ArkaniHamed:2007js}.

This poses a puzzle with which we are yet to fully reconcile. Here we will enlist various points of view that one may adopt, at this point: The simplest perspective is to assume that the Wormholes {\it by an unknown mechanism} produce some of the desired and fine-grained physics of quantum gravity. From a Euclidean quantum gravitational perspective, this is similar to the Gibbons-Hawking analyses of Black Hole thermodynamics. The Euclidean gravitational framework is ill-defined, yet it captures salient features of Black Hole thermodynamics. This necessitates an alternative and a more thorough understanding of the quantum dynamics of gravity in general. Such a framework is lacking at this point. The interested Reader may keep an eye on some recent ideas explored in {\it e.g.}~\cite{Belin:2020hea, Belin:2020jxr} and its subsequent developments.

A second possibility is that Wormholes capture quantum gravitational aspects only in low-dimensional systems, and does not necessarily play any role in higher dimensions. In many cases, specially for explicit examples of AdS/CFT correspondence, higher-dimensional cases are at a much better control, while the lower-dimensional examples are, at least, subtle. So, one logical possibility is that Wormholes are a part of this subtlety in lower dimensional Holography. Nonetheless, it remains unclear from the bulk picture why Wormholes do not contribute at all (on-shell or even off-shell) and cause the factorization problem in higher dimensions. Furthermore, if we declare that {\it by some mechanism} Wormholes never appear in higher dimensions, it remains unclear how the information paradox is then resolved in such cases. It appears fair to state that, despite a lot of exciting developments, these core questions remain as outstanding problems at large.\footnote{We refer the interested Reader to {\it e.g.}~\cite{Betzios:2019rds, Betzios:2021fnm, VanRaamsdonk:2020tlr, VanRaamsdonk:2021qgv} for more exploration of Wormholes in general dimensions, especially correlation functions of such Holographic states and a discussion on the corresponding QFT models. See also \cite{Iqbal:2021ouw} for a discussion on traversable Wormholes based on force-free electrodynamics.}

A third possibility is an admixture of the above two. Depending on the physical question, Wormholes may or may not play a role. The rules of this game is far from clear, at present. In part, a lot of recent research is devoted into exploring and understanding these issues better. Let us collect some thoughts and advances towards one particular such aspect.

As was reviewed earlier, Wormholes naturally provide an interpretation in terms of a path integral that also performs an (ensemble) averaging. This averaging is rooted in the non-factorizability discussed in section 4.2. Most recent works are motivated on this premise: For example, in two-dimensional (quantum) gravity and its corresponding dual one-dimensional quantum mechanical description, the role of ensemble averaging on partition function, R\'{e}nyi-entropy and the late-time behaviour of spectral form factor is analyzed in \cite{Verlinde:2021kgt, Verlinde:2021jwu}. The best known example of this duality exists between the so-called Jackiw-Teitelboim (JT) gravity and the low energy limit of the so-called Sachdev-Ye-Kitaev (SYK) model --- see {\it e.g.}~\cite{Maldacena:2016hyu, Maldacena:2016upp} for a detailed account of this duality --- in which the ensemble averaging is explicitly realized. To contrast this, we also refer to the Reader to the series of works in \cite{Johnson:2019eik, Johnson:2020heh, Johnson:2020exp, Johnson:2020mwi, Johnson:2020lns, Johnson:2021owr, Johnson:2021rsh, Johnson:2021zuo, Johnson:2021tnl, Johnson:2022wsr} which advocate that no ensemble averaging is required if the JT-gravity picture is viewed from a non-perturbative definition of random matrix models.

Much less is explicitly understood in higher dimensions. However, recent works in \cite{Cotler:2020ugk, Cotler:2020hgz, Cotler:2020lxj, Cotler:2021cqa} explore the role of Wormholes in AdS$_3$, {\it i.e.}~in one higher dimension. It has been argued that pure gravity in AdS$_3$, which is otherwise thought to not posses a Holographic dual description, can be thought as the dual of an ensemble average of two-dimensional CFTs. Furthermore, new Wormhole solutions were also obtained in higher dimensions, which contribute non-perturbatively to the quantum gravity path-integral. A systematic and complete understanding of higher dimensional Wormholes, along these directions will certainly be an important progress.

Even in the best understood low-dimensional examples, it is important to understand how stable the relevant Wormhole solutions are. From a {\it bottom-up} approach, stable Wormhole solutions can be constructed in low-energy models, currently no explicit such stable solution exists in a UV-complete gravitational theory, {\it i.e.}~in supergravity models that one obtains in the low-energy limit of string theory. In particular, \cite{Marolf:2021kjc} a thorough analyses have been carried out of various such models, across various dimensions, and it was found that the Wormhole solutions always suffer from D-brane nucleation instabilities. This instability is similar to the classic example of Wormhole constructed by Maldacena-Maoz in \cite{Maldacena:2004rf}. While this approach cannot rule out the possibility of a stable Wormhole solution in a particular UV-complete theory, since it is impossible to exhaust all the possibilities, it can provide a constructive example with some luck and educated guesses. At present, however, an explicit example of a stable Wormhole in a UV-complete description is lacking.

Given that Wormholes cannot be ruled out of consideration, an alternative perspective is to take them seriously to a logical extreme. The (unreasonable) effectiveness of semi-classical (quantum) gravity can be thought of as encoding an averaged description of the underlying microscopic quantum gravitational spectrum. In particular, \cite{Belin:2020hea} proposes a statistical hypothesis on the OPE coefficients of a dual two-dimensional CFT and relates this proposal to the Wormhole contribution in Euclidean AdS$_3$. Furthermore, \cite{Belin:2020jxr} combines a generalized notion of this OPE-statistics in the presence of a global symmetry with Wormholes to arrive at the desired conclusion that global symmetries cannot exist in quantum gravity. At the very least, it appears plausible that even though Wormholes may not be the complete story, they certainly encode some fine-grained quantum information of gravity.

It is worth mentioning that there are explicit examples in which Wormholes do not imply an averaging. For example, \cite{Saad:2021rcu} explicit example of a Wormhole has been constructed in terms of the dual SYK-theory with frozen coupling constants. These turn out to be ``half-Wormholes" in that they do not asymptote to multiple boundaries. Therefore, these Wormhole configurations have no non-factorizable correlators between the CFTs defined on the left and the right boundaries.

From an observational outlook, recent work on traversable Wormholes have brought a fresh perspective. While such Wormholes are not designed for exotic faster-than-speed-of-light travel, they represent a class of examples that can be realized in a low-energy lab and explored further. In particular, a suitable configuration and dynamics in an SYK-like system can be prepared in the lab, which, using the Holographic duality, uncovers the physics of Wormholes in an AdS-geometry. For example, in the SYK-like description in the lab, such traversable Wormholes can be thought of as teleportation protocols in the quantum mechanical model. This aspect brings together aspects of quantum gravity within the realm of quantum simulations on quantum computers. See {\it e.g.}~\cite{Maldacena:2017axo, Susskind:2017nto, Bao:2018msr, vanBreukelen:2017dul, Bak:2018txn, Gao:2019nyj, Caceres:2021nsa}, for a theoretical account and \cite{Landsman:2018jpm} for an experimental outlook.

Let us conclude this review with a speculative remark. It is well-known that Euclidean instanton solution play an important role in understanding the phenomenon of resurgence in quantum field theories, see {\it e.g.}\cite{Dunne:2015eaa}. This is an incredibly curious idea in which perturbative quantum physics around one saddle encodes non-perturbative physics of some other saddle. Whether Wormholes can be relevant in such a structure in quantum dynamics of gravity is an extremely interesting question to address.

\section{Acknowledgements}

AK thanks Avik Banerjee, Aranya Bhattacharya, Elena Caceres, Shouvik Dutta, Bobby Ezhuthachan, Hong Liu, Nirmalya Kajuri, Sandipan Kundu, Alok Laddha, Ayan Mukhopadhyay, Ayan K.~Patra, Augniva Ray, Shibaji Roy, Sanjit Shashi, Harvendra Singh, Julian Sonner for numerous illuminating discussions. We also thank Arthur Hebecker, Suvrat Raju and Thomas van Riet for useful feedback on the manuscript. A special thanks go to Elena Caceres, Ayan K.~Patra and Sanjit Shashi for collaboration that is directly relevant for this review. This work is supported by the Department of Atomic Energy, Govt.~of India and CEFIPRA no 6304-3.

\appendix
\section{Isometries of AdS$_3$}\label{isoads3}

Recall that AdS$_3$ can be embedded in the ${\mathbb R}^{2,2}$: 
\begin{equation}
ds^2 = -d\bar{v}^2 - d\bar{u}^2 + d\bar{x}^2 + d\bar{y}^2 \ ,
\end{equation}
by the hyperboloid surface:
\begin{equation}
-\bar{v}^2 - \bar{u}^2 + \bar{x}^2 + \bar{y}^2 = -\ell^2 \ ,
\end{equation}
where $\ell$ is the AdS curvature scale.

Now, choose a four-vector $\bar{x}^a = (\bar{v},\bar{u},\bar{x},\bar{y})$. AdS$_3$ has six linearly independent Killing vectors that correspond to rotations and boosts in the $(2+2)$ spacetime. These are given by
\begin{equation}
J_{ab} \equiv \bar{x}_b \frac{\partial}{\partial\bar{x}^a} - \bar{x}_a \frac{\partial}{\partial\bar{x}^b} \ .
\end{equation}
The Killing vectors form an so$(2,2)$ algebra:
\begin{equation}
[J_{ab},J_{cd}] = \eta_{ac}J_{bd} - \eta_{ad}J_{bc} - \eta_{bc}J_{ad} + \eta_{bd}J_{ac} \ . \label{comm}
\end{equation}

In the Poincar\'e metric for AdS$_3$:
\begin{equation}
\frac{ds^2}{\ell^2} = \frac{-dt^2 + dx^2 + dy^2}{y^2} \ , \label{Pmet}
\end{equation}
the Killing vectors are as follows:
\begin{align}
J_{01} &= \left(\frac{\ell^2 + t^2 + x^2 + y^2}{2\ell}\right)\partial_t + \frac{tx}{\ell}\partial_x + \frac{ty}{\ell}\partial_y \ , \label{k1} \\
J_{02} &= \left(\frac{-\ell^2 + t^2 + x^2 + y^2}{2\ell}\right)\partial_t + \frac{tx}{\ell}\partial_x + \frac{ty}{\ell}\partial_y \ , \\
J_{03} &= -x\partial_t - t\partial_x \ , \\
J_{12} &= -t\partial_t - x\partial_x - y\partial_y \ , \\
J_{13} &= \left(\frac{\ell^2-t^2-x^2+y^2}{2\ell}\right)\partial_x - \frac{tx}{\ell}\partial_t - xy\partial_y \ , \\
J_{23} &= \left(\frac{-\ell^2-t^2-x^2+y^2}{2\ell}\right)\partial_x - \frac{tx}{\ell}\partial_t - \frac{xy}{\ell}\partial_y \ . \label{k6}
\end{align}

Multi-boundary black holes are typically thought of as quotients of AdS$_3$ by a discrete set of isometries in the group SO$(2,2)$. For example, consider the quotienting by the dilatation generator that acts as: $e^{-2\pi\kappa J_{12}} \cdot (t,x,y) = e^{2\pi\kappa}(t,x,y)$, and hence yields the identification:
\begin{equation}
(t, x, y) \sim e^{-2\pi\kappa J_{12}} \cdot (t,x,y) = e^{2\pi\kappa}(t,x,y) \ , \label{dilAct}
\end{equation}
where $\kappa \in {\mathbb R}$ is a parameter of the identification. Similarly, given a Killing vector $\xi$, and a parameter $\alpha_0$, the identification sub-group is given by $\{e^{\alpha_0 \xi}\}$.

\section{Technical Details of Identification}\label{identads3}

In this appendix we include more details on the identification procedure of section \ref{EWormMulti}. We begin with the Killing vectors in (\ref{k1})-(\ref{k6}), at $t=0$ slice, and redefine the following basis on ${\mathbb H}^2$, subsequently presenting also their lift to the full AdS$_3$-geometry (therefore, defined for all values of $t$):
\begin{eqnarray}
J_{T} \equiv \ell\left(\partial_z + \partial_{\bar{z}}\right) & \longrightarrow& \tilde{J}_{T} \equiv J_{13} - J_{23} \ , \\
J_{D} \equiv z\partial_z + \bar{z}\partial_{\bar{z}} & \longrightarrow& \tilde{J}_{D} \equiv -J_{12} \ , \\
J_{S} \equiv \dfrac{1}{\ell} \left(z^2\partial_z + \bar{z}^2\partial_{\bar{z}}\right) & \longrightarrow& \tilde{J}_{S} \equiv -J_{13} - J_{23} \ .
\end{eqnarray}
Here, the tilde stands for the lifted Killing vectors. It is easy to check that the tilde Killing vectors form an ${\rm sl}(2, {\mathbb R})$-algebra. To identify the precise transformation in terms of the Killing vectors, note that $e^{\kappa J_T}$, $e^{\kappa J_D}$ and $e^{\kappa J_S}$ correspond to translation, dilatation and special conformal transformation. The inversion transformation can be constructed as $\I_a = e^{a J_T} e^{J_S/a} e^{a J_T}$.

To see the effect of the Killing vectors on geodesics on ${\mathbb H}^2$, let us first obtain the geodesics. These are obtained by extremizing the functional:
\begin{eqnarray}
I = \int \frac{dy}{y}\sqrt{1 + \left( \frac{dx}{dy}\right)^2}  \ ,
\end{eqnarray}
that leads to the following equations of motion:
\begin{eqnarray}
\frac{1}{y} \frac{dx}{dy} \frac{1}{\sqrt{1 + (dx/dy)^2}} = \alpha \ . \label{eom}
\end{eqnarray}
Here $\alpha$ is an integral of motion. There are two distinct classes of solutions to (\ref{eom}):
\begin{eqnarray}
&& \alpha = 0 \ , \quad x = {\rm constant} \ , \\
&& \alpha \not = 0 \ , \quad (x-c)^2 + y^2 = \alpha^{-2} \ ,
\end{eqnarray}
\textit{i.e.}, straight lines and semicircles. The action of the transformations discussed above can be determined, on such geodesics, for example:
\begin{eqnarray}
&& e^{aJ_T/\ell} \cdot \left(x_1,\sqrt{R^2 - (x_1 - c)^2}\right) = \left(x_2,\sqrt{R^2 - (x_2 - a - c)^2}\right),\ \ \ x_2 \equiv x_1 + a \ , \nonumber\\
&& e^{aJ_D}\left(x_1,\sqrt{R^2 - (x_1 - c)^2}\right) = \left(x_2,\sqrt{(e^a R)^2 - (x_2 - e^{a}c)^2}\right),\ \ \ x_2 \equiv e^{a}x_1 \ , \nonumber\\
&& \mathcal{I}_a \cdot \left(x_1,\sqrt{R^2 - x_1^2}\right) = \left(x_2,\sqrt{\frac{(a\ell)^4}{R^2} - x_2^2}\right),\ \ \ x_2 \equiv -\frac{(a\ell)^2}{R^2} x_1 \ .
\end{eqnarray}

A crucial combination is the orientation-reversing isometry, which is explicitly given by\cite{Caceres:2019giy}
\begin{eqnarray}
\O = e^{c_2 J_T/\ell} e^{\nu J_D} \I_{R_1/\ell} e^{-c_1 J_T/\ell} \ , \quad \nu = \frac{R_2}{R_1} \ , \label{orientrev}
\end{eqnarray}
which maps two oppositely oriented semi-circles centered at $c_1$ and $c_2$, with radii $R_1$ and $R_2$, into each other.\footnote{Demanding further that the quotient space does not contain any fixed point of $\O$ leads to: $|c_1 - c_2| > R_1 + R_2$, which can be implemented. Note that, without this constraint, the fixed point of $\O$ will result in a curvature singularity at the fixed point.} In view of the discussion on transformations, the figure \ref{figs:3bdrypic} can be generated with quotienting AdS$_3$ by dilatation, followed by orientation-reversing isometry. This is summarized by
\begin{eqnarray}
\tilde{\O} = e^{c_2 \tilde{J}_T/\ell} e^{R \tilde{J}_T/\ell}  e^{\ell \tilde{J}_S/R} e^{R \tilde{J}_T/\ell} e^{-c_1 \tilde{J}_T/\ell} \ , \label{orientrevfig12}
\end{eqnarray}
which can also be written as a single exponential, using the algebra of the generators.


\bibliography{syk}

\providecommand{\href}[2]{#2}\begingroup\raggedright\begin{thebibliography}{100}

\bibitem{EventHorizonTelescope:2021bee}
{\scshape Event Horizon Telescope} collaboration, \emph{{First M87 Event
  Horizon Telescope Results. VII. Polarization of the Ring}},
  \href{https://doi.org/10.3847/2041-8213/abe71d}{\emph{Astrophys. J. Lett.}
  {\bfseries 910} (2021) L12}
  [\href{https://arxiv.org/abs/2105.01169}{{\ttfamily 2105.01169}}].

\bibitem{EventHorizonTelescope:2021srq}
{\scshape Event Horizon Telescope} collaboration, \emph{{First M87 Event
  Horizon Telescope Results. VIII. Magnetic Field Structure near The Event
  Horizon}}, \href{https://doi.org/10.3847/2041-8213/abe4de}{\emph{Astrophys.
  J. Lett.} {\bfseries 910} (2021) L13}
  [\href{https://arxiv.org/abs/2105.01173}{{\ttfamily 2105.01173}}].

\bibitem{LIGOScientific:2016aoc}
{\scshape LIGO Scientific, Virgo} collaboration, \emph{{Observation of
  Gravitational Waves from a Binary Black Hole Merger}},
  \href{https://doi.org/10.1103/PhysRevLett.116.061102}{\emph{Phys. Rev. Lett.}
  {\bfseries 116} (2016) 061102}
  [\href{https://arxiv.org/abs/1602.03837}{{\ttfamily 1602.03837}}].

\bibitem{Einstein:1935tc}
A.~Einstein and N.~Rosen, \emph{{The Particle Problem in the General Theory of
  Relativity}}, \href{https://doi.org/10.1103/PhysRev.48.73}{\emph{Phys. Rev.}
  {\bfseries 48} (1935) 73}.

\bibitem{Misner:1957mt}
C.W.~Misner and J.A.~Wheeler, \emph{{Classical physics as geometry:
  Gravitation, electromagnetism, unquantized charge, and mass as properties of
  curved empty space}},
  \href{https://doi.org/10.1016/0003-4916(57)90049-0}{\emph{Annals Phys.}
  {\bfseries 2} (1957) 525}.

\bibitem{Hebecker:2018ofv}
A.~Hebecker, T.~Mikhail and P.~Soler, \emph{{Euclidean wormholes, baby
  universes, and their impact on particle physics and cosmology}},
  \href{https://doi.org/10.3389/fspas.2018.00035}{\emph{Front. Astron. Space
  Sci.} {\bfseries 5} (2018) 35}
  [\href{https://arxiv.org/abs/1807.00824}{{\ttfamily 1807.00824}}].

\bibitem{VanRiet:2020pcn}
T.~Van~Riet, \emph{{Instantons, Euclidean wormholes and AdS/CFT}},
  \href{https://doi.org/10.22323/1.376.0121}{\emph{PoS} {\bfseries CORFU2019}
  (2020) 121} [\href{https://arxiv.org/abs/2004.08956}{{\ttfamily
  2004.08956}}].

\bibitem{Stephens:1993an}
C.R.~Stephens, G.~'t~Hooft and B.F.~Whiting, \emph{{Black hole evaporation
  without information loss}},
  \href{https://doi.org/10.1088/0264-9381/11/3/014}{\emph{Class. Quant. Grav.}
  {\bfseries 11} (1994) 621}
  [\href{https://arxiv.org/abs/gr-qc/9310006}{{\ttfamily gr-qc/9310006}}].

\bibitem{Susskind:1994vu}
L.~Susskind, \emph{{The World as a hologram}},
  \href{https://doi.org/10.1063/1.531249}{\emph{J. Math. Phys.} {\bfseries 36}
  (1995) 6377} [\href{https://arxiv.org/abs/hep-th/9409089}{{\ttfamily
  hep-th/9409089}}].

\bibitem{Maldacena:1997re}
J.M.~Maldacena, \emph{{The Large N limit of superconformal field theories and
  supergravity}}, \href{https://doi.org/10.1023/A:1026654312961}{\emph{Adv.
  Theor. Math. Phys.} {\bfseries 2} (1998) 231}
  [\href{https://arxiv.org/abs/hep-th/9711200}{{\ttfamily hep-th/9711200}}].

\bibitem{Ryu:2006bv}
S.~Ryu and T.~Takayanagi, \emph{{Holographic derivation of entanglement entropy
  from AdS/CFT}},
  \href{https://doi.org/10.1103/PhysRevLett.96.181602}{\emph{Phys. Rev. Lett.}
  {\bfseries 96} (2006) 181602}
  [\href{https://arxiv.org/abs/hep-th/0603001}{{\ttfamily hep-th/0603001}}].

\bibitem{Ryu:2006ef}
S.~Ryu and T.~Takayanagi, \emph{{Aspects of Holographic Entanglement Entropy}},
  \href{https://doi.org/10.1088/1126-6708/2006/08/045}{\emph{JHEP} {\bfseries
  08} (2006) 045} [\href{https://arxiv.org/abs/hep-th/0605073}{{\ttfamily
  hep-th/0605073}}].

\bibitem{Maldacena:2013xja}
J.~Maldacena and L.~Susskind, \emph{{Cool horizons for entangled black holes}},
  \href{https://doi.org/10.1002/prop.201300020}{\emph{Fortsch. Phys.}
  {\bfseries 61} (2013) 781} [\href{https://arxiv.org/abs/1306.0533}{{\ttfamily
  1306.0533}}].

\bibitem{Harlow:2016vwg}
D.~Harlow, \emph{{The Ryu\textendash{}Takayanagi Formula from Quantum Error
  Correction}}, \href{https://doi.org/10.1007/s00220-017-2904-z}{\emph{Commun.
  Math. Phys.} {\bfseries 354} (2017) 865}
  [\href{https://arxiv.org/abs/1607.03901}{{\ttfamily 1607.03901}}].

\bibitem{Harlow:2018fse}
D.~Harlow, \emph{{TASI Lectures on the Emergence of Bulk Physics in AdS/CFT}},
  \href{https://doi.org/10.22323/1.305.0002}{\emph{PoS} {\bfseries TASI2017}
  (2018) 002} [\href{https://arxiv.org/abs/1802.01040}{{\ttfamily
  1802.01040}}].

\bibitem{Bekenstein:1973ur}
J.D.~Bekenstein, \emph{{Black holes and entropy}},
  \href{https://doi.org/10.1103/PhysRevD.7.2333}{\emph{Phys. Rev. D} {\bfseries
  7} (1973) 2333}.

\bibitem{Hawking:1975vcx}
S.W.~Hawking, \emph{{Particle Creation by Black Holes}},
  \href{https://doi.org/10.1007/BF02345020}{\emph{Commun. Math. Phys.}
  {\bfseries 43} (1975) 199}.

\bibitem{Mathur:2009hf}
S.D.~Mathur, \emph{{The Information paradox: A Pedagogical introduction}},
  \href{https://doi.org/10.1088/0264-9381/26/22/224001}{\emph{Class. Quant.
  Grav.} {\bfseries 26} (2009) 224001}
  [\href{https://arxiv.org/abs/0909.1038}{{\ttfamily 0909.1038}}].

\bibitem{Page:1993wv}
D.N.~Page, \emph{{Information in black hole radiation}},
  \href{https://doi.org/10.1103/PhysRevLett.71.3743}{\emph{Phys. Rev. Lett.}
  {\bfseries 71} (1993) 3743}
  [\href{https://arxiv.org/abs/hep-th/9306083}{{\ttfamily hep-th/9306083}}].

\bibitem{Almheiri:2018xdw}
A.~Almheiri, \emph{{Holographic Quantum Error Correction and the Projected
  Black Hole Interior}},  \href{https://arxiv.org/abs/1810.02055}{{\ttfamily
  1810.02055}}.

\bibitem{Hayden:2018khn}
P.~Hayden and G.~Penington, \emph{{Learning the Alpha-bits of Black Holes}},
  \href{https://doi.org/10.1007/JHEP12(2019)007}{\emph{JHEP} {\bfseries 12}
  (2019) 007} [\href{https://arxiv.org/abs/1807.06041}{{\ttfamily
  1807.06041}}].

\bibitem{Penington:2019npb}
G.~Penington, \emph{{Entanglement Wedge Reconstruction and the Information
  Paradox}},  \href{https://arxiv.org/abs/1905.08255}{{\ttfamily 1905.08255}}.

\bibitem{Almheiri:2019psf}
A.~Almheiri, N.~Engelhardt, D.~Marolf and H.~Maxfield, \emph{{The entropy of
  bulk quantum fields and the entanglement wedge of an evaporating black
  hole}}, \href{https://doi.org/10.1007/JHEP12(2019)063}{\emph{JHEP} {\bfseries
  12} (2019) 063} [\href{https://arxiv.org/abs/1905.08762}{{\ttfamily
  1905.08762}}].

\bibitem{Almheiri:2019hni}
A.~Almheiri, R.~Mahajan, J.~Maldacena and Y.~Zhao, \emph{{The Page curve of
  Hawking radiation from semiclassical geometry}},
  \href{https://doi.org/10.1007/JHEP03(2020)149}{\emph{JHEP} {\bfseries 03}
  (2020) 149} [\href{https://arxiv.org/abs/1908.10996}{{\ttfamily
  1908.10996}}].

\bibitem{Lata}
B.A.~{\it et al}, \emph{{Quantum Information Scrambling: From Holography to
  Quantum Simulators}}, .

\bibitem{Marino:2015yie}
M.~Mari\~no, \emph{{Instantons and Large N}: {An Introduction to
  Non-Perturbative Methods in Quantum Field Theory}}, Cambridge University
  Press (9, 2015).

\bibitem{Gubser:1998bc}
S.S.~Gubser, I.R.~Klebanov and A.M.~Polyakov, \emph{{Gauge theory correlators
  from noncritical string theory}},
  \href{https://doi.org/10.1016/S0370-2693(98)00377-3}{\emph{Phys. Lett. B}
  {\bfseries 428} (1998) 105}
  [\href{https://arxiv.org/abs/hep-th/9802109}{{\ttfamily hep-th/9802109}}].

\bibitem{Witten:1998qj}
E.~Witten, \emph{{Anti-de Sitter space and holography}},
  \href{https://doi.org/10.4310/ATMP.1998.v2.n2.a2}{\emph{Adv. Theor. Math.
  Phys.} {\bfseries 2} (1998) 253}
  [\href{https://arxiv.org/abs/hep-th/9802150}{{\ttfamily hep-th/9802150}}].

\bibitem{Sachdev_1993}
S.~Sachdev and J.~Ye, \emph{Gapless spin-fluid ground state in a random quantum
  heisenberg magnet},
  \href{https://doi.org/10.1103/physrevlett.70.3339}{\emph{Physical Review
  Letters} {\bfseries 70} (1993) 3339–3342}.

\bibitem{Polchinski:2016xgd}
J.~Polchinski and V.~Rosenhaus, \emph{{The Spectrum in the Sachdev-Ye-Kitaev
  Model}}, \href{https://doi.org/10.1007/JHEP04(2016)001}{\emph{JHEP}
  {\bfseries 04} (2016) 001}
  [\href{https://arxiv.org/abs/1601.06768}{{\ttfamily 1601.06768}}].

\bibitem{Maldacena:2016hyu}
J.~Maldacena and D.~Stanford, \emph{{Remarks on the Sachdev-Ye-Kitaev model}},
  \href{https://doi.org/10.1103/PhysRevD.94.106002}{\emph{Phys. Rev. D}
  {\bfseries 94} (2016) 106002}
  [\href{https://arxiv.org/abs/1604.07818}{{\ttfamily 1604.07818}}].

\bibitem{Maldacena:2016upp}
J.~Maldacena, D.~Stanford and Z.~Yang, \emph{{Conformal symmetry and its
  breaking in two dimensional Nearly Anti-de-Sitter space}},
  \href{https://doi.org/10.1093/ptep/ptw124}{\emph{PTEP} {\bfseries 2016}
  (2016) 12C104} [\href{https://arxiv.org/abs/1606.01857}{{\ttfamily
  1606.01857}}].

\bibitem{Klebanov:2002ja}
I.R.~Klebanov and A.M.~Polyakov, \emph{{AdS dual of the critical O(N) vector
  model}}, \href{https://doi.org/10.1016/S0370-2693(02)02980-5}{\emph{Phys.
  Lett. B} {\bfseries 550} (2002) 213}
  [\href{https://arxiv.org/abs/hep-th/0210114}{{\ttfamily hep-th/0210114}}].

\bibitem{Giombi:2009wh}
S.~Giombi and X.~Yin, \emph{{Higher Spin Gauge Theory and Holography: The
  Three-Point Functions}},
  \href{https://doi.org/10.1007/JHEP09(2010)115}{\emph{JHEP} {\bfseries 09}
  (2010) 115} [\href{https://arxiv.org/abs/0912.3462}{{\ttfamily 0912.3462}}].

\bibitem{Vasiliev:2003ev}
M.A.~Vasiliev, \emph{{Nonlinear equations for symmetric massless higher spin
  fields in (A)dS(d)}},
  \href{https://doi.org/10.1016/S0370-2693(03)00872-4}{\emph{Phys. Lett. B}
  {\bfseries 567} (2003) 139}
  [\href{https://arxiv.org/abs/hep-th/0304049}{{\ttfamily hep-th/0304049}}].

\bibitem{Hubeny:2007xt}
V.E.~Hubeny, M.~Rangamani and T.~Takayanagi, \emph{{A Covariant holographic
  entanglement entropy proposal}},
  \href{https://doi.org/10.1088/1126-6708/2007/07/062}{\emph{JHEP} {\bfseries
  07} (2007) 062} [\href{https://arxiv.org/abs/0705.0016}{{\ttfamily
  0705.0016}}].

\bibitem{Rangamani:2016dms}
M.~Rangamani and T.~Takayanagi, \emph{{Holographic Entanglement Entropy}},
  vol.~931, Springer (2017),
  \href{https://doi.org/10.1007/978-3-319-52573-0}{10.1007/978-3-319-52573-0},
  [\href{https://arxiv.org/abs/1609.01287}{{\ttfamily 1609.01287}}].

\bibitem{Engelhardt:2014gca}
N.~Engelhardt and A.C.~Wall, \emph{{Quantum Extremal Surfaces: Holographic
  Entanglement Entropy beyond the Classical Regime}},
  \href{https://doi.org/10.1007/JHEP01(2015)073}{\emph{JHEP} {\bfseries 01}
  (2015) 073} [\href{https://arxiv.org/abs/1408.3203}{{\ttfamily 1408.3203}}].

\bibitem{gibbons1993euclidean}
G.W.~Gibbons, \emph{Euclidean quantum gravity}, World Scientific, Singapore
  River Edge, NJ (1993).

\bibitem{Gibbons:1978ac}
G.W.~Gibbons, S.W.~Hawking and M.J.~Perry, \emph{{Path Integrals and the
  Indefiniteness of the Gravitational Action}},
  \href{https://doi.org/10.1016/0550-3213(78)90161-X}{\emph{Nucl. Phys. B}
  {\bfseries 138} (1978) 141}.

\bibitem{Eguchi:1980jx}
T.~Eguchi, P.B.~Gilkey and A.J.~Hanson, \emph{{Gravitation, Gauge Theories and
  Differential Geometry}},
  \href{https://doi.org/10.1016/0370-1573(80)90130-1}{\emph{Phys. Rept.}
  {\bfseries 66} (1980) 213}.

\bibitem{Lavrelashvili:1987jg}
G.V.~Lavrelashvili, V.A.~Rubakov and P.G.~Tinyakov, \emph{{Disruption of
  Quantum Coherence upon a Change in Spatial Topology in Quantum Gravity}},
  {\emph{JETP Lett.} {\bfseries 46} (1987) 167}.

\bibitem{Hawking:1987mz}
S.W.~Hawking, \emph{{Quantum Coherence Down the Wormhole}},
  \href{https://doi.org/10.1016/0370-2693(87)90028-1}{\emph{Phys. Lett. B}
  {\bfseries 195} (1987) 337}.

\bibitem{Giddings:1988cx}
S.B.~Giddings and A.~Strominger, \emph{{Loss of Incoherence and Determination
  of Coupling Constants in Quantum Gravity}},
  \href{https://doi.org/10.1016/0550-3213(88)90109-5}{\emph{Nucl. Phys. B}
  {\bfseries 307} (1988) 854}.

\bibitem{Coleman:1988cy}
S.R.~Coleman, \emph{{Black Holes as Red Herrings: Topological Fluctuations and
  the Loss of Quantum Coherence}},
  \href{https://doi.org/10.1016/0550-3213(88)90110-1}{\emph{Nucl. Phys. B}
  {\bfseries 307} (1988) 867}.

\bibitem{Giddings:1988wv}
S.B.~Giddings and A.~Strominger, \emph{{Baby Universes, Third Quantization and
  the Cosmological Constant}},
  \href{https://doi.org/10.1016/0550-3213(89)90353-2}{\emph{Nucl. Phys. B}
  {\bfseries 321} (1989) 481}.

\bibitem{Giddings:1987cg}
S.B.~Giddings and A.~Strominger, \emph{{Axion Induced Topology Change in
  Quantum Gravity and String Theory}},
  \href{https://doi.org/10.1016/0550-3213(88)90446-4}{\emph{Nucl. Phys. B}
  {\bfseries 306} (1988) 890}.

\bibitem{Preskill:1988na}
J.~Preskill, \emph{{Wormholes in Space-time and the Constants of Nature}},
  \href{https://doi.org/10.1016/0550-3213(89)90592-0}{\emph{Nucl. Phys. B}
  {\bfseries 323} (1989) 141}.

\bibitem{ArkaniHamed:2007js}
N.~Arkani-Hamed, J.~Orgera and J.~Polchinski, \emph{{Euclidean wormholes in
  string theory}},
  \href{https://doi.org/10.1088/1126-6708/2007/12/018}{\emph{JHEP} {\bfseries
  12} (2007) 018} [\href{https://arxiv.org/abs/0705.2768}{{\ttfamily
  0705.2768}}].

\bibitem{Hertog:2017owm}
T.~Hertog, M.~Trigiante and T.~Van~Riet, \emph{{Axion Wormholes in AdS
  Compactifications}},
  \href{https://doi.org/10.1007/JHEP06(2017)067}{\emph{JHEP} {\bfseries 06}
  (2017) 067} [\href{https://arxiv.org/abs/1702.04622}{{\ttfamily
  1702.04622}}].

\bibitem{Katmadas:2018ksp}
S.~Katmadas, D.~Ruggeri, M.~Trigiante and T.~Van~Riet, \emph{{The holographic
  dual to supergravity instantons in $\rm AdS_5\times S^5/\mathbb{Z}_k$}},
  \href{https://doi.org/10.1007/JHEP10(2019)205}{\emph{JHEP} {\bfseries 10}
  (2019) 205} [\href{https://arxiv.org/abs/1812.05986}{{\ttfamily
  1812.05986}}].

\bibitem{Hertog:2018kbz}
T.~Hertog, B.~Truijen and T.~Van~Riet, \emph{{Euclidean axion wormholes have
  multiple negative modes}},
  \href{https://doi.org/10.1103/PhysRevLett.123.081302}{\emph{Phys. Rev. Lett.}
  {\bfseries 123} (2019) 081302}
  [\href{https://arxiv.org/abs/1811.12690}{{\ttfamily 1811.12690}}].

\bibitem{Loges:2022nuw}
G.J.~Loges, G.~Shiu and N.~Sudhir, \emph{{Complex Saddles and Euclidean
  Wormholes in the Lorentzian Path Integral}},
  \href{https://arxiv.org/abs/2203.01956}{{\ttfamily 2203.01956}}.

\bibitem{Marolf:2021kjc}
D.~Marolf and J.E.~Santos, \emph{{AdS Euclidean wormholes}},
  \href{https://arxiv.org/abs/2101.08875}{{\ttfamily 2101.08875}}.

\bibitem{Almheiri:2020cfm}
A.~Almheiri, T.~Hartman, J.~Maldacena, E.~Shaghoulian and A.~Tajdini,
  \emph{{The entropy of Hawking radiation}},
  \href{https://doi.org/10.1103/RevModPhys.93.035002}{\emph{Rev. Mod. Phys.}
  {\bfseries 93} (2021) 035002}
  [\href{https://arxiv.org/abs/2006.06872}{{\ttfamily 2006.06872}}].

\bibitem{Raju:2020smc}
S.~Raju, \emph{{Lessons from the information paradox}},
  \href{https://doi.org/10.1016/j.physrep.2021.10.001}{\emph{Phys. Rept.}
  {\bfseries 943} (2022) 2187}
  [\href{https://arxiv.org/abs/2012.05770}{{\ttfamily 2012.05770}}].

\bibitem{Maldacena:2001kr}
J.M.~Maldacena, \emph{{Eternal black holes in anti-de Sitter}},
  \href{https://doi.org/10.1088/1126-6708/2003/04/021}{\emph{JHEP} {\bfseries
  04} (2003) 021} [\href{https://arxiv.org/abs/hep-th/0106112}{{\ttfamily
  hep-th/0106112}}].

\bibitem{Maldacena:2018lmt}
J.~Maldacena and X.-L.~Qi, \emph{{Eternal traversable wormhole}},
  \href{https://arxiv.org/abs/1804.00491}{{\ttfamily 1804.00491}}.

\bibitem{Cottrell:2018ash}
W.~Cottrell, B.~Freivogel, D.M.~Hofman and S.F.~Lokhande, \emph{{How to Build
  the Thermofield Double State}},
  \href{https://doi.org/10.1007/JHEP02(2019)058}{\emph{JHEP} {\bfseries 02}
  (2019) 058} [\href{https://arxiv.org/abs/1811.11528}{{\ttfamily
  1811.11528}}].

\bibitem{Takahashi:1996zn}
Y.~Takahashi and H.~Umezawa, \emph{{Thermo field dynamics}},
  \href{https://doi.org/10.1142/S0217979296000817}{\emph{Int. J. Mod. Phys. B}
  {\bfseries 10} (1996) 1755}.

\bibitem{Mathur:2014dia}
S.D.~Mathur, \emph{{What is the dual of two entangled CFTs?}},
  \href{https://arxiv.org/abs/1402.6378}{{\ttfamily 1402.6378}}.

\bibitem{Israel:1976ur}
W.~Israel, \emph{{Thermo field dynamics of black holes}},
  \href{https://doi.org/10.1016/0375-9601(76)90178-X}{\emph{Phys. Lett. A}
  {\bfseries 57} (1976) 107}.

\bibitem{Penington:2019kki}
G.~Penington, S.H.~Shenker, D.~Stanford and Z.~Yang, \emph{{Replica wormholes
  and the black hole interior}},
  \href{https://arxiv.org/abs/1911.11977}{{\ttfamily 1911.11977}}.

\bibitem{Almheiri:2019qdq}
A.~Almheiri, T.~Hartman, J.~Maldacena, E.~Shaghoulian and A.~Tajdini,
  \emph{{Replica Wormholes and the Entropy of Hawking Radiation}},
  \href{https://doi.org/10.1007/JHEP05(2020)013}{\emph{JHEP} {\bfseries 05}
  (2020) 013} [\href{https://arxiv.org/abs/1911.12333}{{\ttfamily
  1911.12333}}].

\bibitem{Chen:2020hmv}
H.Z.~Chen, R.C.~Myers, D.~Neuenfeld, I.A.~Reyes and J.~Sandor, \emph{{Quantum
  Extremal Islands Made Easy, Part II: Black Holes on the Brane}},
  \href{https://doi.org/10.1007/JHEP12(2020)025}{\emph{JHEP} {\bfseries 12}
  (2020) 025} [\href{https://arxiv.org/abs/2010.00018}{{\ttfamily
  2010.00018}}].

\bibitem{Dong:2017xht}
X.~Dong and A.~Lewkowycz, \emph{{Entropy, Extremality, Euclidean Variations,
  and the Equations of Motion}},
  \href{https://doi.org/10.1007/JHEP01(2018)081}{\emph{JHEP} {\bfseries 01}
  (2018) 081} [\href{https://arxiv.org/abs/1705.08453}{{\ttfamily
  1705.08453}}].

\bibitem{Geng:2020qvw}
H.~Geng and A.~Karch, \emph{{Massive islands}},
  \href{https://doi.org/10.1007/JHEP09(2020)121}{\emph{JHEP} {\bfseries 09}
  (2020) 121} [\href{https://arxiv.org/abs/2006.02438}{{\ttfamily
  2006.02438}}].

\bibitem{Geng:2020fxl}
H.~Geng, A.~Karch, C.~Perez-Pardavila, S.~Raju, L.~Randall, M.~Riojas et~al.,
  \emph{{Information Transfer with a Gravitating Bath}},
  \href{https://doi.org/10.21468/SciPostPhys.10.5.103}{\emph{SciPost Phys.}
  {\bfseries 10} (2021) 103}
  [\href{https://arxiv.org/abs/2012.04671}{{\ttfamily 2012.04671}}].

\bibitem{Geng:2021hlu}
H.~Geng, A.~Karch, C.~Perez-Pardavila, S.~Raju, L.~Randall, M.~Riojas et~al.,
  \emph{{Inconsistency of islands in theories with long-range gravity}},
  \href{https://doi.org/10.1007/JHEP01(2022)182}{\emph{JHEP} {\bfseries 01}
  (2022) 182} [\href{https://arxiv.org/abs/2107.03390}{{\ttfamily
  2107.03390}}].

\bibitem{Raju:2021lwh}
S.~Raju, \emph{{Failure of the split property in gravity and the information
  paradox}},  \href{https://arxiv.org/abs/2110.05470}{{\ttfamily 2110.05470}}.

\bibitem{Laddha:2020kvp}
A.~Laddha, S.G.~Prabhu, S.~Raju and P.~Shrivastava, \emph{{The Holographic
  Nature of Null Infinity}},
  \href{https://doi.org/10.21468/SciPostPhys.10.2.041}{\emph{SciPost Phys.}
  {\bfseries 10} (2021) 041}
  [\href{https://arxiv.org/abs/2002.02448}{{\ttfamily 2002.02448}}].

\bibitem{Nayak:2018qej}
P.~Nayak, A.~Shukla, R.M.~Soni, S.P.~Trivedi and V.~Vishal, \emph{{On the
  Dynamics of Near-Extremal Black Holes}},
  \href{https://doi.org/10.1007/JHEP09(2018)048}{\emph{JHEP} {\bfseries 09}
  (2018) 048} [\href{https://arxiv.org/abs/1802.09547}{{\ttfamily
  1802.09547}}].

\bibitem{Almheiri:2014cka}
A.~Almheiri and J.~Polchinski, \emph{{Models of AdS$_{2}$ backreaction and
  holography}}, \href{https://doi.org/10.1007/JHEP11(2015)014}{\emph{JHEP}
  {\bfseries 11} (2015) 014} [\href{https://arxiv.org/abs/1402.6334}{{\ttfamily
  1402.6334}}].

\bibitem{Saad:2021rcu}
P.~Saad, S.H.~Shenker, D.~Stanford and S.~Yao, \emph{{Wormholes without
  averaging}},  \href{https://arxiv.org/abs/2103.16754}{{\ttfamily
  2103.16754}}.

\bibitem{Caceres:2019giy}
E.~Caceres, A.~Kundu, A.K.~Patra and S.~Shashi, \emph{{A Killing Vector
  Treatment of Multiboundary Wormholes}},
  \href{https://doi.org/10.1007/JHEP02(2020)149}{\emph{JHEP} {\bfseries 02}
  (2020) 149} [\href{https://arxiv.org/abs/1912.08793}{{\ttfamily
  1912.08793}}].

\bibitem{Skenderis:2009ju}
K.~Skenderis and B.C.~van Rees, \emph{{Holography and wormholes in 2+1
  dimensions}}, \href{https://doi.org/10.1007/s00220-010-1163-z}{\emph{Commun.
  Math. Phys.} {\bfseries 301} (2011) 583}
  [\href{https://arxiv.org/abs/0912.2090}{{\ttfamily 0912.2090}}].

\bibitem{Balasubramanian:2014hda}
V.~Balasubramanian, P.~Hayden, A.~Maloney, D.~Marolf and S.F.~Ross,
  \emph{{Multiboundary Wormholes and Holographic Entanglement}},
  \href{https://doi.org/10.1088/0264-9381/31/18/185015}{\emph{Class. Quant.
  Grav.} {\bfseries 31} (2014) 185015}
  [\href{https://arxiv.org/abs/1406.2663}{{\ttfamily 1406.2663}}].

\bibitem{Aminneborg:1997pz}
S.~Aminneborg, I.~Bengtsson, D.~Brill, S.~Holst and P.~Peldan, \emph{{Black
  holes and wormholes in (2+1)-dimensions}},
  \href{https://doi.org/10.1088/0264-9381/15/3/013}{\emph{Class. Quant. Grav.}
  {\bfseries 15} (1998) 627}
  [\href{https://arxiv.org/abs/gr-qc/9707036}{{\ttfamily gr-qc/9707036}}].

\bibitem{Aminneborg:1998si}
S.~Aminneborg, I.~Bengtsson and S.~Holst, \emph{{A Spinning anti-de Sitter
  wormhole}}, \href{https://doi.org/10.1088/0264-9381/16/2/004}{\emph{Class.
  Quant. Grav.} {\bfseries 16} (1999) 363}
  [\href{https://arxiv.org/abs/gr-qc/9805028}{{\ttfamily gr-qc/9805028}}].

\bibitem{Krasnov:2001va}
K.~Krasnov, \emph{{Analytic continuation for asymptotically AdS 3-D gravity}},
  \href{https://doi.org/10.1088/0264-9381/19/9/306}{\emph{Class. Quant. Grav.}
  {\bfseries 19} (2002) 2399}
  [\href{https://arxiv.org/abs/gr-qc/0111049}{{\ttfamily gr-qc/0111049}}].

\bibitem{Visser:1995cc}
M.~Visser, \emph{{Lorentzian wormholes: From Einstein to Hawking}} (1995).

\bibitem{Stanford:2014jda}
D.~Stanford and L.~Susskind, \emph{{Complexity and Shock Wave Geometries}},
  \href{https://doi.org/10.1103/PhysRevD.90.126007}{\emph{Phys. Rev. D}
  {\bfseries 90} (2014) 126007}
  [\href{https://arxiv.org/abs/1406.2678}{{\ttfamily 1406.2678}}].

\bibitem{Susskind:2014rva}
L.~Susskind, \emph{{Computational Complexity and Black Hole Horizons}},
  \href{https://doi.org/10.1002/prop.201500092}{\emph{Fortsch. Phys.}
  {\bfseries 64} (2016) 24} [\href{https://arxiv.org/abs/1403.5695}{{\ttfamily
  1403.5695}}].

\bibitem{Brown:2015bva}
A.R.~Brown, D.A.~Roberts, L.~Susskind, B.~Swingle and Y.~Zhao,
  \emph{{Holographic Complexity Equals Bulk Action?}},
  \href{https://doi.org/10.1103/PhysRevLett.116.191301}{\emph{Phys. Rev. Lett.}
  {\bfseries 116} (2016) 191301}
  [\href{https://arxiv.org/abs/1509.07876}{{\ttfamily 1509.07876}}].

\bibitem{Brown:2015lvg}
A.R.~Brown, D.A.~Roberts, L.~Susskind, B.~Swingle and Y.~Zhao,
  \emph{{Complexity, action, and black holes}},
  \href{https://doi.org/10.1103/PhysRevD.93.086006}{\emph{Phys. Rev. D}
  {\bfseries 93} (2016) 086006}
  [\href{https://arxiv.org/abs/1512.04993}{{\ttfamily 1512.04993}}].

\bibitem{Chen:2021lnq}
B.~Chen, B.~Czech and Z.-z.~Wang, \emph{{Quantum Information in Holographic
  Duality}},  \href{https://arxiv.org/abs/2108.09188}{{\ttfamily 2108.09188}}.

\bibitem{Graham:2007va}
N.~Graham and K.D.~Olum, \emph{{Achronal averaged null energy condition}},
  \href{https://doi.org/10.1103/PhysRevD.76.064001}{\emph{Phys. Rev. D}
  {\bfseries 76} (2007) 064001}
  [\href{https://arxiv.org/abs/0705.3193}{{\ttfamily 0705.3193}}].

\bibitem{Gao:2016bin}
P.~Gao, D.L.~Jafferis and A.C.~Wall, \emph{{Traversable Wormholes via a Double
  Trace Deformation}},
  \href{https://doi.org/10.1007/JHEP12(2017)151}{\emph{JHEP} {\bfseries 12}
  (2017) 151} [\href{https://arxiv.org/abs/1608.05687}{{\ttfamily
  1608.05687}}].

\bibitem{Morris:1988cz}
M.S.~Morris and K.S.~Thorne, \emph{{Wormholes in space-time and their use for
  interstellar travel: A tool for teaching general relativity}},
  \href{https://doi.org/10.1119/1.15620}{\emph{Am. J. Phys.} {\bfseries 56}
  (1988) 395}.

\bibitem{Morris:1988tu}
M.S.~Morris, K.S.~Thorne and U.~Yurtsever, \emph{{Wormholes, Time Machines, and
  the Weak Energy Condition}},
  \href{https://doi.org/10.1103/PhysRevLett.61.1446}{\emph{Phys. Rev. Lett.}
  {\bfseries 61} (1988) 1446}.

\bibitem{Visser:1989kh}
M.~Visser, \emph{{Traversable wormholes: Some simple examples}},
  \href{https://doi.org/10.1103/PhysRevD.39.3182}{\emph{Phys. Rev. D}
  {\bfseries 39} (1989) 3182}
  [\href{https://arxiv.org/abs/0809.0907}{{\ttfamily 0809.0907}}].

\bibitem{Visser:1989kg}
M.~Visser, \emph{{Traversable wormholes from surgically modified Schwarzschild
  space-times}},
  \href{https://doi.org/10.1016/0550-3213(89)90100-4}{\emph{Nucl. Phys. B}
  {\bfseries 328} (1989) 203}
  [\href{https://arxiv.org/abs/0809.0927}{{\ttfamily 0809.0927}}].

\bibitem{Poisson:1995sv}
E.~Poisson and M.~Visser, \emph{{Thin shell wormholes: Linearization
  stability}}, \href{https://doi.org/10.1103/PhysRevD.52.7318}{\emph{Phys. Rev.
  D} {\bfseries 52} (1995) 7318}
  [\href{https://arxiv.org/abs/gr-qc/9506083}{{\ttfamily gr-qc/9506083}}].

\bibitem{Barcelo:2000zf}
C.~Barcelo and M.~Visser, \emph{{Scalar fields, energy conditions, and
  traversable wormholes}},
  \href{https://doi.org/10.1088/0264-9381/17/18/318}{\emph{Class. Quant. Grav.}
  {\bfseries 17} (2000) 3843}
  [\href{https://arxiv.org/abs/gr-qc/0003025}{{\ttfamily gr-qc/0003025}}].

\bibitem{Visser:2003yf}
M.~Visser, S.~Kar and N.~Dadhich, \emph{{Traversable wormholes with arbitrarily
  small energy condition violations}},
  \href{https://doi.org/10.1103/PhysRevLett.90.201102}{\emph{Phys. Rev. Lett.}
  {\bfseries 90} (2003) 201102}
  [\href{https://arxiv.org/abs/gr-qc/0301003}{{\ttfamily gr-qc/0301003}}].

\bibitem{Bhawal:1992sz}
B.~Bhawal and S.~Kar, \emph{{Lorentzian wormholes in Einstein-Gauss-Bonnet
  theory}}, \href{https://doi.org/10.1103/PhysRevD.46.2464}{\emph{Phys. Rev. D}
  {\bfseries 46} (1992) 2464}.

\bibitem{Thibeault:2005ha}
M.~Thibeault, C.~Simeone and E.F.~Eiroa, \emph{{Thin-shell wormholes in
  Einstein-Maxwell theory with a Gauss-Bonnet term}},
  \href{https://doi.org/10.1007/s10714-006-0324-z}{\emph{Gen. Rel. Grav.}
  {\bfseries 38} (2006) 1593}
  [\href{https://arxiv.org/abs/gr-qc/0512029}{{\ttfamily gr-qc/0512029}}].

\bibitem{Arias:2010xg}
R.E.~Arias, M.~Botta~Cantcheff and G.A.~Silva, \emph{{Lorentzian AdS, Wormholes
  and Holography}},
  \href{https://doi.org/10.1103/PhysRevD.83.066015}{\emph{Phys. Rev. D}
  {\bfseries 83} (2011) 066015}
  [\href{https://arxiv.org/abs/1012.4478}{{\ttfamily 1012.4478}}].

\bibitem{Chernicoff:2020tvr}
M.~Chernicoff, E.~Garc\'\i{}a, G.~Giribet and E.~Rub\'\i{}n~de Celis,
  \emph{{Thin-shell wormholes in AdS$_5$ and string dioptrics}},
  \href{https://doi.org/10.1007/JHEP10(2020)019}{\emph{JHEP} {\bfseries 10}
  (2020) 019} [\href{https://arxiv.org/abs/2006.07428}{{\ttfamily
  2006.07428}}].

\bibitem{Garcia-Garcia:2019poj}
A.M.~Garc\'\i{}a-Garc\'\i{}a, T.~Nosaka, D.~Rosa and J.J.M.~Verbaarschot,
  \emph{{Quantum chaos transition in a two-site Sachdev-Ye-Kitaev model dual to
  an eternal traversable wormhole}},
  \href{https://doi.org/10.1103/PhysRevD.100.026002}{\emph{Phys. Rev. D}
  {\bfseries 100} (2019) 026002}
  [\href{https://arxiv.org/abs/1901.06031}{{\ttfamily 1901.06031}}].

\bibitem{Garcia-Garcia:2020ttf}
A.M.~Garc\'\i{}a-Garc\'\i{}a and V.~Godet, \emph{{Euclidean wormhole in the
  Sachdev-Ye-Kitaev model}},
  \href{https://doi.org/10.1103/PhysRevD.103.046014}{\emph{Phys. Rev. D}
  {\bfseries 103} (2021) 046014}
  [\href{https://arxiv.org/abs/2010.11633}{{\ttfamily 2010.11633}}].

\bibitem{Freivogel:2019lej}
B.~Freivogel, V.~Godet, E.~Morvan, J.F.~Pedraza and A.~Rotundo, \emph{{Lessons
  on eternal traversable wormholes in AdS}},
  \href{https://doi.org/10.1007/JHEP07(2019)122}{\emph{JHEP} {\bfseries 07}
  (2019) 122} [\href{https://arxiv.org/abs/1903.05732}{{\ttfamily
  1903.05732}}].

\bibitem{Susskind:2017nto}
L.~Susskind and Y.~Zhao, \emph{{Teleportation through the wormhole}},
  \href{https://doi.org/10.1103/PhysRevD.98.046016}{\emph{Phys. Rev. D}
  {\bfseries 98} (2018) 046016}
  [\href{https://arxiv.org/abs/1707.04354}{{\ttfamily 1707.04354}}].

\bibitem{Brown:2019hmk}
A.R.~Brown, H.~Gharibyan, S.~Leichenauer, H.W.~Lin, S.~Nezami, G.~Salton
  et~al., \emph{{Quantum Gravity in the Lab: Teleportation by Size and
  Traversable Wormholes}},  \href{https://arxiv.org/abs/1911.06314}{{\ttfamily
  1911.06314}}.

\bibitem{Nezami:2021yaq}
S.~Nezami, H.W.~Lin, A.R.~Brown, H.~Gharibyan, S.~Leichenauer, G.~Salton
  et~al., \emph{{Quantum Gravity in the Lab: Teleportation by Size and
  Traversable Wormholes, Part II}},
  \href{https://arxiv.org/abs/2102.01064}{{\ttfamily 2102.01064}}.

\bibitem{Maldacena:2018gjk}
J.~Maldacena, A.~Milekhin and F.~Popov, \emph{{Traversable wormholes in four
  dimensions}},  \href{https://arxiv.org/abs/1807.04726}{{\ttfamily
  1807.04726}}.

\bibitem{Bintanja:2021xfs}
S.~Bintanja, R.~Esp\'\i{}ndola, B.~Freivogel and D.~Nikolakopoulou, \emph{{How
  to Make Traversable Wormholes: Eternal AdS$_4$ Wormholes from Coupled
  CFT's}},  \href{https://arxiv.org/abs/2102.06628}{{\ttfamily 2102.06628}}.

\bibitem{Emparan:2020ldj}
R.~Emparan, B.~Grado-White, D.~Marolf and M.~Tomasevic, \emph{{Multi-mouth
  Traversable Wormholes}},
  \href{https://doi.org/10.1007/JHEP05(2021)032}{\emph{JHEP} {\bfseries 05}
  (2021) 032} [\href{https://arxiv.org/abs/2012.07821}{{\ttfamily
  2012.07821}}].

\bibitem{AlBalushi:2020kso}
A.~Al~Balushi, Z.~Wang and D.~Marolf, \emph{{Traversability of Multi-Boundary
  Wormholes}}, \href{https://doi.org/10.1007/JHEP04(2021)083}{\emph{JHEP}
  {\bfseries 04} (2021) 083}
  [\href{https://arxiv.org/abs/2012.04635}{{\ttfamily 2012.04635}}].

\bibitem{Fu:2018oaq}
Z.~Fu, B.~Grado-White and D.~Marolf, \emph{{A perturbative perspective on
  self-supporting wormholes}},
  \href{https://doi.org/10.1088/1361-6382/aafcea}{\emph{Class. Quant. Grav.}
  {\bfseries 36} (2019) 045006}
  [\href{https://arxiv.org/abs/1807.07917}{{\ttfamily 1807.07917}}].

\bibitem{Caceres:2018ehr}
E.~Caceres, A.S.~Misobuchi and M.-L.~Xiao, \emph{{Rotating traversable
  wormholes in AdS}},
  \href{https://doi.org/10.1007/JHEP12(2018)005}{\emph{JHEP} {\bfseries 12}
  (2018) 005} [\href{https://arxiv.org/abs/1807.07239}{{\ttfamily
  1807.07239}}].

\bibitem{Marolf:2019ojx}
D.~Marolf and S.~McBride, \emph{{Simple Perturbatively Traversable Wormholes
  from Bulk Fermions}},
  \href{https://doi.org/10.1007/JHEP11(2019)037}{\emph{JHEP} {\bfseries 11}
  (2019) 037} [\href{https://arxiv.org/abs/1908.03998}{{\ttfamily
  1908.03998}}].

\bibitem{Fu:2019vco}
Z.~Fu, B.~Grado-White and D.~Marolf, \emph{{Traversable Asymptotically Flat
  Wormholes with Short Transit Times}},
  \href{https://doi.org/10.1088/1361-6382/ab56e4}{\emph{Class. Quant. Grav.}
  {\bfseries 36} (2019) 245018}
  [\href{https://arxiv.org/abs/1908.03273}{{\ttfamily 1908.03273}}].

\bibitem{Horowitz:2019hgb}
G.T.~Horowitz, D.~Marolf, J.E.~Santos and D.~Wang, \emph{{Creating a
  Traversable Wormhole}},
  \href{https://doi.org/10.1088/1361-6382/ab436f}{\emph{Class. Quant. Grav.}
  {\bfseries 36} (2019) 205011}
  [\href{https://arxiv.org/abs/1904.02187}{{\ttfamily 1904.02187}}].

\bibitem{Maldacena:2019ufo}
J.~Maldacena and A.~Milekhin, \emph{{SYK wormhole formation in real time}},
  \href{https://doi.org/10.1007/JHEP04(2021)258}{\emph{JHEP} {\bfseries 04}
  (2021) 258} [\href{https://arxiv.org/abs/1912.03276}{{\ttfamily
  1912.03276}}].

\bibitem{Maldacena:2017axo}
J.~Maldacena, D.~Stanford and Z.~Yang, \emph{{Diving into traversable
  wormholes}}, \href{https://doi.org/10.1002/prop.201700034}{\emph{Fortsch.
  Phys.} {\bfseries 65} (2017) 1700034}
  [\href{https://arxiv.org/abs/1704.05333}{{\ttfamily 1704.05333}}].

\bibitem{Freivogel:2019whb}
B.~Freivogel, D.A.~Galante, D.~Nikolakopoulou and A.~Rotundo,
  \emph{{Traversable wormholes in AdS and bounds on information transfer}},
  \href{https://doi.org/10.1007/JHEP01(2020)050}{\emph{JHEP} {\bfseries 01}
  (2020) 050} [\href{https://arxiv.org/abs/1907.13140}{{\ttfamily
  1907.13140}}].

\bibitem{Berenstein:2019yfv}
D.~Berenstein, \emph{{Quenches on thermofield double states and time reversal
  symmetry}}, \href{https://doi.org/10.1103/PhysRevD.100.066022}{\emph{Phys.
  Rev. D} {\bfseries 100} (2019) 066022}
  [\href{https://arxiv.org/abs/1906.08292}{{\ttfamily 1906.08292}}].

\bibitem{Fallows:2020ugr}
S.~Fallows and S.F.~Ross, \emph{{Making near-extremal wormholes traversable}},
  \href{https://doi.org/10.1007/JHEP12(2020)044}{\emph{JHEP} {\bfseries 12}
  (2020) 044} [\href{https://arxiv.org/abs/2008.07946}{{\ttfamily
  2008.07946}}].

\bibitem{Balasubramanian:2020ffd}
V.~Balasubramanian, M.~Decross and G.~S\'arosi, \emph{{Knitting Wormholes by
  Entanglement in Supergravity}},
  \href{https://doi.org/10.1007/JHEP11(2020)167}{\emph{JHEP} {\bfseries 11}
  (2020) 167} [\href{https://arxiv.org/abs/2009.08980}{{\ttfamily
  2009.08980}}].

\bibitem{Maldacena:2020sxe}
J.~Maldacena and A.~Milekhin, \emph{{Humanly traversable wormholes}},
  \href{https://doi.org/10.1103/PhysRevD.103.066007}{\emph{Phys. Rev. D}
  {\bfseries 103} (2021) 066007}
  [\href{https://arxiv.org/abs/2008.06618}{{\ttfamily 2008.06618}}].

\bibitem{Chen:2020tes}
Y.~Chen, V.~Gorbenko and J.~Maldacena, \emph{{Bra-ket wormholes in
  gravitationally prepared states}},
  \href{https://doi.org/10.1007/JHEP02(2021)009}{\emph{JHEP} {\bfseries 02}
  (2021) 009} [\href{https://arxiv.org/abs/2007.16091}{{\ttfamily
  2007.16091}}].

\bibitem{Karch:2007pd}
A.~Karch and A.~O'Bannon, \emph{{Metallic AdS/CFT}},
  \href{https://doi.org/10.1088/1126-6708/2007/09/024}{\emph{JHEP} {\bfseries
  09} (2007) 024} [\href{https://arxiv.org/abs/0705.3870}{{\ttfamily
  0705.3870}}].

\bibitem{Albash:2007bq}
T.~Albash, V.G.~Filev, C.V.~Johnson and A.~Kundu, \emph{{Quarks in an external
  electric field in finite temperature large N gauge theory}},
  \href{https://doi.org/10.1088/1126-6708/2008/08/092}{\emph{JHEP} {\bfseries
  08} (2008) 092} [\href{https://arxiv.org/abs/0709.1554}{{\ttfamily
  0709.1554}}].

\bibitem{Alam:2012fw}
M.S.~Alam, V.S.~Kaplunovsky and A.~Kundu, \emph{{Chiral Symmetry Breaking and
  External Fields in the Kuperstein-Sonnenschein Model}},
  \href{https://doi.org/10.1007/JHEP04(2012)111}{\emph{JHEP} {\bfseries 04}
  (2012) 111} [\href{https://arxiv.org/abs/1202.3488}{{\ttfamily 1202.3488}}].

\bibitem{Kim:2011qh}
K.-Y.~Kim, J.P.~Shock and J.~Tarrio, \emph{{The open string membrane paradigm
  with external electromagnetic fields}},
  \href{https://doi.org/10.1007/JHEP06(2011)017}{\emph{JHEP} {\bfseries 06}
  (2011) 017} [\href{https://arxiv.org/abs/1103.4581}{{\ttfamily 1103.4581}}].

\bibitem{Sonner:2012if}
J.~Sonner and A.G.~Green, \emph{{Hawking Radiation and Non-equilibrium Quantum
  Critical Current Noise}},
  \href{https://doi.org/10.1103/PhysRevLett.109.091601}{\emph{Phys. Rev. Lett.}
  {\bfseries 109} (2012) 091601}
  [\href{https://arxiv.org/abs/1203.4908}{{\ttfamily 1203.4908}}].

\bibitem{Kundu:2013eba}
A.~Kundu and S.~Kundu, \emph{{Steady-state Physics, Effective Temperature
  Dynamics in Holography}},
  \href{https://doi.org/10.1103/PhysRevD.91.046004}{\emph{Phys. Rev. D}
  {\bfseries 91} (2015) 046004}
  [\href{https://arxiv.org/abs/1307.6607}{{\ttfamily 1307.6607}}].

\bibitem{Banerjee:2015cvy}
A.~Banerjee, A.~Kundu and S.~Kundu, \emph{{Flavour Fields in Steady State:
  Stress Tensor and Free Energy}},
  \href{https://doi.org/10.1007/JHEP02(2016)102}{\emph{JHEP} {\bfseries 02}
  (2016) 102} [\href{https://arxiv.org/abs/1512.05472}{{\ttfamily
  1512.05472}}].

\bibitem{Kundu:2015qda}
A.~Kundu, \emph{{Effective Temperature in Steady-state Dynamics from
  Holography}}, \href{https://doi.org/10.1007/JHEP09(2015)042}{\emph{JHEP}
  {\bfseries 09} (2015) 042}
  [\href{https://arxiv.org/abs/1507.00818}{{\ttfamily 1507.00818}}].

\bibitem{Banerjee:2016qeu}
A.~Banerjee, A.~Kundu and S.~Kundu, \emph{{Emergent Horizons and Causal
  Structures in Holography}},
  \href{https://doi.org/10.1007/JHEP09(2016)166}{\emph{JHEP} {\bfseries 09}
  (2016) 166} [\href{https://arxiv.org/abs/1605.07368}{{\ttfamily
  1605.07368}}].

\bibitem{Kundu:2018sof}
A.~Kundu, \emph{{Effective Thermal Physics in Holography: A Brief Review}},
  \href{https://arxiv.org/abs/1812.09447}{{\ttfamily 1812.09447}}.

\bibitem{Kundu:2019ull}
A.~Kundu, \emph{{Steady States, Thermal Physics, and Holography}},
  \href{https://doi.org/10.1155/2019/2635917}{\emph{Adv. High Energy Phys.}
  {\bfseries 2019} (2019) 2635917}.

\bibitem{Karch:2002sh}
A.~Karch and E.~Katz, \emph{{Adding flavor to AdS / CFT}},
  \href{https://doi.org/10.1088/1126-6708/2002/06/043}{\emph{JHEP} {\bfseries
  06} (2002) 043} [\href{https://arxiv.org/abs/hep-th/0205236}{{\ttfamily
  hep-th/0205236}}].

\bibitem{Marolf:2003vf}
D.~Marolf, L.~Martucci and P.J.~Silva, \emph{{Actions and Fermionic symmetries
  for D-branes in bosonic backgrounds}},
  \href{https://doi.org/10.1088/1126-6708/2003/07/019}{\emph{JHEP} {\bfseries
  07} (2003) 019} [\href{https://arxiv.org/abs/hep-th/0306066}{{\ttfamily
  hep-th/0306066}}].

\bibitem{Sonner:2013mba}
J.~Sonner, \emph{{Holographic Schwinger Effect and the Geometry of
  Entanglement}},
  \href{https://doi.org/10.1103/PhysRevLett.111.211603}{\emph{Phys. Rev. Lett.}
  {\bfseries 111} (2013) 211603}
  [\href{https://arxiv.org/abs/1307.6850}{{\ttfamily 1307.6850}}].

\bibitem{Jensen:2014bpa}
K.~Jensen, A.~Karch and B.~Robinson, \emph{{Holographic dual of a Hawking pair
  has a wormhole}},
  \href{https://doi.org/10.1103/PhysRevD.90.064019}{\emph{Phys. Rev. D}
  {\bfseries 90} (2014) 064019}
  [\href{https://arxiv.org/abs/1405.2065}{{\ttfamily 1405.2065}}].

\bibitem{Jensen:2014lua}
K.~Jensen and J.~Sonner, \emph{{Wormholes and entanglement in holography}},
  \href{https://doi.org/10.1142/S0218271814420036}{\emph{Int. J. Mod. Phys. D}
  {\bfseries 23} (2014) 1442003}
  [\href{https://arxiv.org/abs/1405.4817}{{\ttfamily 1405.4817}}].

\bibitem{deBoer:2017xdk}
J.~de~Boer, E.~Llabr\'es, J.F.~Pedraza and D.~Vegh, \emph{{Chaotic strings in
  AdS/CFT}}, \href{https://doi.org/10.1103/PhysRevLett.120.201604}{\emph{Phys.
  Rev. Lett.} {\bfseries 120} (2018) 201604}
  [\href{https://arxiv.org/abs/1709.01052}{{\ttfamily 1709.01052}}].

\bibitem{Murata:2017rbp}
K.~Murata, \emph{{Fast scrambling in holographic Einstein-Podolsky-Rosen
  pair}}, \href{https://doi.org/10.1007/JHEP11(2017)049}{\emph{JHEP} {\bfseries
  11} (2017) 049} [\href{https://arxiv.org/abs/1708.09493}{{\ttfamily
  1708.09493}}].

\bibitem{Banerjee:2018twd}
A.~Banerjee, A.~Kundu and R.R.~Poojary, \emph{{Strings, Branes, Schwarzian
  Action and Maximal Chaos}},
  \href{https://arxiv.org/abs/1809.02090}{{\ttfamily 1809.02090}}.

\bibitem{Banerjee:2018kwy}
A.~Banerjee, A.~Kundu and R.~Poojary, \emph{{Maximal Chaos from Strings, Branes
  and Schwarzian Action}},
  \href{https://doi.org/10.1007/JHEP06(2019)076}{\emph{JHEP} {\bfseries 06}
  (2019) 076} [\href{https://arxiv.org/abs/1811.04977}{{\ttfamily
  1811.04977}}].

\bibitem{Banerjee:2019vff}
A.~Banerjee, A.~Kundu and R.R.~Poojary, \emph{{Rotating black holes in AdS
  spacetime, extremality, and chaos}},
  \href{https://doi.org/10.1103/PhysRevD.102.106013}{\emph{Phys. Rev. D}
  {\bfseries 102} (2020) 106013}
  [\href{https://arxiv.org/abs/1912.12996}{{\ttfamily 1912.12996}}].

\bibitem{Gao:2018yzk}
P.~Gao and H.~Liu, \emph{{Regenesis and quantum traversable wormholes}},
  \href{https://doi.org/10.1007/JHEP10(2019)048}{\emph{JHEP} {\bfseries 10}
  (2019) 048} [\href{https://arxiv.org/abs/1810.01444}{{\ttfamily
  1810.01444}}].

\bibitem{Lieb:1972wy}
E.H.~Lieb and D.W.~Robinson, \emph{{The finite group velocity of quantum spin
  systems}}, \href{https://doi.org/10.1007/BF01645779}{\emph{Commun. Math.
  Phys.} {\bfseries 28} (1972) 251}.

\bibitem{Giddings:2020yes}
S.B.~Giddings and G.J.~Turiaci, \emph{{Wormhole calculus, replicas, and
  entropies}}, \href{https://doi.org/10.1007/JHEP09(2020)194}{\emph{JHEP}
  {\bfseries 09} (2020) 194}
  [\href{https://arxiv.org/abs/2004.02900}{{\ttfamily 2004.02900}}].

\bibitem{Marolf:2020xie}
D.~Marolf and H.~Maxfield, \emph{{Transcending the ensemble: baby universes,
  spacetime wormholes, and the order and disorder of black hole information}},
  \href{https://doi.org/10.1007/JHEP08(2020)044}{\emph{JHEP} {\bfseries 08}
  (2020) 044} [\href{https://arxiv.org/abs/2002.08950}{{\ttfamily
  2002.08950}}].

\bibitem{Belin:2020hea}
A.~Belin and J.~de~Boer, \emph{{Random statistics of OPE coefficients and
  Euclidean wormholes}},
  \href{https://doi.org/10.1088/1361-6382/ac1082}{\emph{Class. Quant. Grav.}
  {\bfseries 38} (2021) 164001}
  [\href{https://arxiv.org/abs/2006.05499}{{\ttfamily 2006.05499}}].

\bibitem{Belin:2020jxr}
A.~Belin, J.~De~Boer, P.~Nayak and J.~Sonner, \emph{{Charged Eigenstate
  Thermalization, Euclidean Wormholes and Global Symmetries in Quantum
  Gravity}},  \href{https://arxiv.org/abs/2012.07875}{{\ttfamily 2012.07875}}.

\bibitem{Betzios:2019rds}
P.~Betzios, E.~Kiritsis and O.~Papadoulaki, \emph{{Euclidean Wormholes and
  Holography}}, \href{https://doi.org/10.1007/JHEP06(2019)042}{\emph{JHEP}
  {\bfseries 06} (2019) 042}
  [\href{https://arxiv.org/abs/1903.05658}{{\ttfamily 1903.05658}}].

\bibitem{Betzios:2021fnm}
P.~Betzios, E.~Kiritsis and O.~Papadoulaki, \emph{{Interacting systems and
  wormholes}}, \href{https://doi.org/10.1007/JHEP02(2022)126}{\emph{JHEP}
  {\bfseries 02} (2022) 126}
  [\href{https://arxiv.org/abs/2110.14655}{{\ttfamily 2110.14655}}].

\bibitem{VanRaamsdonk:2020tlr}
M.~Van~Raamsdonk, \emph{{Comments on wormholes, ensembles, and cosmology}},
  \href{https://doi.org/10.1007/JHEP12(2021)156}{\emph{JHEP} {\bfseries 12}
  (2021) 156} [\href{https://arxiv.org/abs/2008.02259}{{\ttfamily
  2008.02259}}].

\bibitem{VanRaamsdonk:2021qgv}
M.~Van~Raamsdonk, \emph{{Cosmology from confinement?}},
  \href{https://doi.org/10.1007/JHEP03(2022)039}{\emph{JHEP} {\bfseries 03}
  (2022) 039} [\href{https://arxiv.org/abs/2102.05057}{{\ttfamily
  2102.05057}}].

\bibitem{Iqbal:2021ouw}
N.~Iqbal and S.F.~Ross, \emph{{Towards traversable wormholes from force-free
  plasmas}}, \href{https://doi.org/10.21468/SciPostPhys.12.3.086}{\emph{SciPost
  Phys.} {\bfseries 12} (2022) 086}
  [\href{https://arxiv.org/abs/2103.01920}{{\ttfamily 2103.01920}}].

\bibitem{Verlinde:2021kgt}
H.~Verlinde, \emph{{Wormholes in Quantum Mechanics}},
  \href{https://arxiv.org/abs/2105.02129}{{\ttfamily 2105.02129}}.

\bibitem{Verlinde:2021jwu}
H.~Verlinde, \emph{{Deconstructing the Wormhole: Factorization, Entanglement
  and Decoherence}},  \href{https://arxiv.org/abs/2105.02142}{{\ttfamily
  2105.02142}}.

\bibitem{Johnson:2019eik}
C.V.~Johnson, \emph{{Nonperturbative Jackiw-Teitelboim gravity}},
  \href{https://doi.org/10.1103/PhysRevD.101.106023}{\emph{Phys. Rev. D}
  {\bfseries 101} (2020) 106023}
  [\href{https://arxiv.org/abs/1912.03637}{{\ttfamily 1912.03637}}].

\bibitem{Johnson:2020heh}
C.V.~Johnson, \emph{{Jackiw-Teitelboim supergravity, minimal strings, and
  matrix models}},
  \href{https://doi.org/10.1103/PhysRevD.103.046012}{\emph{Phys. Rev. D}
  {\bfseries 103} (2021) 046012}
  [\href{https://arxiv.org/abs/2005.01893}{{\ttfamily 2005.01893}}].

\bibitem{Johnson:2020exp}
C.V.~Johnson, \emph{{Explorations of nonperturbative Jackiw-Teitelboim gravity
  and supergravity}},
  \href{https://doi.org/10.1103/PhysRevD.103.046013}{\emph{Phys. Rev. D}
  {\bfseries 103} (2021) 046013}
  [\href{https://arxiv.org/abs/2006.10959}{{\ttfamily 2006.10959}}].

\bibitem{Johnson:2020mwi}
C.V.~Johnson, \emph{{Low Energy Thermodynamics of JT Gravity and
  Supergravity}},  \href{https://arxiv.org/abs/2008.13120}{{\ttfamily
  2008.13120}}.

\bibitem{Johnson:2020lns}
C.V.~Johnson and F.~Rosso, \emph{{Solving Puzzles in Deformed JT Gravity: Phase
  Transitions and Non-Perturbative Effects}},
  \href{https://doi.org/10.1007/JHEP04(2021)030}{\emph{JHEP} {\bfseries 04}
  (2021) 030} [\href{https://arxiv.org/abs/2011.06026}{{\ttfamily
  2011.06026}}].

\bibitem{Johnson:2021owr}
C.V.~Johnson, F.~Rosso and A.~Svesko, \emph{{Jackiw-Teitelboim supergravity as
  a double-cut matrix model}},
  \href{https://doi.org/10.1103/PhysRevD.104.086019}{\emph{Phys. Rev. D}
  {\bfseries 104} (2021) 086019}
  [\href{https://arxiv.org/abs/2102.02227}{{\ttfamily 2102.02227}}].

\bibitem{Johnson:2021rsh}
C.V.~Johnson, \emph{{On the Quenched Free Energy of JT Gravity and
  Supergravity}},  \href{https://arxiv.org/abs/2104.02733}{{\ttfamily
  2104.02733}}.

\bibitem{Johnson:2021zuo}
C.V.~Johnson, \emph{{Quantum Gravity Microstates from Fredholm Determinants}},
  \href{https://doi.org/10.1103/PhysRevLett.127.181602}{\emph{Phys. Rev. Lett.}
  {\bfseries 127} (2021) 181602}
  [\href{https://arxiv.org/abs/2106.09048}{{\ttfamily 2106.09048}}].

\bibitem{Johnson:2021tnl}
C.V.~Johnson, \emph{{Consistency Conditions for Non-Perturbative Completions of
  JT Gravity}},  \href{https://arxiv.org/abs/2112.00766}{{\ttfamily
  2112.00766}}.

\bibitem{Johnson:2022wsr}
C.V.~Johnson, \emph{{The Microstate Physics of JT Gravity and Supergravity}},
  \href{https://arxiv.org/abs/2201.11942}{{\ttfamily 2201.11942}}.

\bibitem{Cotler:2020ugk}
J.~Cotler and K.~Jensen, \emph{{AdS$_{3}$ gravity and random CFT}},
  \href{https://doi.org/10.1007/JHEP04(2021)033}{\emph{JHEP} {\bfseries 04}
  (2021) 033} [\href{https://arxiv.org/abs/2006.08648}{{\ttfamily
  2006.08648}}].

\bibitem{Cotler:2020hgz}
J.~Cotler and K.~Jensen, \emph{{AdS$_3$ wormholes from a modular bootstrap}},
  \href{https://doi.org/10.1007/JHEP11(2020)058}{\emph{JHEP} {\bfseries 11}
  (2020) 058} [\href{https://arxiv.org/abs/2007.15653}{{\ttfamily
  2007.15653}}].

\bibitem{Cotler:2020lxj}
J.~Cotler and K.~Jensen, \emph{{Gravitational Constrained Instantons}},
  \href{https://arxiv.org/abs/2010.02241}{{\ttfamily 2010.02241}}.

\bibitem{Cotler:2021cqa}
J.~Cotler and K.~Jensen, \emph{{Wormholes and black hole microstates in
  AdS/CFT}}, \href{https://doi.org/10.1007/JHEP09(2021)001}{\emph{JHEP}
  {\bfseries 09} (2021) 001}
  [\href{https://arxiv.org/abs/2104.00601}{{\ttfamily 2104.00601}}].

\bibitem{Maldacena:2004rf}
J.M.~Maldacena and L.~Maoz, \emph{{Wormholes in AdS}},
  \href{https://doi.org/10.1088/1126-6708/2004/02/053}{\emph{JHEP} {\bfseries
  02} (2004) 053} [\href{https://arxiv.org/abs/hep-th/0401024}{{\ttfamily
  hep-th/0401024}}].

\bibitem{Bao:2018msr}
N.~Bao, A.~Chatwin-Davies, J.~Pollack and G.N.~Remmen, \emph{{Traversable
  Wormholes as Quantum Channels: Exploring CFT Entanglement Structure and
  Channel Capacity in Holography}},
  \href{https://doi.org/10.1007/JHEP11(2018)071}{\emph{JHEP} {\bfseries 11}
  (2018) 071} [\href{https://arxiv.org/abs/1808.05963}{{\ttfamily
  1808.05963}}].

\bibitem{vanBreukelen:2017dul}
R.~van Breukelen and K.~Papadodimas, \emph{{Quantum teleportation through
  time-shifted AdS wormholes}},
  \href{https://doi.org/10.1007/JHEP08(2018)142}{\emph{JHEP} {\bfseries 08}
  (2018) 142} [\href{https://arxiv.org/abs/1708.09370}{{\ttfamily
  1708.09370}}].

\bibitem{Bak:2018txn}
D.~Bak, C.~Kim and S.-H.~Yi, \emph{{Bulk view of teleportation and traversable
  wormholes}}, \href{https://doi.org/10.1007/JHEP08(2018)140}{\emph{JHEP}
  {\bfseries 08} (2018) 140}
  [\href{https://arxiv.org/abs/1805.12349}{{\ttfamily 1805.12349}}].

\bibitem{Gao:2019nyj}
P.~Gao and D.L.~Jafferis, \emph{{A traversable wormhole teleportation protocol
  in the SYK model}},
  \href{https://doi.org/10.1007/JHEP07(2021)097}{\emph{JHEP} {\bfseries 07}
  (2021) 097} [\href{https://arxiv.org/abs/1911.07416}{{\ttfamily
  1911.07416}}].

\bibitem{Caceres:2021nsa}
E.~Caceres, A.~Misobuchi and R.~Pimentel, \emph{{Sparse SYK and traversable
  wormholes}},  \href{https://arxiv.org/abs/2108.08808}{{\ttfamily
  2108.08808}}.

\bibitem{Landsman:2018jpm}
K.A.~Landsman, C.~Figgatt, T.~Schuster, N.M.~Linke, B.~Yoshida, N.Y.~Yao
  et~al., \emph{{Verified Quantum Information Scrambling}},
  \href{https://doi.org/10.1038/s41586-019-0952-6}{\emph{Nature} {\bfseries
  567} (2019) 61} [\href{https://arxiv.org/abs/1806.02807}{{\ttfamily
  1806.02807}}].

\bibitem{Dunne:2015eaa}
G.V.~Dunne and M.~\"Unsal, \emph{{What is QFT? Resurgent trans-series,
  Lefschetz thimbles, and new exact saddles}},
  \href{https://doi.org/10.22323/1.251.0010}{\emph{PoS} {\bfseries LATTICE2015}
  (2016) 010} [\href{https://arxiv.org/abs/1511.05977}{{\ttfamily
  1511.05977}}].

\end{thebibliography}\endgroup
\bibliographystyle{JHEP}

\end{document}